\documentclass[conference, draftcls, onecolumn]{IEEEtran}

\usepackage{ifpdf}

\usepackage[noadjust]{cite}

\ifCLASSINFOpdf
  \usepackage[pdftex]{graphicx}
\else
  \usepackage[dvips]{graphicx}
\fi

\usepackage{amsmath}
\usepackage{array}

\ifCLASSOPTIONcompsoc
  \usepackage[caption=false,font=normalsize,labelfont=sf,textfont=sf]{subfig}
\else
  \usepackage[caption=false,font=footnotesize]{subfig}
\fi

\usepackage{fixltx2e}

\usepackage{url}

\usepackage{mathtools}
\usepackage{amssymb}
\usepackage{amsthm}
\theoremstyle{plain}
\newtheorem{definition}{Definition}
\newtheorem{lemma}{Lemma}

\newtheorem{theorem}{Theorem}
\newtheorem{corollary}{Corollary}

\newtheorem{fact}{Fact}

\usepackage{xcolor}
\definecolor{burgundy}{rgb}{0.545098,0,0}
\definecolor{navyblue}{rgb}{0.0, 0.0, 0.5}
\definecolor{leafgreen}{rgb}{0.290196, 0.470588, 0.0}
\definecolor{bluegreen}{rgb}{0, 0.470588, 0.415686}
\definecolor{zuhl}{rgb}{0.1875, 0.26171875, 0.46484375}
\definecolor{orange}{rgb}{1, 0.6470588235, 0}
\definecolor{red}{rgb}{1, 0, 0}

\usepackage{overpic}
\usepackage{pict2e}

\newcommand{\bvec}[1]{\boldsymbol{#1}}

\newcommand{\sgn}{\operatorname{sgn}}
\newcommand{\supp}{\operatorname{supp}}

\newcommand{\figref}[1]{Fig.~\ref{#1}}
\newcommand{\factref}[1]{Fact~\ref{#1}}
\newcommand{\lemref}[1]{Lemma~\ref{#1}}
\newcommand{\thref}[1]{Theorem~\ref{#1}}
\newcommand{\defref}[1]{Definition~\ref{#1}}
\newcommand{\corref}[1]{Corollary~\ref{#1}}
\newcommand{\sectref}[1]{Section~\ref{#1}}

\newcommand{\appref}[1]{Appendix~\ref{#1}}

\allowdisplaybreaks[3]

\usepackage{hyperref}
\hypersetup{
    colorlinks,
    linkcolor={red!60!black},
    citecolor={blue!60!black},
    urlcolor={blue!50!black}
}
\usepackage{microtype}

\begin{document}

\title{Sharp Bounds on Arimoto's Conditional \\ R\'{e}nyi Entropies Between Two Distinct Orders}

\IEEEoverridecommandlockouts

\author{%
\IEEEauthorblockN{%
Yuta~Sakai and Ken-ichi~Iwata%
\thanks{This work was partially supported by the Ministry of Education, Science, Sports and Culture, Grant-in-Aid for Scientific Research through the Japan Society for the Promotion of Science under Grant 26420352.}%
}
\IEEEauthorblockA{Graduate School of Engineering,
University of Fukui,
Japan,
Email: \{y-sakai, k-iwata\}@u-fukui.ac.jp}%
}

\maketitle

\begin{abstract}
This study examines sharp bounds on Arimoto's conditional R\'{e}nyi entropy of order $\beta$ with a fixed another one of distinct order $\alpha \neq \beta$.
Arimoto inspired the relation between the R\'{e}nyi entropy and the $\ell_{r}$-norm of probability distributions, and he introduced a conditional version of the R\'{e}nyi entropy.
From this perspective, we analyze the $\ell_{r}$-norms of particular distributions. As results, we identify specific probability distributions whose achieve our sharp bounds on the conditional R\'{e}nyi entropy.
The sharp bounds derived in this study can be applicable to other information measures, e.g., the minimum average probability of error, the Bhattacharyya parameter, Gallager's reliability function $E_{0}$, and Sibson's $\alpha$-mutual information, whose are strictly monotone functions of the conditional R\'{e}nyi entropy.
\end{abstract}

\IEEEpeerreviewmaketitle

\section{Introduction}

In information theory, the Shannon entropy $H(X)$ and the conditional Shannon entropy $H(X \mid Y)$ \cite{shannon} are traditional information measures of random variables (RVs) $X$ and $Y$, whose characterize several theoretical limits for information transmission.
Later, the R\'{e}nyi entropy $H_{\alpha}( X )$ \cite{renyi} was axiomatically proposed as a generalized Shannon entropy with order $\alpha$.
For a discrete RV%
\footnote{An RV $X$ with distribution $P$ is denoted by $X \sim P$.}
$X \sim P$, the R\'{e}nyi entropy of order $\alpha \in [0, \infty]$ is defined by
\begin{align}
H_{\alpha}( X )
=
H_{\alpha}( P )
\coloneqq
\lim_{r \to \alpha} \frac{ r }{ 1 - r } \ln \| P \|_{r} ,
\label{def:renyi}
\end{align}
where $\ln$ denotes the natural logarithm, the $\ell_{r}$-norm of a discrete probability distribution $P$ is defined by
\begin{align}
\| P \|_{r}
\coloneqq
\bigg( \sum_{x \in \supp(P)} P( x )^{r} \bigg)^{1 / r}
\label{def:norm}
\end{align}
for $r \in \mathbb{R}$, and $\supp( P ) \coloneqq \{ x \in \mathcal{X} \mid P( x ) > 0 \}$ denotes the support of a distribution $P$ on a countable alphabet $\mathcal{X}$.
Note that \eqref{def:renyi} is well-defined since the limiting value exists for each $\alpha \in [0, \infty]$ as follows:
\begin{align}
H_{\alpha}( X )
& =
\frac{ \alpha }{ 1 - \alpha } \ln \| P \|_{\alpha}
\qquad \mathrm{for} \ \alpha \in (0, 1) \cup (1, \infty) ,
\\
H_{0}( X )
&
=
\ln | \! \supp( P ) | ,
\label{eq:max-entropy} \\
H_{1}( X )
& =
\mathbb{E}[ - \ln P( X ) ]
\eqqcolon
H( X ) ,
\label{def:shannon} \\
H_{\infty}( X)
& =
- \ln \| P \|_{\infty} ,
\label{def:min_entropy}
\end{align}
where $| \cdot |$ denotes the cardinality%
\footnote{In this study, suppose that $| \mathcal{S} | = \infty$ if $\mathcal{S}$ is a countably infinite set; thus, note in \eqref{eq:max-entropy} that $H_{0}(X) = \infty$ if $\supp( P )$ is countably infinite.}
of the countable set, $\mathbb{E}[ \cdot ]$ denotes the expectation of the RV, and $\| P \|_{\infty} \coloneqq \lim_{r \to \infty} \| P \|_{r} = \max_{x \in \supp(P)} P(x)$ denotes the $\ell_{\infty}$-norm of $P$.
Moreover, Arimoto \cite{arimoto} proposed a conditional version%
\footnote{There are many definition of conditional R\'{e}nyi entropy (cf. \cite{fehr, teixeira, tomamichel}).
In this paper, the conditional R\'{e}nyi entropy means Arimoto's definition unless otherwise noted.}
of the R\'{e}nyi entropy $H_{\alpha}(X \mid Y)$ as a generalized conditional Shannon entropy with order $\alpha$.
For a pair of RVs $(X, Y) \sim P_{X|Y} P_{Y}$, the conditional R\'{e}nyi entropy \cite{arimoto} of order $\alpha \in [0, \infty]$ is defined by
\begin{align}
H_{\alpha}(X \mid Y)
\coloneqq
\lim_{r \to \alpha} \frac{ r }{ 1 - r } \ln N_{r}(X \mid Y) ,
\label{def:A_renyi}
\end{align}
where the expectation of $\ell_{r}$-norm is denoted by
\begin{align}
N_{r}(X \mid Y)
\coloneqq
\mathbb{E} \big[ \| P_{X|Y}(\cdot \mid Y) \|_{r} \big]
\label{def:expect_norm}
\end{align}
for $r \in (0, \infty]$, and note in \eqref{def:expect_norm} that $Y \sim P_{Y}$.
In this study, the RV $Y$ can be considered to be either discrete or continuous.
By convention, we write $H_{\alpha}(X \mid Y = y) \coloneqq H_{\alpha}( P_{X|Y}(\cdot \mid y) )$ for $y \in \supp(P_{Y})$.
As with the unconditional R\'{e}nyi entropy \eqref{def:renyi}, note that \eqref{def:A_renyi} is also well-defined since the limiting value also exists for each $\alpha \in [0, \infty]$ as follows:%
\footnote{Proofs of \eqref{eq:cond_0}--\eqref{eq:cond_infty} can be found in, e.g., \cite[Propositions~1 and~2]{fehr}.}
\begin{align}
H_{\alpha}(X \mid Y)
& =
\frac{ \alpha }{ 1 - \alpha } \ln N_{\alpha}(X \mid Y)
\quad \mathrm{for} \ \alpha \in (0, 1) \cup (1, \infty) ,
\\
H_{0}(X \mid Y)
& =
\sup_{y \in \supp(P_{Y})} H_{0}(X \mid Y = y) ,
\label{eq:cond_0} \\
H_{1}(X \mid Y)
& =
\mathbb{E}[ - \ln P_{X|Y}(X \mid Y) ]
\eqqcolon
H(X \mid Y) ,
\label{def:cond_shannon} \\
H_{\infty}(X \mid Y)
& =
- \ln N_{\infty}(X \mid Y) .
\label{eq:cond_infty}
\end{align}
Note that Arimoto \cite{arimoto} proposed $H_{\alpha}(X \mid Y)$ in terms of the relation between the $\ell_{r}$-norm and the unconditional R\'{e}nyi entropy, shown in \eqref{def:renyi} (see also \cite[Section~II-A]{ita}).
As shown in \eqref{def:shannon} and \eqref{def:cond_shannon}, R\'{e}nyi's information measures can be reduced to Shannon's information measures as $\alpha \to 1$.
In many situations, R\'{e}nyi's information measures derive stronger results than Shannon's information measures (cf. \cite{arikan, bunte, campbell, csiszar, tomamichel}).
In addition, the quantity $H_{\alpha}(X \mid Y)$ is closely related to Gallager's reliability function $E_{0}$ \cite[Eq.~(5.6.14)]{gallager} and Sibson's $\alpha$-mutual information \cite{sibson, ho}; and thus, coding theorems with them can be written by $H_{\alpha}(X \mid Y)$ (cf. \cite{arimoto, ita}).
Many basic properties of $H_{\alpha}(X \mid Y)$ were studied by Fehr and Berens \cite{fehr}.

Bounds on information measures are crucial tools in several engineering fields, e.g., information theory, coding theory, cryptology, machine learning, statistics, etc.
In this paper, a bound is said to be \emph{sharp} if there is no tighter bound than it in the same situation.
One of well-known sharp bounds is Fano's inequality \cite{fano}, which bounds the conditional Shannon entropy $H(X \mid Y)$ from above for fixed (i) average probability of error $\Pr(X \neq f(Y))$ and (ii) size of support $|\!\supp( P_{X} )|$, where the function $f$ is an estimator of $X$ given $Y$.
As related bounds, the reverse of Fano's inequality, i.e., sharp lower bounds on $H(X \mid Y)$ with a fixed minimum average probability of error
\begin{align}
P_{\mathrm{e}}(X \mid Y)
\coloneqq
\min_{f} \Pr(X \neq f(Y)) ,
\label{def:Pe}
\end{align}
were established by Kovalevsky \cite{kovalevsky} and Tebbe and Dwyer \cite{tebbe} (see also \cite{feder}).
Ho and Verd\'{u} \cite{verdu} generalized Fano's inequality by relaxing its constraints from fixed number $|\!\supp( P_{X} )|$ to fixed distribution $P_{X}$.
Very recently, Sason and Verd\'{u} \cite{sason} generalized Fano's inequality and the reverse of it to sharp bounds on the conditional R\'{e}nyi entropy $H_{\alpha}(X \mid Y)$.
In their study \cite{sason}, interplay between $H_{\alpha}(X \mid Y)$ and $P_{\mathrm{e}}(X \mid Y)$ was investigated with broad applications and comparisons to related works.
On the other hand, we \cite{part2} derived sharp bounds on $H(X \mid Y)$ with a fixed $H_{\alpha}(X \mid Y)$, and vice versa, by analyzing interplay between $H(X \mid Y)$ and $N_{r}(X \mid Y)$ (cf. \eqref{def:A_renyi}).
Unconditional versions of its results \cite{part2} were also examined in \cite{part1}.

In this study, we further generalize interplay between Shannon's information measure and R\'{e}nyi's information measure of our results \cite{part1, part2} to interplay between two R\'{e}nyi's information measures with distinct orders $\alpha \neq \beta$.
We start to analyze extremal probability distributions $\bvec{v}_{n}( \cdot )$ and $\bvec{w}( \cdot )$ defined in \sectref{sect:v_w}, where the extremal distributions means that our sharp bounds on $H_{\alpha}(X)$ can be achieved by them (cf. \sectref{sect:unconditional}).
To utilize the nature of the expectation $N_{r}(X \mid Y)$ of $\ell_{r}$-norm, our analyses of this study are concentrated on the $\ell_{r}$-norm of extremal distributions.
Main results of this study are shown in \sectref{sect:cond}, which show sharp bounds on $H_{\beta}(X \mid Y)$ with a fixed another one $H_{\alpha}(X \mid Y)$, $\alpha \neq \beta$, in several situation.
In this study, we represent our bounds via specific distributions to ensure sharpnesses of the bounds.
The main results of this study are organized as follows:
\begin{itemize}
\item
\sectref{sect:v} shows sharp bounds on $H_{\beta}(X \mid Y)$ with fixed $H_{\alpha}(X \mid Y)$ and the cardinality $|\!\supp( P_{X} )| < \infty$ for distinct orders $\alpha \neq \beta$ as follows:
\begin{itemize}
\item
\thref{th:A_renyi_alpha_inf} gives bounds on $H_{\alpha}(X \mid Y)$ with a fixed $H_{\infty}(X \mid Y)$ for $\alpha \in (0, \infty)$, and vice versa.
\item
\thref{th:A_renyi_2} gives bounds on $H_{\beta}(X \mid Y)$ with a fixed $H_{\alpha}(X \mid Y)$ for $\alpha, \beta \in [1/2, \infty]$ and $|\!\supp( P_{X} )| \le 2$.
\item
\thref{th:ST} gives bounds on $H_{\beta}(X \mid Y)$ with a fixed $H_{\alpha}(X \mid Y)$ for $\alpha, \beta \in [1/2, \infty]$ and $|\!\supp( P_{X} )| \ge 3$.
\end{itemize}
\item
\sectref{sect:w} shows sharp bounds on $H_{\beta}(X \mid Y)$ with a fixed $H_{\alpha}(X \mid Y)$ for two orders $\alpha \in (0, 1) \cup (1, \infty]$ and $\beta \in (0, \infty]$, as shown in \thref{th:UV}.
Note that unlike Theorems~\ref{th:A_renyi_alpha_inf}--\ref{th:ST}, \thref{th:UV} has no constraint of the support $\supp( P_{X} )$.
\end{itemize}
Finally, \sectref{sect:appl} shows some applications of sharp bounds on the conditional R\'{e}nyi entropy to other related information measures, whose are strictly monotone functions of the conditional R\'{e}nyi entropy.

\section{Extremal Distributions $\bvec{v}_{n}( \cdot )$ and $\bvec{w}( \cdot )$ and Their Properties}
\label{sect:v_w}

In this subsection, we introduce the probability distributions $\bvec{v}_{n}( \cdot )$ and $\bvec{w}( \cdot )$, which play significant roles in this study.
In addition, the $\ell_{r}$-norms and R\'{e}nyi entropy of them are investigated.
Until \sectref{sect:unconditional}, we defer to show extremality of these distributions $\bvec{v}_{n}( \cdot )$ and $\bvec{w}( \cdot )$ in terms of the $\ell_{r}$-norm and R\'{e}nyi entropy.

For each $n \in \mathbb{N}$ and $p \in [1/n, 1]$, we define the $n$-dimensional probability vector%
\footnote{Note in \cite[Eq.~(3)]{part1} that the probability vector $\bvec{v}_{n}( \cdot )$ is defined by another form;
however, a simple change of variables immediately shows that these are essentially equavalent.}
\begin{align}
\bvec{v}_{n}( p )
\coloneqq
( v_{0}, v_{1}, v_{2}, \dots, v_{n-1} ) ,
\label{def:v}
\end{align}
where $\mathbb{N}$ denotes the set of positive integers and $v_{i}$ is chosen so that
\begin{align}
v_{i}
\coloneqq
\begin{dcases}
p
& \mathrm{if} \ i = 0 ,
\\
\frac{ 1 - p }{ n - 1 }
& \mathrm{otherwise}
\end{dcases}
\end{align}
for each $i \in \{ 0, 1, 2, \dots, n-1 \}$.
In addition, for $p \in (0, 1]$, we define the infinite-dimensional probability vector
\begin{align}
\bvec{w}( p )
\coloneqq
( w_{0}, w_{1}, w_{2}, \dots ) ,
\label{def:w}
\end{align}
where  $w_{i}$ is chosen so that
\begin{align}
w_{i}
\coloneqq
\begin{cases}
p
& \mathrm{if} \ 0 \le i < \lfloor 1/p \rfloor ,
\\
1 - \lfloor 1 / p \rfloor \, p
& \mathrm{if} \ i = \lfloor 1/p \rfloor ,
\\
0
& \mathrm{if} \ i > \lfloor 1/p \rfloor
\end{cases}
\end{align}
for each $i \in \{ 0, 1, 2 \dots \}$, and $\lfloor x \rfloor \coloneqq \max \{ z \in \mathbb{Z} \mid z \le x \}$ denotes the floor function of $x \in \mathbb{R}$.
Note that $| \! \supp( \bvec{v}_{n}( p ) ) | = n$ for every $p \in [1/n, 1)$, and $| \! \supp( \bvec{w}( p ) )| = m+1$ for every $m \in \mathbb{N}$ and $p \in [1/(m+1), 1/m)$, i.e., these are discrete probability distributions with finite supports.
Since $\bvec{v}_{1}( 1 )$ has only one probability mass $1$ whenever $n = 1$, we omit its trivial case in our analyses; and assume that $n \in \mathbb{N}_{\ge 2}$ in this study, where $\mathbb{N}_{\ge k}$ denotes the set of integers $n$ satisfying $n \ge k$.
By the definition \eqref{def:norm} of $\ell_{r}$-norm, for each $r \in (0, \infty)$, the $\ell_{r}$-norms of these distributions $\bvec{v}_{n}( \cdot )$ and $\bvec{w}( \cdot )$ can be calculated as follows:
\begin{align}
\| \bvec{v}_{n}( p ) \|_{r}
& =
\Big( p^{r} + (n-1)^{1-r} \, (1-p)^{r} \Big)^{1/r} ,
\label{eq:norm_v} \\
\| \bvec{w}( p ) \|_{r}
& =
\bigg( \bigg\lfloor \frac{1}{p} \bigg\rfloor \, p^{r} + \bigg( 1 - \bigg\lfloor \frac{1}{p} \bigg\rfloor \, p \bigg)^{r} \bigg)^{1/r} ,
\label{eq:norm_w}
\end{align}
respectively.
In particular, the $\ell_{\infty}$-norms are $\| \bvec{v}_{n}( p ) \|_{\infty} = p$ for $p \in [1/n, 1]$ and $\| \bvec{w}( p ) \|_{\infty} = p$ for $p \in (0, 1]$.
Substituting \eqref{eq:norm_v} and \eqref{eq:norm_w} into \eqref{def:renyi}, the R\'{e}nyi entropies of the distributions $\bvec{v}_{n}( \cdot )$ and $\bvec{w}( \cdot )$, respectively, can also be calculated as follows:
\begin{align}
H_{\alpha}( \bvec{v}_{n}( p ) )
& =
\frac{ 1 }{ 1 - \alpha } \ln \Big( p^{\alpha} + (n-1)^{1-\alpha} \, (1-p)^{\alpha} \Big) ,
\\
H_{\alpha}( \bvec{w}( p ) )
& =
\frac{ 1 }{ 1 - \alpha } \ln \bigg( \bigg\lfloor \frac{1}{p} \bigg\rfloor \, p^{\alpha} + \bigg( 1 - \bigg\lfloor \frac{1}{p} \bigg\rfloor \, p \bigg)^{\alpha} \bigg) ,
\end{align}
respectively.
We first show the monotonicities of $\ell_{r}$-norm of the distributions $\bvec{v}_{n}( \cdot )$ and $\bvec{w}( \cdot )$ in the following lemma.

\begin{lemma}
\label{lem:mono}
Let $r \in (0, 1) \cup (1, \infty]$ and $n \in \mathbb{N}_{\ge 2}$ be fixed numbers.
If $r \in (0, 1)$, then both $\ell_{r}$-norms $p_{v} \mapsto \| \bvec{v}_{n}( p_{v} ) \|_{r}$ and $p_{w} \mapsto \| \bvec{w}( p_{w} ) \|_{r}$ are strictly decreasing functions of $p_{v} \in [1/n, 1]$ and $p_{w} \in (0, 1]$, respectively.
Conversely, if $r \in (1, \infty]$, then both $\ell_{r}$-norms $p_{v} \mapsto \| \bvec{v}_{n}( p_{v} ) \|_{r}$ and $p_{w} \mapsto \| \bvec{w}( p_{w} ) \|_{r}$ are strictly increasing functions of $p_{v} \in [1/n, 1]$ and $p_{w} \in (0, 1]$, respectively.
\end{lemma}

\begin{IEEEproof}[Proof of \lemref{lem:mono}]
Since $\| \bvec{v}_{n}( p_{v} ) \|_{\infty} = p_{v}$ and $\| \bvec{w}( p_{w} ) \|_{\infty} = p_{w}$ for $p_{v} \in [1/n, 1]$ and $p_{w} \in (0, 1]$, respectively,
\lemref{lem:mono} is trivial if $r = \infty$.
Hence, it suffices to consider the $\ell_{r}$-norm for $r \in (0, 1) \cup (1, \infty)$.

We first verify the monotonicity of the function $p \mapsto \| \bvec{v}_{n}( p ) \|_{r}$.
A direct calculation shows
\begin{align}
\frac{ \partial \| \bvec{v}_{n}( p ) \|_{r} }{ \partial p }
& \overset{\eqref{eq:norm_v}}{=}
\frac{ \partial }{ \partial p } \Big( p^{r} + (n-1)^{1-r} \, (1-p)^{r} \Big)^{1/r}
\\
& =
\frac{1}{r} \Big( p^{r} + (n-1)^{1-r} \, (1-p)^{r} \Big)^{(1/r)-1} \bigg( \frac{ \partial }{ \partial p } \Big( p^{r} + (n-1)^{1-r} \, (1-p)^{r} \Big) \bigg)
\\
& =
\frac{1}{r} \Big( p^{r} + (n-1)^{1-r} \, (1-p)^{r} \Big)^{(1/r)-1} \Big( r \, p^{r-1} - r \, (n-1)^{1-r} \, (1-p)^{r-1} \Big)
\\
& =
\Big( p^{r} + (n-1)^{1-r} \, (1-p)^{r} \Big)^{(1/r)-1} \Big( p^{r-1} - (n-1)^{1-r} \, (1-p)^{r-1} \Big) .
\label{diff1:norm_v}
\end{align}
If we define the sign function as
\begin{align}
\sgn( x )
\coloneqq
\begin{cases}
- 1
& \mathrm{if} \ x < 0 ,
\\
0
& \mathrm{if} \ x = 0 ,
\\
1
& \mathrm{if} \ x > 0
\end{cases}
\end{align}
for $x \in \mathbb{R}$, then it follows that
\begin{align}
\sgn \bigg( \frac{ \partial \| \bvec{v}_{n}( p ) \|_{r} }{ \partial p } \bigg)
& \overset{\eqref{diff1:norm_v}}{=}
\underbrace{ \sgn\bigg( \Big( p^{r} + (n-1)^{1-r} \, (1-p)^{r} \Big)^{(1/r)-1} \bigg) }_{=1} \, \sgn\Big( p^{r-1} - (n-1)^{1-r} \, (1-p)^{r-1} \Big)
\\
& =
\sgn\Big( p^{r-1} - (n-1)^{1-r} \, (1-p)^{r-1} \Big)
\\
& =
\begin{cases}
-1
& \mathrm{if} \ r < 1 ,
\\
0
& \mathrm{if} \ r = 1 ,
\\
1
& \mathrm{if} \ r > 1
\end{cases}
\label{eq:sgn_norm_v_diff1}
\end{align}
for every $n \in \mathbb{N}_{\ge 2}$, $p \in (1/n, 1)$, and $r \in (0, \infty)$.
This implies that for any fixed $n \in \mathbb{N}_{\ge 2}$,
\begin{itemize}
\item
if $r \in (0, 1)$, then $p \mapsto \| \bvec{v}_{n}( p ) \|_{r}$ is strictly decreasing for $p \in [1/n, 1]$,
\item
if $r \in (1, \infty)$, then $p \mapsto \| \bvec{v}_{n}( p ) \|_{r}$ is strictly increasing for $p \in [1/n, 1]$;
\end{itemize}
and therefore, the assertion of \lemref{lem:mono} holds for $p \mapsto \| \bvec{v}_{n}( p ) \|_{r}$.

We next verify the monotonicity of the function $p \mapsto \| \bvec{w}( p ) \|_{r}$.
Since $\lfloor 1/p \rfloor = m$ for each $p \in (1/(m+1), 1/m]$ and $m \in \mathbb{N}$, we readily see that
\begin{align}
\frac{ \partial \| \bvec{w}( p ) \|_{r} }{ \partial p }
& \overset{\eqref{eq:norm_w}}{=}
\frac{ \partial }{ \partial p } \Big( m \, p^{r} + \big( 1 - m \, p \big)^{r} \Big)^{1/r}
\\
& =
\frac{1}{r} \Big( m \, p^{r} + \big( 1 - m \, p \big)^{r} \Big)^{(1/r) - 1} \bigg( \frac{ \partial }{ \partial p } \Big( m \, p^{r} + \big( 1 - m \, p \big)^{r} \Big) \bigg)
\\
& =
\frac{1}{r} \Big( m \, p^{r} + \big( 1 - m \, p \big)^{r} \Big)^{(1/r) - 1} \Big( r \, m \, p^{r-1} - r \, m \big( 1 - m \, p \big)^{r-1} \Big)
\\
& =
m \, \Big( m \, p^{r} + \big( 1 - m \, p \big)^{r} \Big)^{(1/r) - 1} \Big( p^{r-1} - \big( 1 - m \, p \big)^{r-1} \Big)
\label{diff1:norm_w}
\end{align}
for every $m \in \mathbb{N}$, $p \in (1/(m+1), 1/m)$, and $r \in (0, \infty)$.
Hence, we obtain
\begin{align}
\sgn \bigg( \frac{ \partial \| \bvec{w}( p ) \|_{r} }{ \partial p } \bigg)
& \overset{\eqref{diff1:norm_w}}{=}
\underbrace{ \sgn \bigg( m\, \Big( m \, p^{r} + \big( 1 - m \, p \big)^{r} \Big)^{(1/r) - 1} \bigg) }_{= 1} \, \sgn \Big( p^{r-1} - \big( 1 - m \, p \big)^{r-1} \Big)
\\
& =
\Big( p^{r-1} - \big( 1 - m \, p \big)^{r-1} \Big)
\\
& =
\begin{cases}
-1
& \mathrm{if} \ r < 1 ,
\\
0
& \mathrm{if} \ r = 1 ,
\\
1
& \mathrm{if} \ r > 1
\end{cases}
\label{eq:sgn_diff1_norm_w}
\end{align}
for every $m \in \mathbb{N}$, $p \in (1/(m+1), 1/m)$, and $r \in (0, \infty)$.
This implies that for each fixed $m \in \mathbb{N}$ and $r \in (0, 1) \cup (1, \infty)$,
\begin{itemize}
\item
if $r \in (0, 1)$, then $p \mapsto \| \bvec{w}( p ) \|_{r}$ is strictly decreasing for $p \in (1/(m+1), 1/m]$,
\item
if $r \in (1, \infty)$, then $p \mapsto \| \bvec{w}( p ) \|_{r}$ is strictly increasing for $p \in (1/(m+1), 1/m]$.
\end{itemize}
Finally, it follows that
\begin{align}
\lim_{p \to (1/m)^{+}} \| \bvec{w}( p ) \|_{r}
& =
\lim_{p \to (1/m)^{+}} \Big( \big\lfloor 1/p \big\rfloor \, p^{r} + \big( 1 - \big\lfloor 1/p \big\rfloor \, p \big)^{r} \Big)^{1/r}
\\
& =
\Big( (m-1) \, \big( 1/m \big)^{r} + \Big( 1 - (m-1) \, \big( 1/m \big) \Big)^{r} \Big)^{1/r}
\\
& =
\Big( (m-1) \, m^{-r} + m^{-r} \Big)^{1/r}
\\
& =
m^{(1-r)/r}
\label{eq:w_1/m} \\
& =
\| \bvec{w}( 1/m ) \|_{r}
\end{align}
for each $m \in \mathbb{N}_{\ge 2}$ and $r \in (0, 1)$,
which implies that $p \mapsto \| \bvec{w}( p ) \|_{r}$ is continuous on $p \in (0, 1]$;
therefore, the monotonicity of $p \mapsto \| \bvec{w}( p ) \|_{r}$ come from \eqref{eq:sgn_diff1_norm_w} can be improved as follows:
\begin{itemize}
\item
if $r \in (0, 1)$, then $p \mapsto \| \bvec{w}( p ) \|_{r}$ is strictly decreasing for $p \in (0, 1]$,
\item
if $r \in (1, \infty)$, then $p \mapsto \| \bvec{w}( p ) \|_{r}$ is strictly increasing for $p \in (0, 1]$.
\end{itemize}
This completes the proof of \lemref{lem:mono}.
\end{IEEEproof}

\lemref{lem:mono} implies the existences of inverse functions.
Let%
\footnote{Eq.~\eqref{def:theta} is defined to fulfill $\theta( \infty ) = -1$.}
\begin{align}
\theta( r )
\coloneqq
\lim_{t \to r} \frac{ 1 - t }{ t } ,
\label{def:theta}
\end{align}
and let $\mathcal{I}_{n}( r )$ and $\mathcal{J}( r )$ be real intervals defined by
\begin{align}
\mathcal{I}_{n}( r )
& \coloneqq
\begin{cases}
\big[ 1, n^{\theta( r )} \big]
& \mathrm{if} \ 0 < r < 1 ,
\\
\big[ n^{\theta( r )}, 1 \big]
& \mathrm{if} \ 1 < r \le \infty ,
\end{cases}
\label{def:I} \\
\mathcal{J}( r )
& \coloneqq
\begin{cases}
[1, \infty)
& \mathrm{if} \ 0 < r < 1 ,
\\
(0, 1]
& \mathrm{if} \ 1 < r \le \infty
\end{cases}
\label{def:J}
\end{align}
for each $n \in \mathbb{N}_{\ge 2}$ and $r \in (0, 1) \cup (1, \infty]$, respectively.
For each $r \in (0, 1) \cup (1, \infty]$ and $n \in \mathbb{N}_{\ge 2}$, we denote by
\begin{align}
&
N_{r}^{-1}(\bvec{v}_{n} : \cdot) : \mathcal{I}_{n}( r ) \to [1/n, 1] ,
\label{def:inv_Nv} \\
&
N_{r}^{-1}(\bvec{w} : \cdot) : \mathcal{J}( r ) \to (0, 1]
\label{def:inv_Nw}
\end{align}
inverse functions of $p_{v} \mapsto \| \bvec{v}_{n}( p_{v} ) \|_{r}$ and $p_{w} \mapsto \| \bvec{w}( p_{w} ) \|_{r}$, respectively.
As simple instances of them, if $r = \infty$, then $N_{\infty}^{-1}( \bvec{v}_{n} : t_{v} ) = t_{v}$ and $N_{\infty}^{-1}( \bvec{w} : t_{w} ) = t_{w}$ for $t_{v} \in [1/n, 1]$ and $t_{w} \in (0, 1]$, respectively, because $\| \bvec{v}_{n}( p_{v} ) \|_{\infty} = p_{v}$ and $\| \bvec{w}( p_{w} ) \|_{\infty} = p_{w}$ for $p_{v} \in [1/n, 1]$ and $p_{w} \in (0, 1]$, respectively.

Since logarithm functions are strictly monotone, it also follows from \eqref{def:renyi} and \lemref{lem:mono} that both R\'{e}nyi entropies $p_{v} \mapsto H_{\alpha}( \bvec{v}_{n}( p_{v} ) )$ and $p_{w} \mapsto H_{\alpha}( \bvec{w}( p_{w} ) )$ also have inverse functions for every%
\footnote{If $\alpha = 1$, i.e., if these are Shannon entropies, these inverse functions also exist due to \cite[Lemma~1]{part1}.}
$\alpha \in (0, \infty]$,
as with \eqref{def:inv_Nv} and \eqref{def:inv_Nw}.
For each $n \in \mathbb{N}_{\ge 2}$ and $\alpha \in (0, \infty]$, we denote by
\begin{align}
&
H_{\alpha}^{-1}( \bvec{v}_{n} : \cdot ) : [0, \ln n] \to [1/n, 1] ,
\label{def:inv_Hv} \\
&
H_{\alpha}^{-1}( \bvec{w} : \cdot ) : [0, \infty) \to (0, 1]
\label{def:inv_Hw}
\end{align}
inverse functions of $p_{v} \mapsto H_{\alpha}( \bvec{v}_{n}( p_{v} ) )$ and $p_{w} \mapsto H_{\alpha}( \bvec{w}( p_{w} ) )$, respectively.
By convention of the Shannon entropy, we write $H^{-1}( \bvec{v}_{n} : \cdot )$ and $H^{-1}( \bvec{w} : \cdot )$ as the inverse functions $H_{1}^{-1}( \bvec{v}_{n} : \cdot )$ and $H_{1}^{-1}( \bvec{w} : \cdot )$ with $\alpha = 1$, respectively.
In general, these inverse functions are hard-to-express in closed-forms, as with the inverse function of the binary entropy function $h_{2} : t \mapsto - t \ln t - (1-t) \ln (1-t)$.
As special cases of them, we give the following specific closed-forms.

\begin{fact}
\label{ex:renyi}
If $\alpha = 1/2$, $\alpha = 2$, or $\alpha = \infty$, then the inverse functions \eqref{def:inv_Hv} and \eqref{def:inv_Hw} can be expressed in the following closed-forms:
\begin{align}
H_{1/2}^{-1}( \bvec{v}_{n} : \mu )
& =
\frac{ n \, (n - 1) - (n-2) \, \mathrm{e}^{\mu} + 2 \, \sqrt{ \mathrm{e}^{\mu} \, (n-1) \, (n-\mathrm{e}^{\mu}) } }{ n^{2} }
&& \mathrm{for} \ n \in \mathbb{N} \ \mathrm{and} \ \mu \in [0, \ln n] ,
\\
H_{1/2}^{-1}( \bvec{w} : \mu )
& =
\frac{ (m+1) + (m-1) \, \mathrm{e}^{\mu} + 2 \, \sqrt{ \mathrm{e}^{\mu} \, m \, (1 + m - \mathrm{e}^{\mu})} }{ m \, (1+m)^{2} }
&& \mathrm{with} \ m = \lfloor \mathrm{e}^{\mu} \rfloor \ \mathrm{for} \ \mu \in [0, \infty) ,
\\
H_{2}^{-1}( \bvec{v}_{n} : \mu )
& =
\frac{ 1 + \sqrt{ \mathrm{e}^{-\mu} \, (n-1) \, (n-\mathrm{e}^{\mu}) } }{ n }
&& \mathrm{for} \ n \in \mathbb{N} \ \mathrm{and} \ \mu \in [0, \ln n] ,
\\
H_{2}^{-1}( \bvec{w} : \mu )
& =
\frac{ m + \sqrt{ \mathrm{e}^{-\mu} \, m \, (1 + m - \mathrm{e}^{\mu}) } }{ m \, (1+m) }
&& \mathrm{with} \ m = \lfloor \mathrm{e}^{\mu} \rfloor \ \mathrm{for} \ \mu \in [0, \infty) ,
\\
H_{\infty}^{-1}( \bvec{v}_{n} : \mu )
& =
\mathrm{e}^{-\mu}
&& \mathrm{for} \ n \in \mathbb{N} \ \mathrm{and} \ \mu \in [0, \ln n] ,
\label{eq:inv_Hv_infty} \\
H_{\infty}^{-1}( \bvec{w} : \mu )
& =
\mathrm{e}^{-\mu}
&& \mathrm{for} \ \mu \in [0, \infty) ,
\end{align}
where $\mathrm{e}$ denotes the base of natural logarithm.
\end{fact}

\begin{figure}[!t]
\centering
\subfloat[{Plot of $H_{\alpha}^{-1}( \bvec{v}_{n} : \mu )$ with $n = 4$ for $\mu \in [0, \ln 4]$.}]{
\begin{overpic}[width = 0.475\hsize, clip]{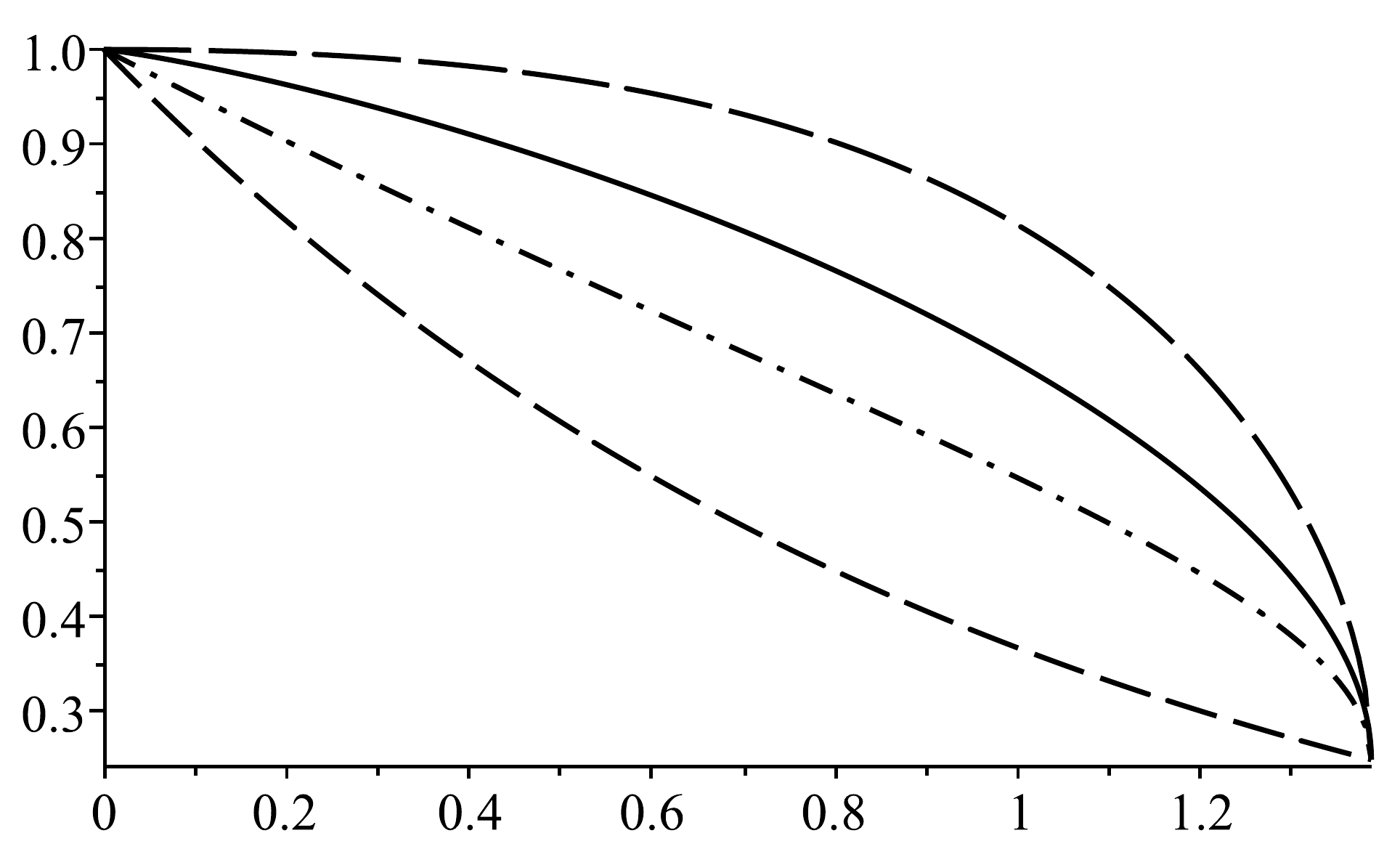}
\put(42, -1){$\mu = H_{\alpha}( \bvec{v}_{n}( p ) )$}
\put(-5, 18){\rotatebox{90}{$p = H_{\alpha}^{-1}( \bvec{v}_{n} : \mu )$}}
\put(96, 2){\scriptsize [nats]}
\put(15, 25){$\alpha = \infty$}
\put(19, 28){\vector(1, 1){10}}
\put(35, 15){$\alpha = 2$}
\put(38, 18){\vector(1, 2){10}}
\put(65, 55){$\alpha = 1$ (Shannon)}
\put(69, 54){\vector(-2, -3){7.5}}
\put(85, 48){$\alpha = 1/2$}
\put(89, 47){\vector(-1, -1){7}}
\end{overpic}
}\hfill
\subfloat[{Plot of $H_{\alpha}^{-1}( \bvec{w} : \mu )$ for $\mu \in [0, \ln 4]$.}]{
\begin{overpic}[width = 0.475\hsize, clip]{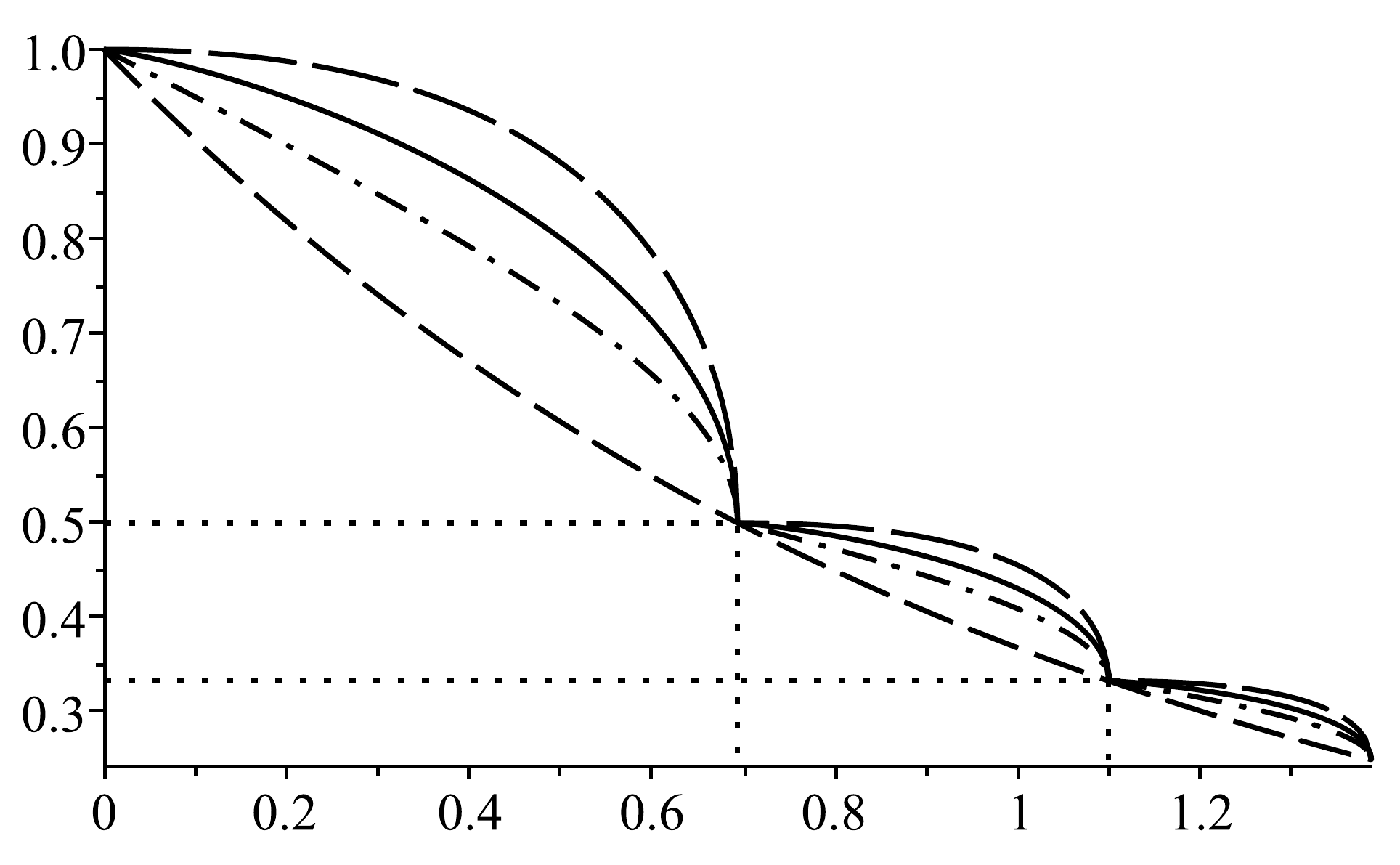}
\put(42, -1){$\mu = H_{\alpha}( \bvec{w}( p ) )$}
\put(-5, 18){\rotatebox{90}{$p = H_{\alpha}^{-1}( \bvec{w} : \mu )$}}
\put(96, 2){\scriptsize [nats]}
\put(10, 37){$\alpha = \infty$}
\put(13, 40){\vector(1, 1){6}}
\put(23, 29){$\alpha = 2$}
\put(26, 32){\vector(1, 2){6}}
\put(35, 58){$\alpha = 1$ (Shannon)}
\put(38, 57){\vector(-1, -1){6.5}}
\put(52, 50){$\alpha = 1/2$}
\put(53, 49){\vector(-2, -1){7}}
\put(63, 34){$(\mu, p) = (\ln 2, 1/2)$}
\put(62.5, 34){\vector(-1, -1){9}}
\put(90, 20){$(\ln 3, 1/3)$}
\put(89.5, 20){\vector(-3, -2){9}}
\end{overpic}
}
\caption{%
Plots of the inverse functions \eqref{def:inv_Hv} and \eqref{def:inv_Hw} with $\alpha = 1/2$, $\alpha = 1$, $\alpha = 2$, and $\alpha = \infty$.
The horizontal axes denote the R\'{e}nyi entropy of distributions $\bvec{v}_{n}( p )$ and $\bvec{w}( p )$, i.e., the arguments of the inverse functions \eqref{def:inv_Hv} and \eqref{def:inv_Hw}.
The vertical axes denote the parameter $p$ of distributions $\bvec{v}_{n}( p )$ and $\bvec{w}( p )$, i.e., the values of the inverse functions \eqref{def:inv_Hv} and \eqref{def:inv_Hw}.
\factref{ex:renyi} is used to plot them for the cases: $\alpha = 1/2$, $\alpha = 2$, and $\alpha = \infty$.}
\label{fig:inv_H}
\end{figure}

\factref{ex:renyi} can be verified by the quadratic formula in the case of%
\footnote{If $\alpha = \infty$, then \factref{ex:renyi} is almost trivial from the definition \eqref{def:min_entropy}.}
$\alpha = 1/2$ and $\alpha = 2$.
In \figref{fig:inv_H}, we illustrate instances of the inverse functions of \factref{ex:renyi}, along with the inverse functions $H^{-1}( \bvec{v}_{n} : \cdot )$ and $H^{-1}( \bvec{w} : \cdot )$ of the Shannon entropies.
As with \factref{ex:renyi}, from the relation between the R\'{e}nyi entropy and $\ell_{r}$-norm (cf. \eqref{def:renyi}), the inverse functions $N_{r}^{-1}( \bvec{v}_{n} : \cdot )$ of \eqref{def:inv_Nv} and $N_{r}^{-1}( \bvec{w} : \cdot )$ of \eqref{def:inv_Nw} can also be expressed in closed-forms if $r = 1/2$, $r = 2$, or $r = \infty$.
By \factref{ex:renyi}, sharp bounds established in this paper can be expressed in closed-forms in some situations.

Using the inverse functions $H^{-1}( \bvec{v}_{n} : \cdot )$ and $H^{-1}( \bvec{w} : \cdot )$ of the Shannon entropies, we introduce relations of the convexity/concavity of the $\ell_{r}$-norm with respect to the Shannon entropy of distributions $\bvec{v}_{n}( \cdot )$ and $\bvec{w}( \cdot )$ in Lemmas~\ref{lem:Hconvex_v} and~\ref{lem:Hconvex_w}, respectively.

\begin{lemma}[{\cite[Lemma~2]{part2}}]
\label{lem:Hconvex_v}
If $n = 2$, for each $r \in (0, 1) \cup (1, \infty)$, the $\ell_{r}$-norm $\mu \mapsto \| \bvec{v}_{2}( H^{-1}( \bvec{v}_{2} : \mu ) ) \|_{r}$ is strictly concave in $\mu \in [0, \ln 2]$. In addition%
\footnote{This concavity is shown in not \cite[Lemma~2]{part2} but the below paragraph of \cite[Lemma~2]{part2}.},
for each $n \in \mathbb{N}_{\ge 2}$, the $\ell_{\infty}$-norm $\mu \mapsto \| \bvec{v}_{n}( H^{-1}( \bvec{v}_{n} : \mu ) ) \|_{\infty}$ is strictly concave in $\mu \in [0, \ln n]$.
Moreover, for each $n \in \mathbb{N}_{\ge 3}$ and $r \in [1/2, 1) \cup (1, \infty)$, there exists an inflection point $\chi_{n}( r ) \in (0, \ln n)$ such that satisfies the following:
\begin{itemize}
\item
the $\ell_{r}$-norm $\mu \mapsto \| \bvec{v}_{n}( H^{-1}( \bvec{v}_{n} : \mu ) ) \|_{r}$ is strictly concave in $\mu \in [0, \chi_{n}( r )]$,
\item
the $\ell_{r}$-norm $\mu \mapsto \| \bvec{v}_{n}( H^{-1}( \bvec{v}_{n} : \mu ) ) \|_{r}$ is strictly convex in $\mu \in [\chi_{n}( r ), \ln n]$.
\end{itemize}
\end{lemma}

\begin{lemma}[{\cite[Lemma~3]{part2}\footnote{In \cite[Lemma~3]{part2}, the case $r = \infty$ is not considered; however, it can also be proved by the fact that $\| \bvec{w}( p ) \|_{\infty} = p$ for $p \in (0, 1]$, as with \lemref{lem:Hconvex_v}.}}]
\label{lem:Hconvex_w}
For each $m \in \mathbb{N}$ and $r \in (0, 1) \cup (1, \infty]$, the $\ell_{r}$-norm $\mu \mapsto \| \bvec{w}( H^{-1}( \bvec{w} : \mu ) ) \|_{r}$ is strictly concave in $\mu \in [\ln m, \ln(m+1)]$.
\end{lemma}

In \cite{part2}, Lemmas~\ref{lem:Hconvex_v} and~\ref{lem:Hconvex_w} were used to derive sharp bounds on the conditional Shannon entropy $H(X \mid Y)$ with a fixed conditional R\'{e}nyi entropy $H_{\alpha}(X \mid Y)$, and vice versa, from perspectives of the expectation of \eqref{def:expect_norm} and \eqref{def:cond_shannon}.
In this study, we establish sharp bounds on $H_{\beta}(X \mid Y)$ with a fixed $H_{\alpha}(X \mid Y)$ for distinct orders $\alpha \neq \beta$ in a similar manner to \cite{part2}, i.e., the property of expectation \eqref{def:expect_norm} are employed.
To this end, we further examine the convexity/concavity of $\ell_{r}$-norms with respect to $\ell_{s}$-norm for distributions $\bvec{v}_{n}( \cdot )$ and $\bvec{w}( \cdot )$, as with Lemmas~\ref{lem:Hconvex_v} and~\ref{lem:Hconvex_w}, respectively.
To derive such convexity/concavity lemmas, we now give the following \lemref{lem:sgn_g}.

\begin{lemma}
\label{lem:sgn_g}
We define the function
\begin{align}
g(n, z; r, s)
\coloneqq
\big( z^{r} + (n-1) \big) \ln_{r} z - \big( z^{s} + (n-1) \big) \ln_{s} z
\label{def:g}
\end{align}
for each $n \in \mathbb{N}_{\ge 2}$, $z \in (0, \infty)$, and $r, s \in (0, \infty)$, where
the $q$-logarithm function%
\footnote{Note that the limiting value $\lim_{q \to 1} (x^{1-q} - 1)/(1-q) = \ln x$ can be verified by L'H\^{o}pital's rule.}
\cite{tsallis}
is defined by
\begin{align}
\ln_{q} x
\coloneqq
\begin{dcases}
\ln x
& \mathrm{if} \ q = 1 ,
\\
\frac{ x^{1-q} - 1 }{ 1 - q }
& \mathrm{if} \ q \neq 1
\end{dcases}
\label{def:qlog}
\end{align}
for $x > 0$ and $q \in \mathbb{R}$.
Then, the following three assertions hold:
\begin{itemize}
\item
For any $n \in \mathbb{N}_{\ge 2}$, any $z \in (0, 1)$, and any $0 < r < s < \infty$, it holds that
\begin{align}
g(n, z; r, s)
=
- g(n, z; s, r)
>
0 ,
\label{eq:sgn_g_1}
\end{align}
\item
if $n = 2$, then for any $z \in (1, \infty)$ and $1/2 \le r < s < \infty$, it holds that
\begin{align}
g(2, z; r, s)
=
- g(2, z; s, r)
<
0 ,
\label{eq:sgn_g_2}
\end{align}
\item
for any $n \ge \mathbb{N}_{\ge 3}$ and any $1/2 \le r < s < \infty$, there exists $\zeta(n; r, s) \in (1, \infty)$ such that
\begin{align}
\sgn\Big( g(n, z; r, s) \Big)
=
- \sgn\Big( g(n, z; s, r) \Big)
=
\begin{cases}
-1
& \mathrm{if} \ \zeta(n; r, s) < z < \infty ,
\\
0
& \mathrm{if} \ z = 1 \ \mathrm{or} \ z = \zeta(n; r, s) ,
\\
1
& \mathrm{if} \ 1 < z < \zeta(n; r, s)
\end{cases}
\label{eq:sgn_g_3}
\end{align}
for every $z \in (1, \infty)$.
\end{itemize}
\end{lemma}

\lemref{lem:sgn_g} is proved in \appref{app:sgn_g}.
Defining%
\footnote{In \eqref{def:gamma}, suppose that $\gamma(\infty, \infty) = 1$.}
\begin{align}
\gamma(r, s)
& \coloneqq
\lim_{(a, b) \to (r, s)} \frac{ 1 - a }{ 1 - b } ,
\label{def:gamma}
\end{align}
we present the convexity/concavity of $\ell_{s}$-norms of $\bvec{v}_{n}( \cdot )$ and $\bvec{w}( \cdot )$ with respect to $\ell_{r}$-norms of them, $r \neq s$, in Lemmas~\ref{lem:convex_v} and~\ref{lem:convex_w}, respectively.
We emphasize that \lemref{lem:sgn_g} is a key lemma for deriving Lemmas~\ref{lem:convex_v} and~\ref{lem:convex_w}.

\begin{lemma}
\label{lem:convex_v}
For each $n \in \mathbb{N}_{\ge 2}$ and $r, s \in (0, 1) \cup (1, \infty)$, it holds that
\begin{itemize}
\item
the $\ell_{\infty}$-norm $t \mapsto \| \bvec{v}_{n}( N_{r}^{-1}( \bvec{v}_{n} : t ) ) \|_{\infty}$ is strictly concave in $t \in \mathcal{I}_{n}( r )$,
\item
if $s \in (0, 1)$, then $t \mapsto \| \bvec{v}_{n}( N_{\infty}^{-1}( \bvec{v}_{n} : t ) ) \|_{s}$ is strictly concave in $t \in [1/n, 1]$,
\item
if $s \in (1, \infty)$, then $t \mapsto \| \bvec{v}_{n}( N_{\infty}^{-1}( \bvec{v}_{n} : t ) ) \|_{s}$ is strictly convex in $t \in [1/n, 1]$.
\end{itemize}
Moreover, if $n = 2$, then it holds that for any distinct $r, s \in [1/2, 1) \cup (1, \infty)$,
\begin{itemize}
\item
if $\gamma( r, s ) > 1$, then $t \mapsto \| \bvec{v}_{n}( N_{r}^{-1}( \bvec{v}_{n} : t ) ) \|_{s}$ is strictly convex in $t \in \mathcal{I}_{n}( r )$,
\item
if $\gamma( r, s ) < 1$, then $t \mapsto \| \bvec{v}_{n}( N_{r}^{-1}( \bvec{v}_{n} : t ) ) \|_{s}$ is strictly concave in $t \in \mathcal{I}_{n}( r )$.
\end{itemize}
Furthermore, for each $n \in \mathbb{N}_{\ge 3}$ and distinct $r, s \in [1/2, 1) \cup (1, \infty)$, there exists an inflection point $\tau(n; r, s) \in \mathcal{I}_{n}( r ) \setminus \{ 1, n^{\theta( r )} \}$ such that
\begin{itemize}
\item
if $\gamma(r, s) > 1$, then
\begin{itemize}
\item
the $\ell_{s}$-norm $t \mapsto \| \bvec{v}_{n}( N_{r}^{-1}( \bvec{v}_{n} : t ) ) \|_{s}$ is strictly convex in $t \in \mathcal{I}_{n}^{(1)}( r, s )$,
\item
the $\ell_{s}$-norm $t \mapsto \| \bvec{v}_{n}( N_{r}^{-1}( \bvec{v}_{n} : t ) ) \|_{s}$ is strictly concave in $t \in \mathcal{I}_{n}^{(2)}( r, s )$,
\end{itemize}
\item
if $\gamma(r, s) < 1$, then
\begin{itemize}
\item
the $\ell_{s}$-norm $t \mapsto \| \bvec{v}_{n}( N_{r}^{-1}( \bvec{v}_{n} : t ) ) \|_{s}$ is strictly convex in $t \in \mathcal{I}_{n}^{(2)}( r, s )$,
\item
the $\ell_{s}$-norm $t \mapsto \| \bvec{v}_{n}( N_{r}^{-1}( \bvec{v}_{n} : t ) ) \|_{s}$ is strictly concave in $t \in \mathcal{I}_{n}^{(1)}( r, s )$,
\end{itemize}
\end{itemize}
where real intervals $\mathcal{I}_{n}^{(1)}( r, s )$ and $\mathcal{I}_{n}^{(2)}( r, s )$ are defined by
\begin{align}
\mathcal{I}_{n}^{(1)}( r, s )
& \coloneqq
\begin{cases}
\big[ 1, \tau(n; r, s) \big]
& \mathrm{if} \ r \in (0, 1) ,
\\
\big[ \tau(n; r, s), 1 \big]
& \mathrm{if} \ r \in (1, \infty) ,
\end{cases}
\label{def:In_1}\\
\mathcal{I}_{n}^{(2)}( r, s )
& \coloneqq
\begin{cases}
\big[ \tau(n; r, s), n^{\theta( r )} \big]
& \mathrm{if} \ r \in (0, 1) ,
\\
\big[ n^{\theta( r )}, \tau(n; r, s) \big]
& \mathrm{if} \ r \in (1, \infty) ,
\end{cases}
\label{def:In_2}
\end{align}
respectively.
\end{lemma}

\begin{figure}[!t]
\centering
\begin{overpic}[width = 0.5\hsize, clip]{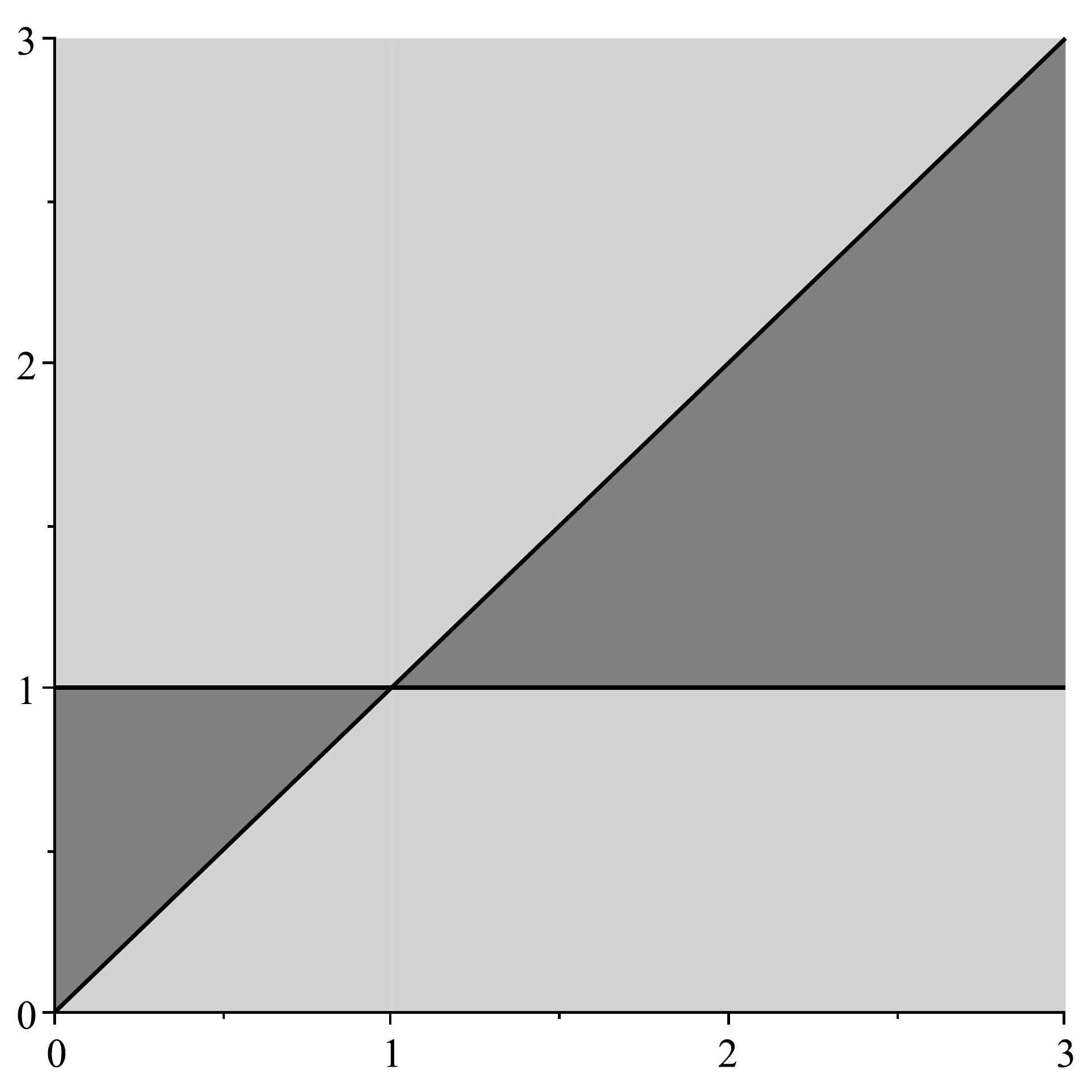}
\put(50, 0){\Large $r$}
\put(-3, 50){\rotatebox{90}{\Large $s$}}
\put(15, 55){$(r, s) = (1, 1)$}
\put(30.5, 53){\vector(1, -3){5}}
\end{overpic}
\caption{Plot of two regions of pairs $(r, s)$.
The dark gray region fulfills $\gamma(r, s) > 1$; and
the light gray region fulfills $\gamma(r, s) < 1$, where $\gamma(r, s)$ is defined in \eqref{def:gamma}.}
\label{fig:gamma}
\end{figure}

Note that the convexity and the concavity of \lemref{lem:convex_v} are switched each other according to either $\gamma(r, s) > 1$ or $\gamma(r, s) < 1$.
We illustrate two regions of pairs $(r, s)$ which fulfill $\gamma(r, s) > 1$ and $\gamma(r, s) < 1$, respectively, in \figref{fig:gamma}.

\begin{IEEEproof}[Proof of \lemref{lem:convex_v}]
In a similar way to the proofs of \cite[Lemma~1]{fabregas} and \cite[Lemma~2]{part2}, we prove this lemma by verifying signs of derivatives.
A simple calculation yields
\begin{align}
\frac{ \partial^{2} \| \bvec{v}_{n}( p ) \|_{r} }{ \partial p^{2} }
& \overset{\eqref{diff1:norm_v}}{=}
\frac{ \partial }{ \partial p } \bigg( \Big( p^{r} + (n-1)^{1-r} \, (1-p)^{r} \Big)^{(1/r)-1} \Big( p^{r-1} - (n-1)^{1-r} \, (1-p)^{r-1} \Big) \bigg)
\\
& =
\bigg( \frac{ \partial }{ \partial p } \Big( p^{r} + (n-1)^{1-r} \, (1-p)^{r} \Big)^{(1/r) - 1} \bigg) \Big( p^{r-1} - (n-1)^{1-r} \, (1-p)^{r-1} \Big)
\notag \\
& \qquad
+ \Big( p^{r} + (n-1)^{1-r} \, (1-p)^{r} \Big)^{(1/r) - 1} \bigg( \frac{ \partial }{ \partial p } \Big( p^{r-1} - (n-1)^{1-r} \, (1-p)^{r-1} \Big) \bigg)
\\
& =
\bigg( \frac{ 1 - r }{ r } \Big( p^{r} + (n-1)^{1-r} \, (1-p)^{r} \Big)^{(1/r) - 2} \bigg( \frac{ \partial }{ \partial p } \Big( p^{r} + (n-1)^{1-r} \, (1-p)^{r} \Big) \bigg) \bigg) \Big( p^{r-1} - (n-1)^{1-r} \, (1-p)^{r-1} \Big)
\notag \\
& \qquad
+ \Big( p^{r} + (n-1)^{1-r} \, (1-p)^{r} \Big)^{(1/r) - 1} \Big( (r-1) \, p^{r-2} + (r-1) \, (n-1)^{1-r} \, (1-p)^{r-2} \Big)
\\
& =
\frac{ 1 - r }{ r } \Big( p^{r} + (n-1)^{1-r} \, (1-p)^{r} \Big)^{(1/r) - 2} \Big( r \, p^{r-1} - r \, (n-1)^{1-r} \, (1-p)^{r-1} \Big) \Big( p^{r-1} - (n-1)^{1-r} \, (1-p)^{r-1} \Big)
\notag \\
& \qquad
+ (r-1) \, \Big( p^{r} + (n-1)^{1-r} \, (1-p)^{r} \Big)^{(1/r) - 1} \Big( p^{r-2} + (n-1)^{1-r} \, (1-p)^{r-2} \Big)
\\
& =
(1-r) \, \Big( p^{r} + (n-1)^{1-r} \, (1-p)^{r} \Big)^{(1/r) - 2} \Big( p^{r-1} - (n-1)^{1-r} \, (1-p)^{r-1} \Big)^{2}
\notag \\
& \qquad
+ (r-1) \, \Big( p^{r} + (n-1)^{1-r} \, (1-p)^{r} \Big)^{(1/r) - 1} \Big( p^{r-2} + (n-1)^{1-r} \, (1-p)^{r-2} \Big)
\\
& =
(1 - r) \, \Big( p^{r} + (n-1)^{1-r} \, (1-p)^{r} \Big)^{(1/r) - 2}
\notag \\
& \quad \times
\bigg[ \Big( p^{r-1} - (n-1)^{1-r} \, (1-p)^{r-1} \Big)^{2} - \Big( p^{r} + (n-1)^{1-r} \, (1-p)^{r} \Big) \Big( p^{r-2} + (n-1)^{1-r} \, (1-p)^{r-2} \Big) \bigg]
\\
& \overset{\text{(a)}}{=}
(1 - r) \, \Psi_{1}(n, p, r) \, \Big( p^{r} + (n-1)^{1-r} \, (1-p)^{r} \Big)^{(1/r) - 2}
\\
& \overset{\text{(b)}}{=}
(r-1) \, (n-1)^{1-r} \, \big( p \, (1-p) \big)^{r-2} \Big( p^{r} + (n-1)^{1-r} \, (1-p)^{r} \Big)^{(1/r) - 2} ,
\label{eq:norm_v_diff2}
\end{align}
where (a) follows by the definition
\begin{align}
\Psi_{1}(n, p, r)
\coloneqq
\Big( p^{r-1} - (n-1)^{1-r} \, (1-p)^{r-1} \Big)^{2} - \Big( p^{r} + (n-1)^{1-r} \, (1-p)^{r} \Big) \Big( p^{r-2} + (n-1)^{1-r} \, (1-p)^{r-2} \Big) ,
\end{align}
and (b) follows from the fact that
\begin{align}
\Psi_{1}(n, p, r)
& =
\Big( p^{2(r-1)} - 2 \, (n-1)^{1-r} \, p^{r-1} \, (1-p)^{r-1} + (n-1)^{2(1-r)} \, (1-p)^{2(r-1)} \Big)
\notag \\
& \qquad
- \Big( p^{2(r-1)} + (n-1)^{1-r} \, p^{r} \, (1-p)^{r-2} + (n-1)^{1-r} \, p^{r-2} \, (1-p)^{r} + (n-1)^{2(1-r)} \, (1-p)^{2(r-1)} \Big)
\\
& =
- 2 \, (n-1)^{1-r} \, p^{r-1} \, (1-p)^{r-1} - (n-1)^{1-r} \, p^{r} \, (1-p)^{r-2} - (n-1)^{1-r} \, p^{r-2} \, (1-p)^{r}
\\
& =
- (n-1)^{1-r} \, \big( p \, (1 - p) \big)^{r-2} \Big( 2 \, p \, (1-p) + p^{2} + (1-p)^{2} \Big)
\\
& =
- (n-1)^{1-r} \, \big( p \, (1 - p) \big)^{r-2} \big( p + (1-p) \big)^{2}
\\
& =
- (n-1)^{1-r} \, \big( p \, (1 - p) \big)^{r-2} .
\end{align}
Then, we obtain
\begin{align}
\sgn \bigg( \frac{ \partial^{2} \| \bvec{v}_{n}( p ) \|_{r} }{ \partial p^{2} } \bigg)
& \overset{\eqref{eq:norm_v_diff2}}{=}
\sgn(r-1) \, \underbrace{ \sgn\Big( (n-1)^{1-r} \Big) }_{=1} \, \underbrace{ \sgn\Big( \big( p \, (1-p) \big)^{r-2} \Big) }_{=1} \, \underbrace{ \sgn\bigg( \Big( p^{r} + (n-1)^{1-r} \, (1-p)^{r} \Big)^{(1/r) - 2} \bigg) }_{=1}
\\
& =
\sgn(r-1)
\\
& =
\begin{cases}
-1
& \mathrm{if} \ r < 1 ,
\\
0
& \mathrm{if} \ r = 1 ,
\\
1
& \mathrm{if} \ r > 1
\end{cases}
\label{eq:sgn_norm_v_diff2}
\end{align}
for every $n \in \mathbb{N}_{\ge 2}$, $p \in (1/n, 1)$, and $r \in (0, \infty)$.
By the inverse function theorem, we have
\begin{align}
\frac{ \partial N_{r}^{-1}( \bvec{v}_{n} : t ) }{ \partial t }
& =
\bigg( \frac{ \partial \| \bvec{v}_{n}( p ) \|_{r} }{ \partial p } \bigg)^{-1} ,
\label{eq:inverse_norm_v_diff1} \\
\frac{ \partial^{2} N_{r}^{-1}( \bvec{v}_{n} : t ) }{ \partial t^{2} }
& =
- \frac{ \partial^{2} \| \bvec{v}_{n}( p ) \|_{r} }{ \partial p^{2} } \, \bigg( \frac{ \partial \| \bvec{v}_{n}( p ) \|_{r} }{ \partial p } \bigg)^{-3}
\label{eq:inverse_norm_v_diff2}
\end{align}
for every $n \in \mathbb{N}_{\ge 2}$, $r \in (0, 1) \cup (1, \infty)$, and $t \in \mathcal{I}_{n}( r ) \setminus \{ 1, n^{\theta( r )} \}$, where $\mathcal{I}_{n}( \cdot )$ is defined in \eqref{def:I}, and the variables $t$ and $p$ are chosen to satisfy $\| \bvec{
v}_{n}( p ) \|_{r} = t$ (cf. the definition \eqref{def:inv_Nv} of $N_{r}^{-1}( \bvec{v}_{n} : \cdot )$), i.e.,
\begin{align}
1/n < p < 1
\iff
\min\{ 1, n^{\theta( r )} \} < t < \max\{ 1, n^{\theta( r )} \} .
\label{eq:p_t_In}
\end{align}
In particular, since $\| \bvec{v}_{n}( p ) \|_{\infty} = p$ for $p \in [1/n, 1]$, it follows from \eqref{eq:sgn_norm_v_diff1} and \eqref{eq:sgn_norm_v_diff2} that
\begin{align}
\sgn\bigg( \frac{ \partial^{2} \| \bvec{v}_{n}( N_{r}^{-1}( \bvec{v}_{n} : t ) ) \|_{\infty} }{ \partial t^{2} } \bigg)
& =
\sgn\bigg( \frac{ \partial^{2} N_{r}^{-1}( \bvec{v}_{n} : t ) }{ \partial t^{2} } \bigg)
\\
& \overset{\eqref{eq:inverse_norm_v_diff2}}{=}
- \sgn\bigg( \frac{ \partial^{2} \| \bvec{v}_{n}( p ) \|_{r} }{ \partial p^{2} } \bigg) \, \sgn\bigg( \bigg( \frac{ \partial \| \bvec{v}_{n}( p ) \|_{r} }{ \partial p } \bigg)^{-3} \bigg)
\\
& =
- \sgn\bigg( \frac{ \partial^{2} \| \bvec{v}_{n}( p ) \|_{r} }{ \partial p^{2} } \bigg) \, \sgn\bigg( \frac{ \partial \| \bvec{v}_{n}( p ) \|_{r} }{ \partial p } \bigg)
\\
& =
- 1
\label{eq:sgn_norm_v_diff2_r_inf}
\end{align}
for every $n \in \mathbb{N}_{\ge 2}$, $r \in (0, 1) \cup (1, \infty)$, and $t \in \mathcal{I}_{n}( r ) \setminus \{ 1, n^{\theta( r )} \}$.
Moreover, since $N_{\infty}^{-1}( \bvec{v}_{n} : t ) = t$ for $t \in [1/n, 1]$, we also get
\begin{align}
\sgn\bigg( \frac{ \partial^{2} \| \bvec{v}_{n}( N_{\infty}^{-1}( \bvec{v}_{n} : t ) ) \|_{s} }{ \partial t^{2} } \bigg)
& =
\sgn\bigg( \frac{ \partial^{2} \| \bvec{v}_{n}( p ) \|_{s} }{ \partial p^{2} } \bigg)
\\
& \overset{\eqref{eq:sgn_norm_v_diff2}}{=}
\begin{cases}
-1
& \mathrm{if} \ s < 1 ,
\\
0
& \mathrm{if} \ s = 1 ,
\\
1
& \mathrm{if} \ s > 1
\end{cases}
\label{eq:sgn_norm_v_diff2_inf_s}
\end{align}
for every $n \in \mathbb{N}_{\ge 2}$, $t \in [1/n, 1]$, and $s \in (0, \infty)$.
Therefore, it follows from \eqref{eq:sgn_norm_v_diff2_r_inf} and \eqref{eq:sgn_norm_v_diff2_inf_s} that
\begin{itemize}
\item
for each $n \in \mathbb{N}_{\ge 2}$ and $r \in (0, 1) \cup (1, \infty)$, the $\ell_{\infty}$-norm $t \mapsto \| \bvec{v}_{n}( N_{r}^{-1}( \bvec{v}_{n} : t ) ) \|_{\infty}$ is strictly concave in $t \in \mathcal{I}_{n}( r )$,
\item
for each $n \in \mathbb{N}_{\ge 2}$ and $s \in (0, 1)$, the $\ell_{s}$-norm $t \mapsto \| \bvec{v}_{n}( N_{\infty}^{-1}( \bvec{v}_{n} : t ) ) \|_{s}$ is strictly concave in $t \in [1/n, 1]$,
\item
for each $n \in \mathbb{N}_{\ge 2}$ and $s \in (1, \infty)$, the $\ell_{s}$-norm $t \mapsto \| \bvec{v}_{n}( N_{\infty}^{-1}( \bvec{v}_{n} : t ) ) \|_{s}$ is strictly convex in $t \in [1/n, 1]$.
\end{itemize}

Henceforth, we consider the convexity/concavity of $t \mapsto \| \bvec{v}_{n}( N_{r}^{-1}( \bvec{v}_{n} : t ) ) \|_{s}$ with respect to $t \in \mathcal{I}_{n}( r )$ for each distinct $r, s \in (0, 1) \cup (1, \infty)$.
By the chain rule of derivatives, we have
\begin{align}
\frac{ \partial^{2} \| \bvec{v}_{n}( N_{r}^{-1}( \bvec{v}_{n} : t ) ) \|_{s} }{ \partial t^{2} }
& =
\frac{ \partial^{2} \| \bvec{v}_{n}( p ) \|_{s} }{ \partial p^{2} } \bigg( \frac{ \partial N_{r}^{-1}( \bvec{v}_{n} : t ) }{ \partial t } \bigg)^{2} + \frac{ \partial \| \bvec{v}_{n}( p ) \|_{s} }{ \partial p } \frac{ \partial^{2} N_{r}^{-1}( \bvec{v}_{n} : t ) }{ \partial t^{2} }
\\
& \overset{\eqref{eq:inverse_norm_v_diff1}}{=}
\frac{ \partial^{2} \| \bvec{v}_{n}( p ) \|_{s} }{ \partial p^{2} } \bigg( \frac{ \partial \| \bvec{v}_{n}( p ) \|_{r} }{ \partial p } \bigg)^{-2} + \frac{ \partial \| \bvec{v}_{n}( p ) \|_{s} }{ \partial p } \frac{ \partial^{2} N_{r}^{-1}( \bvec{v}_{n} : t ) }{ \partial t^{2} }
\\
& \overset{\eqref{eq:inverse_norm_v_diff2}}{=}
\frac{ \partial^{2} \| \bvec{v}_{n}( p ) \|_{s} }{ \partial p^{2} } \bigg( \frac{ \partial \| \bvec{v}_{n}( p ) \|_{r} }{ \partial p } \bigg)^{-2} - \frac{ \partial \| \bvec{v}_{n}( p ) \|_{s} }{ \partial p } \frac{ \partial^{2} \| \bvec{v}_{n}( p ) \|_{r} }{ \partial p^{2} } \bigg( \frac{ \partial \| \bvec{v}_{n}( p ) \|_{r} }{ \partial p } \bigg)^{-3}
\\
& =
\frac{ \partial^{2} \| \bvec{v}_{n}( p ) \|_{r} }{ \partial p^{2} } \frac{ \partial^{2} \| \bvec{v}_{n}( p ) \|_{s} }{ \partial p^{2} } \bigg( \frac{ \partial \| \bvec{v}_{n}( p ) \|_{r} }{ \partial p } \bigg)^{-3}
\notag \\
& \qquad \qquad \qquad \qquad \times
\Bigg[ \frac{ \partial \| \bvec{v}_{n}( p ) \|_{r} }{ \partial p } \bigg( \frac{ \partial^{2} \| \bvec{v}_{n}( p ) \|_{r} }{ \partial p^{2} } \bigg)^{-1} - \frac{ \partial \| \bvec{v}_{n}( p ) \|_{s} }{ \partial p } \bigg( \frac{ \partial^{2} \| \bvec{v}_{n}( p ) \|_{s} }{ \partial p^{2} } \bigg)^{-1} \Bigg]
\\
& \overset{\text{(a)}}{=}
\frac{ \partial^{2} \| \bvec{v}_{n}( p ) \|_{r} }{ \partial p^{2} } \frac{ \partial^{2} \| \bvec{v}_{n}( p ) \|_{s} }{ \partial p^{2} } \bigg( \frac{ \partial \| \bvec{v}_{n}( p ) \|_{r} }{ \partial p } \bigg)^{-3}
\notag \\
& \qquad \qquad \qquad \qquad \times
\frac{ p \, (1-p)^{2} }{ n-1 } \, \bigg[ \big( z^{r} + (n-1) \big) \ln_{r} z - \big( z^{s} + (n-1) \big) \ln_{s} z \bigg]
\\
& \overset{\eqref{def:g}}{=}
g(n, z; r, s) \, \frac{ p \, (1-p)^{2} }{ n-1 } \, \frac{ \partial^{2} \| \bvec{v}_{n}( p ) \|_{r} }{ \partial p^{2} } \frac{ \partial^{2} \| \bvec{v}_{n}( p ) \|_{s} }{ \partial p^{2} } \bigg( \frac{ \partial \| \bvec{v}_{n}( p ) \|_{r} }{ \partial p } \bigg)^{-3} ,
\label{eq:diff2_2norm_v}
\end{align}
where (a) follows from
\begin{itemize}
\item
the change of variables as
\begin{align}
z = z(n, p) \coloneqq (n-1) \, \frac{ p }{ 1-p } ,
\label{def:z}
\end{align}
\item
and the fact that
\begin{align}
\frac{ \partial \| \bvec{v}_{n}( p ) \|_{r} }{ \partial p } \bigg( \frac{ \partial^{2} \| \bvec{v}_{n}( p ) \|_{r} }{ \partial p^{2} } \bigg)^{-1}
& \overset{\eqref{diff1:norm_v}}{=}
\Big( p^{r} + (n-1)^{1-r} \, (1-p)^{r} \Big)^{(1/r)-1} \Big( p^{r-1} - (n-1)^{1-r} \, (1-p)^{r-1} \Big) \notag \\
& \qquad \quad \times
\bigg( \frac{ \partial^{2} \| \bvec{v}_{n}( p ) \|_{r} }{ \partial p^{2} } \bigg)^{-1}
\\
& \overset{\eqref{eq:norm_v_diff2}}{=}
\Big( p^{r} + (n-1)^{1-r} \, (1-p)^{r} \Big)^{(1/r)-1} \Big( p^{r-1} - (n-1)^{1-r} \, (1-p)^{r-1} \Big)
\notag \\
& \qquad \quad \times
(r-1)^{-1} \, (n-1)^{r-1} \, \big( p \, (1-p) \big)^{2-r} \Big( p^{r} + (n-1)^{1-r} \, (1-p)^{r} \Big)^{2-(1/r)}
\\
& =
(r-1)^{-1} \, (n-1)^{r-1} \, \big( p \, (1-p) \big)^{2-r}
\notag \\
& \qquad \qquad \qquad \times
\Big( p^{r} + (n-1)^{1-r} \, (1-p)^{r} \Big) \, \Big( p^{r-1} - (n-1)^{1-r} \, (1-p)^{r-1} \Big)
\\
& =
\frac{ p \, (1-p) }{ r - 1 } \, \Big( p^{r} + (n-1)^{1-r} \, (1-p)^{r} \Big) \, \Big( (n-1)^{r-1} \, (1-p)^{1-r} - p^{1-r} \Big)
\\
& =
\frac{ p \, (1-p) }{ r - 1 } \, \Big( (n-1)^{r-1} \, p^{r} \, (1-p)^{1-r} - p + (1-p) - (n-1)^{1-r} \, p^{1-r} \, (1-p)^{r} \Big)
\\
& =
\frac{ p \, (1-p) }{ r - 1 } \, \bigg( (1-2p) +  p \, \Big( \frac{ p \, (n-1) }{ 1-p } \Big)^{r-1} - (1-p) \, \Big( \frac{ p \, (n-1) }{ 1-p } \Big)^{1-r} \bigg)
\\
& \overset{\eqref{def:z}}{=}
\frac{ p \, (1-p) }{ r - 1 } \, \Big( (1-2p) +  p \, z^{r-1} - (1-p) \, z^{1-r} \Big)
\\
& =
\frac{ p \, (1-p) }{ r - 1 } \, \Big( (1 - z^{1-r}) + p \, z^{r-1} \, (1 - 2 \, z^{1-r} + z^{2(1-r)}) \Big)
\\
& =
\frac{ p \, (1-p) }{ r - 1 } \, \Big( (1 - z^{1-r}) + p \, z^{r-1} \, (1 - z^{1-r})^{2} \Big)
\\
& =
\frac{ p \, (1-p) }{ r - 1 } \, (1 - z^{1-r}) \, \Big( 1 + p \, z^{r-1} \, (1-z^{1-r}) \Big)
\\
& =
p \, (1-p) \, \Big( 1 + p \, (z^{r-1}-1) \Big) \, \frac{ z^{1-r} - 1 }{ 1 - r }
\\
& \overset{\eqref{def:qlog}}{=}
p \, (1-p) \, \Big( 1 + p \, (z^{r-1}-1) \Big) \, (\ln_{r} z)
\\
& \overset{\eqref{def:z}}{=}
p \, (1-p) \, \bigg( 1 + \frac{ z }{ (n-1) + z } \, (z^{r-1}-1) \bigg) \, (\ln_{r} z)
\\
& =
\frac{ p \, (1-p) }{ (n-1) + z } \, \big( (n-1) + z + z^{r} - z \big) \, (\ln_{r} z)\\
& =
\frac{ p \, (1-p) }{ (n-1) + z } \, \big( (n-1) + z^{r} \big) \, (\ln_{r} z)
\\
& \overset{\eqref{def:z}}{=}
\frac{ p \, (1-p)^{2} }{ (n-1) \, (1-p) + p \, (n-1) } \, \big( (n-1) + z^{r} \big) \, (\ln_{r} z)
\\
& =
\frac{ p \, (1-p)^{2} }{ n-1 } \, \big( (n-1) + z^{r} \big) \, (\ln_{r} z) .
\end{align}
\end{itemize}
Since $p \in (1/n, 1)$ for $t \in \mathcal{I}_{n}( r ) \setminus \{ 1, n^{\theta( r )} \}$ (cf. \eqref{eq:p_t_In}), it suffices to consider the range of variable $z$ of \eqref{def:z} on $z \in (1, \infty)$.
A further calculation derives
\begin{align}
&
\sgn \bigg( \frac{ \partial^{2} \| \bvec{v}_{n}( N_{r}^{-1}( \bvec{v}_{n} : t ) ) \|_{s} }{ \partial t^{2} } \bigg)
\notag \\
& \qquad \overset{\eqref{eq:diff2_2norm_v}}{=}
\sgn\Big( g(n, z; r, s) \Big) \, \underbrace{ \sgn\Big( \frac{ p \, (1-p)^{2} }{ n-1 } \Big) }_{ = 1 } \, \sgn\bigg( \frac{ \partial^{2} \| \bvec{v}_{n}( p ) \|_{r} }{ \partial p^{2} } \bigg) \, \sgn\bigg( \frac{ \partial^{2} \| \bvec{v}_{n}( p ) \|_{s} }{ \partial p^{2} } \bigg) \, \sgn \bigg( \bigg( \frac{ \partial \| \bvec{v}_{n}( p ) \|_{r} }{ \partial p } \bigg)^{-3} \bigg)
\\
& \qquad =
\sgn\Big( g(n, z; r, s) \Big) \, \sgn\bigg( \frac{ \partial^{2} \| \bvec{v}_{n}( p ) \|_{r} }{ \partial p^{2} } \bigg) \, \sgn\bigg( \frac{ \partial^{2} \| \bvec{v}_{n}( p ) \|_{s} }{ \partial p^{2} } \bigg) \, \sgn \bigg( \frac{ \partial \| \bvec{v}_{n}( p ) \|_{r} }{ \partial p } \bigg)
\\
& \qquad \overset{\eqref{eq:sgn_norm_v_diff2_r_inf}}{=}
\sgn\Big( g(n, z; r, s) \Big) \, \sgn\bigg( \frac{ \partial^{2} \| \bvec{v}_{n}( p ) \|_{s} }{ \partial p^{2} } \bigg)
\\
& \qquad \overset{\eqref{eq:sgn_norm_v_diff2}}{=}
\begin{cases}
- \sgn \Big( g(n, z; r, s) \Big)
& \mathrm{if} \ s < 1 ,
\\
\sgn \Big( g(n, z; r, s) \Big)
& \mathrm{if} \ s > 1
\end{cases}
\label{eq:sgn_diff2_norm_v_r_s}
\end{align}
for every $n \in \mathbb{N}_{\ge 2}$, distinct $r, s \in (0, 1) \cup (1, \infty)$, and $t \in \mathcal{I}_{n}( r ) \setminus \{ 1, n^{\theta( r )} \}$.
That is, the convexity/concavity of $t \mapsto \| \bvec{v}_{n}( N_{r}^{-1}( \bvec{v}_{n} : t ) ) \|_{s}$ with respect to $t \in \mathcal{I}_{n}( r )$ depend on the sign of $g(n, z; r, s)$.
If $n = 2$, we have from \eqref{eq:sgn_g_2} of \lemref{lem:sgn_g} and \eqref{eq:sgn_diff2_norm_v_r_s} that
\begin{align}
\sgn \bigg( \frac{ \partial^{2} \| \bvec{v}_{2}( N_{r}^{-1}( \bvec{v}_{2} : t ) ) \|_{s} }{ \partial t^{2} } \bigg)
& =
\begin{cases}
-1
& \mathrm{if} \ r < 1 < s \ \mathrm{or} \ s < r < 1 \ \mathrm{or} \ s < 1 < r \ \mathrm{or} \ 1 < r < s ,
\\
1
& \mathrm{if} \ r < s < 1 \ \mathrm{or} \ 1 < s < r
\end{cases}
\\
& =
\begin{cases}
-1
& \mathrm{if} \ \gamma( r, s ) < 1 ,
\\
1
& \mathrm{if} \ \gamma( r, s ) > 1
\end{cases}
\end{align}
for every distinct $r, s \in [1/2, 1) \cup (1, \infty)$ and $t \in \mathcal{I}_{2}( r ) \setminus \{ 1, n^{\theta( r )} \}$, where $\gamma( r, s )$ is defined in \eqref{def:gamma}.
This implies the assertion of \lemref{lem:convex_v} for $n = 2$.

Furthermore, we verify the assertion of \lemref{lem:convex_v} for $n \in \mathbb{N}_{\ge 3}$.
It is clear from \eqref{def:z} that $p \mapsto z(n, p)$ is strictly increasing for $p \in [1/n, 1)$.
Moreover, it follows from \eqref{eq:sgn_norm_v_diff1} that
\begin{itemize}
\item
if $r \in (0, 1)$, then $p \mapsto \| \bvec{v}_{n}( p ) \|_{r}$ is strictly decreasing for $p \in [1/n, 1]$,
\item
if $r \in (1, \infty)$, then $p \mapsto \| \bvec{v}_{n}( p ) \|_{r}$ is strictly increasing for $p \in [1/n, 1]$.
\end{itemize}
Hence, we observe from the relation $N_{r}^{-1}( \bvec{v}_{n} : t ) = p$ that
\begin{itemize}
\item
it holds that $\lim_{t \to 1} z(n, N_{r}^{-1}( \bvec{v}_{n} : t )) = \lim_{p \to 1} z(n, p) = \infty$,
\item
it holds that $z(n, N_{r}^{-1}( \bvec{v}_{n} : n^{\theta( r )} )) = z(n, 1/n) = 1$,
\item
if $r \in (0, 1)$, then $t \mapsto z(n, N_{r}^{-1}( \bvec{v}_{n} : t ))$ is strictly decreasing for $t \in \mathcal{I}_{n}( r ) \setminus \{ 1 \}$,
\item
if $r \in (1, \infty)$, then $t \mapsto z(n, N_{r}^{-1}( \bvec{v}_{n} : t ))$ is strictly increasing for $t \in \mathcal{I}_{n}( r ) \setminus \{ 1 \}$.
\end{itemize}
Therefore, it follows from \eqref{eq:sgn_g_3} of \lemref{lem:sgn_g} and \eqref{eq:sgn_diff2_norm_v_r_s} that for any $n \in \mathbb{N}_{\ge 3}$ and distinct $r, s \in [1/2, 1) \cup (1, \infty)$, there exists $\tau(n; r, s) \in \mathcal{I}_{n}( r ) \setminus \{ 1, n^{\theta( r )} \}$ such that satisfies the following:
\begin{itemize}
\item
if $r < s < 1$, then
\begin{align}
\sgn \bigg( \frac{ \partial^{2} \| \bvec{v}_{2}( N_{r}^{-1}( \bvec{v}_{2} : t ) ) \|_{s} }{ \partial t^{2} } \bigg)
& =
\begin{cases}
-1
& \mathrm{if} \ \tau(n; r, s) < t < n^{\theta( r )} ,
\\
0
& \mathrm{if} \ t = \tau(n; r, s) ,
\\
1
& \mathrm{if} \ 1 < t < \tau(n; r, s)
\end{cases}
\label{eq:sgn_diff2_norm_r_s_v_1}
\end{align}
for every $t \in \mathcal{I}_{n}( r ) \setminus \{ 1, n^{\theta( r )} \}$,
\item
if $r < 1 < s$, then
\begin{align}
\sgn \bigg( \frac{ \partial^{2} \| \bvec{v}_{2}( N_{r}^{-1}( \bvec{v}_{2} : t ) ) \|_{s} }{ \partial t^{2} } \bigg)
& =
\begin{cases}
-1
& \mathrm{if} \ 1 < t < \tau(n; r, s) ,
\\
0
& \mathrm{if} \ t = \tau(n; r, s) ,
\\
1
& \mathrm{if} \ \tau(n; r, s) < t < n^{\theta( r )}
\end{cases}
\label{eq:sgn_diff2_norm_r_s_v_2}
\end{align}
for every $t \in \mathcal{I}_{n}( r ) \setminus \{ 1, n^{\theta( r )} \}$,
\item
if $1 < r < s$, then
\begin{align}
\sgn \bigg( \frac{ \partial^{2} \| \bvec{v}_{2}( N_{r}^{-1}( \bvec{v}_{2} : t ) ) \|_{s} }{ \partial t^{2} } \bigg)
& =
\begin{cases}
-1
& \mathrm{if} \ \tau(n; r, s) < t < 1 ,
\\
0
& \mathrm{if} \ t = \tau(n; r, s) ,
\\
1
& \mathrm{if} \ n^{\theta( r )} < t < \tau(n; r, s)
\end{cases}
\label{eq:sgn_diff2_norm_r_s_v_3}
\end{align}
for every $t \in \mathcal{I}_{n}( r ) \setminus \{ 1, n^{\theta( r )} \}$,
\item
if $s < r < 1$, then
\begin{align}
\sgn \bigg( \frac{ \partial^{2} \| \bvec{v}_{2}( N_{r}^{-1}( \bvec{v}_{2} : t ) ) \|_{s} }{ \partial t^{2} } \bigg)
& =
\begin{cases}
-1
& \mathrm{if} \ 1 < t < \tau(n; r, s) ,
\\
0
& \mathrm{if} \ t = \tau(n; r, s) ,
\\
1
& \mathrm{if} \ \tau(n; r, s) < t < n^{\theta( r )}
\end{cases}
\label{eq:sgn_diff2_norm_r_s_v_4}
\end{align}
for every $t \in \mathcal{I}_{n}( r ) \setminus \{ 1, n^{\theta( r )} \}$,
\item
if $s < 1 < r$, then
\begin{align}
\sgn \bigg( \frac{ \partial^{2} \| \bvec{v}_{2}( N_{r}^{-1}( \bvec{v}_{2} : t ) ) \|_{s} }{ \partial t^{2} } \bigg)
& =
\begin{cases}
-1
& \mathrm{if} \ \tau(n; r, s) < t < 1 ,
\\
0
& \mathrm{if} \ t = \tau(n; r, s) ,
\\
1
& \mathrm{if} \ n^{\theta( r )} < t < \tau(n; r, s)
\end{cases}
\label{eq:sgn_diff2_norm_r_s_v_5}
\end{align}
for every $t \in \mathcal{I}_{n}( r ) \setminus \{ 1, n^{\theta( r )} \}$,
\item
if $1 < s < r$, then
\begin{align}
\sgn \bigg( \frac{ \partial^{2} \| \bvec{v}_{2}( N_{r}^{-1}( \bvec{v}_{2} : t ) ) \|_{s} }{ \partial t^{2} } \bigg)
& =
\begin{cases}
-1
& \mathrm{if} \ n^{\theta( r )} < t < \tau(n; r, s) ,
\\
0
& \mathrm{if} \ t = \tau(n; r, s) ,
\\
1
& \mathrm{if} \ \tau(n; r, s) < t < 1
\end{cases}
\label{eq:sgn_diff2_norm_r_s_v_6}
\end{align}
for every $t \in \mathcal{I}_{n}( r ) \setminus \{ 1, n^{\theta( r )} \}$,
\end{itemize}
Combining \eqref{eq:sgn_diff2_norm_r_s_v_1}--\eqref{eq:sgn_diff2_norm_r_s_v_6}, we obtain that for every $n \in \mathbb{N}_{\ge 3}$, distinct $r, s \in [1/2, 1) \cup (1, \infty)$, and $t \in \mathcal{I}_{n}( r )$,
\begin{itemize}
\item
if $\gamma( r, s ) > 1$, then
\begin{align}
\sgn \bigg( \frac{ \partial^{2} \| \bvec{v}_{2}( N_{r}^{-1}( \bvec{v}_{2} : t ) ) \|_{s} }{ \partial t^{2} } \bigg)
& =
\begin{cases}
-1
& \mathrm{if} \ t \in \mathcal{I}_{n}^{(2)}( r, s ) ,
\\
0
& \mathrm{if} \ t = \tau(n; r, s) ,
\\
1
& \mathrm{if} \ t \in \mathcal{I}_{n}^{(1)}( r, s ) ,
\end{cases}
\end{align}
\item
if $\gamma( r, s ) < 1$, then
\begin{align}
\sgn \bigg( \frac{ \partial^{2} \| \bvec{v}_{2}( N_{r}^{-1}( \bvec{v}_{2} : t ) ) \|_{s} }{ \partial t^{2} } \bigg)
& =
\begin{cases}
-1
& \mathrm{if} \ t \in \mathcal{I}_{n}^{(1)}( r, s ) ,
\\
0
& \mathrm{if} \ t = \tau(n; r, s) ,
\\
1
& \mathrm{if} \ t \in \mathcal{I}_{n}^{(2)}( r, s ) ,
\end{cases}
\end{align}
\end{itemize}
where $\mathcal{I}_{n}^{(1)}( r, s )$ and $\mathcal{I}_{n}^{(2)}( r, s )$ are defined in \eqref{def:In_1} and \eqref{def:In_2}, respectively.
This completes the proof of \lemref{lem:convex_v}.
\end{IEEEproof}

\begin{lemma}
\label{lem:convex_w}
Define the real interval $\mathcal{J}_{m}( r )$ by%
\footnote{Note in \eqref{def:Jm} that for every $m \in \mathbb{N}$, it holds that $m^{\theta( r )} < (m+1)^{\theta( r )}$ if $r \in (0, 1)$, and $(m+1)^{\theta( r )} < m^{\theta( r )}$ if $r \in (1, \infty]$.}
\begin{align}
\mathcal{J}_{m}( r )
& \coloneqq
\begin{cases}
\big[ m^{\theta( r )}, (m+1)^{\theta( r )} \big]
& \mathrm{if} \ 0 < r < 1 ,
\\
\big[ (m+1)^{\theta( r )}, m^{\theta( r )} \big]
& \mathrm{if} \ 1 < r \le \infty .
\end{cases}
\label{def:Jm}
\end{align}
For each $m \in \mathbb{N}$ and distinct $r, s \in (0, 1) \cup (1, \infty]$, the following convexity/concavity holds:
\begin{itemize}
\item
if $\gamma( r, s ) > 1$, then $t \mapsto \| \bvec{w}( N_{r}^{-1}( \bvec{w} : t ) ) \|_{s}$ is strictly convex in $t \in \mathcal{J}_{m}( r )$,
\item
if $\gamma( r, s ) < 1$, then $t \mapsto \| \bvec{w}( N_{r}^{-1}( \bvec{w} : t ) ) \|_{s}$ is strictly concave in $t \in \mathcal{J}_{m}( r )$.
\end{itemize}
\end{lemma}

\begin{IEEEproof}[Proof of \lemref{lem:convex_w}]
In a similar manner to the proof of \cite[Lemma~3]{part2}, we also prove this lemma by verifying signs of derivatives, as with the proof of \lemref{lem:convex_v}.
A simple calculation yields
\begin{align}
\frac{ \partial^{2} \| \bvec{w}( p ) \|_{r} }{ \partial p^{2} }
& \overset{\eqref{diff1:norm_w}}{=}
\frac{ \partial }{ \partial p } \bigg( m \, \Big( m \, p^{r} + \big( 1 - m \, p \big)^{r} \Big)^{(1/r) - 1} \Big( p^{r-1} - \big( 1 - m \, p \big)^{r-1} \Big) \bigg)
\\
& =
m \, \bigg( \frac{ \partial }{ \partial p } \Big( m \, p^{r} + \big( 1 - m \, p \big)^{r} \Big)^{(1/r) - 1} \bigg) \Big( p^{r-1} - \big( 1 - m \, p \big)^{r-1} \Big)
\notag \\
& \qquad \qquad
+ m \, \Big( m \, p^{r} + \big( 1 - m \, p \big)^{r} \Big)^{(1/r) - 1} \bigg( \frac{ \partial }{ \partial p } \Big( p^{r-1} - \big( 1 - m \, p \big)^{r-1} \Big) \bigg)
\\
& =
m \, \bigg( \frac{ 1 - r }{ r } \Big( m \, p^{r} + \big( 1 - m \, p \big)^{r} \Big)^{(1/r) - 2} \bigg( \frac{ \partial }{ \partial p } \Big( m \, p^{r} + \big( 1 - m \, p \big)^{r} \Big) \bigg) \bigg) \Big( p^{r-1} - \big( 1 - m \, p \big)^{r-1} \Big)
\notag \\
& \qquad \qquad
+ m \, \Big( m \, p^{r} + \big( 1 - m \, p \big)^{r} \Big)^{(1/r) - 1} \Big( (r-1) \, p^{r-2} + m \, (r-1) \big( 1 - m \, p \big)^{r-2} \Big)
\\
& =
m \, \frac{ 1 - r }{ r } \Big( m \, p^{r} + \big( 1 - m \, p \big)^{r} \Big)^{(1/r) - 2} \Big( m \, r \, p^{r-1} - m \, r \, \big( 1 - m \, p \big)^{r-1} \Big) \Big( p^{r-1} - \big( 1 - m \, p \big)^{r-1} \Big)
\notag \\
& \qquad \qquad
+ m \, (r-1) \, \Big( m \, p^{r} + \big( 1 - m \, p \big)^{r} \Big)^{(1/r) - 1} \Big( p^{r-2} + m \, \big( 1 - m \, p \big)^{r-2} \Big)
\\
& =
m^{2} \, (1-r) \, \Big( m \, p^{r} + \big( 1 - m \, p \big)^{r} \Big)^{(1/r) - 2} \Big( p^{r-1} - \big( 1 - m \, p \big)^{r-1} \Big)^{2}
\notag \\
& \qquad \qquad
+ m \, (r-1) \, \Big( m \, p^{r} + \big( 1 - m \, p \big)^{r} \Big)^{(1/r) - 1} \Big( p^{r-2} + m \, \big( 1 - m \, p \big)^{r-2} \Big)
\\
& =
m \, (1-r) \, \Big( m \, p^{r} + \big( 1 - m \, p \big)^{r} \Big)^{(1/r) - 2}
\notag \\
& \qquad \times
\bigg[ m \, \Big( p^{r-1} - \big( 1 - m \, p \big)^{r-1} \Big)^{2} - \Big( m \, p^{r} + \big( 1 - m \, p \big)^{r} \Big) \Big( p^{r-2} + m \, \big( 1 - m \, p \big)^{r-2} \Big) \bigg]
\\
& \overset{\text{(a)}}{=}
m \, (1-r) \, \Psi_{2}(m, p, r) \, \Big( m \, p^{r} + \big( 1 - m \, p \big)^{r} \Big)^{(1/r) - 2}
\\
& \overset{\text{(b)}}{=}
(r-1) \, m \, p^{r-2} \, \big( 1 - m \, p \big)^{r-2} \, \Big( m \, p^{r} + \big( 1 - m \, p \big)^{r} \Big)^{(1/r) - 2}
\label{eq:diff2_norm_w}
\end{align}
for every $m \in \mathbb{N}$, $r \in (0, \infty)$, and $p \in (1/(m+1), 1/m)$, where (a) follows by the definition
\begin{align}
\Psi_{2}(m, p, r)
\coloneqq
m \, \Big( p^{r-1} - \big( 1 - m \, p \big)^{r-1} \Big)^{2} - \Big( m \, p^{r} + \big( 1 - m \, p \big)^{r} \Big) \Big( p^{r-2} + m \, \big( 1 - m \, p \big)^{r-2} \Big) ,
\end{align}
and (b) follows from the fact that
\begin{align}
\Psi_{2}(m, p, r)
& =
\Big[ m \, p^{2(r-1)} - 2 \, m \, p^{r-1} \, \big( 1 - m \, p \big)^{r-1} + m \, \big( 1 - m \, p \big)^{2(r-1)} \Big]
\notag \\
& \qquad \qquad
- \Big[ m \, p^{2(r-1)} + m^{2} \, p^{r} \, \big( 1 - m \, p \big)^{r-2} + p^{r-2} \, \big( 1 - m \, p \big)^{r} + m \, \big( 1 - m \, p \big)^{2(r-1)} \Big]
\\
& =
- 2 \, m \, p^{r-1} \, \big( 1 - m \, p \big)^{r-1} - m^{2} \, p^{r} \, \big( 1 - m \, p \big)^{r-2} - p^{r-2} \, \big( 1 - m \, p \big)^{r}
\\
& =
- p^{r-2} \, \big( 1 - m \, p \big)^{r-2} \, \Big( 2 \, m \, p \, \big( 1 - m \, p \big) + m^{2} \, p^{2} + \big( 1 - m \, p \big)^{2} \Big)
\\
& =
- p^{r-2} \, \big( 1 - m \, p \big)^{r-2} \, \big( m \, p + \big( 1 - m \, p \big) \big)^{2}
\\
& =
- p^{r-2} \, \big( 1 - m \, p \big)^{r-2} .
\end{align}
Then, we obtain
\begin{align}
\sgn\bigg( \frac{ \partial^{2} \| \bvec{w}( p ) \|_{r} }{ \partial p^{2} } \bigg)
& \overset{\eqref{eq:diff2_norm_w}}{=}
\sgn(r-1) \, \underbrace{ \sgn\Big( m \, p^{r-2} \, \big( 1 - m \, p \big)^{r-2} \Big) }_{=1} \, \underbrace{ \sgn\bigg( \Big( m \, p^{r} + \big( 1 - m \, p \big)^{r} \Big)^{(1/r) - 2} \bigg) }_{=1}
\\
& =
\sgn(r-1)
\\
& =
\begin{cases}
-1
& \mathrm{if} \ r < 1 ,
\\
0
& \mathrm{if} \ r = 1 ,
\\
1
& \mathrm{if} \ r > 1
\end{cases}
\label{eq:sgn_norm_w_diff2}
\end{align}
for every $m \in \mathbb{N}$, $p \in (1/(m+1), 1/m)$, and $r \in (0, \infty)$.
By the inverse function theorem, we have
\begin{align}
\frac{ \partial N_{r}^{-1}( \bvec{w} : t ) }{ \partial t }
& =
\bigg( \frac{ \partial \| \bvec{w}( p ) \|_{r} }{ \partial p } \bigg)^{-1} ,
\label{eq:inverse_norm_w_diff1} \\
\frac{ \partial^{2} N_{r}^{-1}( \bvec{w} : t ) }{ \partial t^{2} }
& =
- \frac{ \partial^{2} \| \bvec{w}( p ) \|_{r} }{ \partial p^{2} } \, \bigg( \frac{ \partial \| \bvec{w}( p ) \|_{r} }{ \partial p } \bigg)^{-3}
\label{eq:inverse_norm_w_diff2}
\end{align}
for every $m \in \mathbb{N}$, $r \in (0, 1) \cup (1, \infty)$, and $t \in \mathcal{J}_{m}( r ) \setminus \{ m^{\theta( r )}, (m+1)^{\theta( r )} \}$, where $\mathcal{J}_{m}( \cdot )$ is defined in \eqref{def:Jm}, and the variables $t$ and $p$ are chosen to satisfy $\| \bvec{
w}( p ) \|_{r} = t$ (cf. the definition \eqref{def:inv_Nw} of $N_{r}^{-1}( \bvec{w} : \cdot )$), i.e.,
\begin{align}
1/(m+1) < p < 1/m
\iff
\min\{ m^{\theta( r )}, (m+1)^{\theta( r )} \} < t < \max\{ m^{\theta( r )}, (m+1)^{\theta( r )} \} .
\label{eq:p_t_Jm}
\end{align}
In particular, since $\| \bvec{w}( p ) \|_{\infty} = p$ for $p \in (0, 1]$, it follows from \eqref{eq:sgn_diff1_norm_w} and \eqref{eq:sgn_norm_w_diff2} that
\begin{align}
\sgn\bigg( \frac{ \partial^{2} \| \bvec{w}( N_{r}^{-1}( \bvec{w} : t ) ) \|_{\infty} }{ \partial t^{2} } \bigg)
& =
\sgn\bigg( \frac{ \partial^{2} N_{r}^{-1}( \bvec{w} : t ) }{ \partial t^{2} } \bigg)
\\
& \overset{\eqref{eq:inverse_norm_w_diff2}}{=}
- \sgn\bigg( \frac{ \partial^{2} \| \bvec{w}( p ) \|_{r} }{ \partial p^{2} } \bigg) \, \sgn\bigg( \bigg( \frac{ \partial \| \bvec{w}( p ) \|_{r} }{ \partial p } \bigg)^{-3} \bigg)
\\
& =
- \sgn\bigg( \frac{ \partial^{2} \| \bvec{w}( p ) \|_{r} }{ \partial p^{2} } \bigg) \, \sgn\bigg( \frac{ \partial \| \bvec{w}( p ) \|_{r} }{ \partial p } \bigg)
\\
& =
- 1
\label{eq:sgn_norm_w_diff2_r_inf}
\end{align}
for every $m \in \mathbb{N}$, $r \in (0, 1) \cup (1, \infty)$, and $t \in \mathcal{J}_{m}( r ) \setminus \{ m^{\theta( r )}, (m+1)^{\theta( r )} \}$.
Moreover, since $N_{\infty}^{-1}( \bvec{w} : t ) = t$ for $t \in (0, 1]$, we also get
\begin{align}
\sgn\bigg( \frac{ \partial^{2} \| \bvec{w}( N_{\infty}^{-1}( \bvec{w} : t ) ) \|_{s} }{ \partial t^{2} } \bigg)
& =
\sgn\bigg( \frac{ \partial^{2} \| \bvec{w}( p ) \|_{s} }{ \partial p^{2} } \bigg)
\\
& \overset{\eqref{eq:sgn_norm_w_diff2}}{=}
\begin{cases}
-1
& \mathrm{if} \ s < 1 ,
\\
0
& \mathrm{if} \ s = 1 ,
\\
1
& \mathrm{if} \ s > 1
\end{cases}
\label{eq:sgn_norm_w_diff2_inf_s}
\end{align}
for every $m \in \mathbb{N}$, $t \in (1/(m+1), 1/m)$, and $s \in (0, 1) \cup (1, \infty)$.
Therefore, it follows from \eqref{eq:sgn_norm_w_diff2_r_inf} and \eqref{eq:sgn_norm_w_diff2_inf_s} that
\begin{itemize}
\item
for each $m \in \mathbb{N}$ and $r \in (0, 1) \cup (1, \infty)$, the $\ell_{\infty}$-norm $t \mapsto \| \bvec{w}( N_{r}^{-1}( \bvec{w} : t ) ) \|_{\infty}$ is strictly concave in $t \in \mathcal{J}_{m}( r )$,
\item
for each $m \in \mathbb{N}$ and $s \in (0, 1)$, the $\ell_{s}$-norm $t \mapsto \| \bvec{w}( N_{\infty}^{-1}( \bvec{w} : t ) ) \|_{s}$ is strictly concave in $t \in [1/(m+1), 1/m]$,
\item
for each $m \in \mathbb{N}$ and $s \in (1, \infty)$, the $\ell_{s}$-norm $t \mapsto \| \bvec{w}( N_{\infty}^{-1}( \bvec{w} : t ) ) \|_{s}$ is strictly convex in $t \in [1/(m+1), 1/m]$.
\end{itemize}

Henceforth, we consider the convexity/concavity of $t \mapsto \| \bvec{w}( N_{r}^{-1}( \bvec{w} : t ) ) \|_{s}$ with respect to $t \in \mathcal{J}_{m}( r )$ for each distinct $r, s \in (0, 1) \cup (1, \infty)$.
By the chain rule of derivatives, we have
\begin{align}
\frac{ \partial^{2} \| \bvec{w}( N_{r}^{-1}( \bvec{w} : t ) ) \|_{s} }{ \partial t^{2} }
& =
\frac{ \partial^{2} \| \bvec{w}( p ) \|_{s} }{ \partial p^{2} } \bigg( \frac{ \partial N_{r}^{-1}( \bvec{w} : t ) }{ \partial t } \bigg)^{2} + \frac{ \partial \| \bvec{w}( p ) \|_{s} }{ \partial p } \frac{ \partial^{2} N_{r}^{-1}( \bvec{w} : t ) }{ \partial t^{2} }
\\
& \overset{\eqref{eq:inverse_norm_w_diff1}}{=}
\frac{ \partial^{2} \| \bvec{w}( p ) \|_{s} }{ \partial p^{2} } \bigg( \frac{ \partial \| \bvec{w}( p ) \|_{r} }{ \partial p } \bigg)^{-2} + \frac{ \partial \| \bvec{w}( p ) \|_{s} }{ \partial p } \frac{ \partial^{2} N_{r}^{-1}( \bvec{w} : t ) }{ \partial t^{2} }
\\
& \overset{\eqref{eq:inverse_norm_w_diff2}}{=}
\frac{ \partial^{2} \| \bvec{w}( p ) \|_{s} }{ \partial p^{2} } \bigg( \frac{ \partial \| \bvec{w}( p ) \|_{r} }{ \partial p } \bigg)^{-2} - \frac{ \partial \| \bvec{w}( p ) \|_{s} }{ \partial p } \frac{ \partial^{2} \| \bvec{w}( p ) \|_{r} }{ \partial p^{2} } \bigg( \frac{ \partial \| \bvec{w}( p ) \|_{r} }{ \partial p } \bigg)^{-3}
\\
& =
\frac{ \partial^{2} \| \bvec{w}( p ) \|_{r} }{ \partial p^{2} } \frac{ \partial^{2} \| \bvec{w}( p ) \|_{s} }{ \partial p^{2} } \bigg( \frac{ \partial \| \bvec{w}( p ) \|_{r} }{ \partial p } \bigg)^{-3}
\notag \\
& \qquad \qquad \qquad \qquad \times
\Bigg[ \frac{ \partial \| \bvec{w}( p ) \|_{r} }{ \partial p } \bigg( \frac{ \partial^{2} \| \bvec{w}( p ) \|_{r} }{ \partial p^{2} } \bigg)^{-1} - \frac{ \partial \| \bvec{w}( p ) \|_{s} }{ \partial p } \bigg( \frac{ \partial^{2} \| \bvec{w}( p ) \|_{s} }{ \partial p^{2} } \bigg)^{-1} \Bigg]
\\
& \overset{\text{(a)}}{=}
\frac{ \partial^{2} \| \bvec{w}( p ) \|_{r} }{ \partial p^{2} } \frac{ \partial^{2} \| \bvec{w}( p ) \|_{s} }{ \partial p^{2} } \bigg( \frac{ \partial \| \bvec{w}( p ) \|_{r} }{ \partial p } \bigg)^{-3}
\notag \\
& \qquad \qquad \qquad \qquad \times
p^{2} \, (1 - m \, p) \, \bigg[ \big( z^{s} + (n-1) \big) \ln_{s} z - \big( z^{r} + (n-1) \big) \ln_{r} z \bigg]
\\
& \overset{\eqref{def:g}}{=}
- p^{2} \, (1 - m \, p) \, g(m+1, z; r, s) \, \frac{ \partial^{2} \| \bvec{w}( p ) \|_{r} }{ \partial p^{2} } \frac{ \partial^{2} \| \bvec{w}( p ) \|_{s} }{ \partial p^{2} } \bigg( \frac{ \partial \| \bvec{w}( p ) \|_{r} }{ \partial p } \bigg)^{-3} ,
\label{eq:diff2_2norm_w}
\end{align}
where (a) follows from
\begin{itemize}
\item
the change of variables as
\begin{align}
z = z(m, p) \coloneqq \frac{ 1 - m \, p }{ p } ,
\label{def:z2}
\end{align}
\item
the fact that
\begin{align}
\frac{ \partial \| \bvec{w}( p ) \|_{r} }{ \partial p } \bigg( \frac{ \partial^{2} \| \bvec{w}( p ) \|_{r} }{ \partial p^{2} } \bigg)^{-1}
& \overset{\eqref{diff1:norm_w}}{=}
m \, \Big( m \, p^{r} + \big( 1 - m \, p \big)^{r} \Big)^{(1/r) - 1} \Big( p^{r-1} - \big( 1 - m \, p \big)^{r-1} \Big) \, \bigg( \frac{ \partial^{2} \| \bvec{w}( p ) \|_{r} }{ \partial p^{2} } \bigg)^{-1}
\\
& \overset{\eqref{eq:diff2_norm_w}}{=}
m \, \Big( m \, p^{r} + \big( 1 - m \, p \big)^{r} \Big)^{(1/r) - 1} \Big( p^{r-1} - \big( 1 - m \, p \big)^{r-1} \Big)
\notag \\
& \qquad \quad \times
(r-1)^{-1} \, m^{-1} \, p^{2-r} \, \big( 1 - m \, p \big)^{2-r} \, \Big( m \, p^{r} + \big( 1 - m \, p \big)^{r} \Big)^{2-(1/r)}
\\
& =
(r-1)^{-1} \, \big( p \, (1 - m \, p) \big)^{2-r} \Big( m \, p^{r} + \big( 1 - m \, p \big)^{r} \Big) \, \Big( p^{r-1} - \big( 1 - m \, p \big)^{r-1} \Big)
\\
& =
\frac{ p \, (1 - m \, p) }{ r - 1 } \, \Big( m \, p^{r} + ( 1 - m \, p )^{r} \Big) \, \Big( (1 - m \, p)^{1-r} - p^{1-r} \Big)
\\
& =
\frac{ p \, (1 - m \, p) }{ r - 1 } \, \Big( m \, p^{r} \, (1 - m \, p)^{1-r} - m \, p + (1 - m \, p) - p^{1-r} \, (1 - m \, p)^{r} \Big)
\\
& =
\frac{ p \, (1 - m \, p) }{ r - 1 } \, \bigg( (1 - 2 \, m \, p) + m \, p \, \Big( \frac{ 1 - m \, p }{ p } \Big)^{1-r} - (1 - m \, p) \, \Big( \frac{ 1 - m \, p }{ p } \Big)^{r-1} \bigg)
\\
& \overset{\eqref{def:z2}}{=}
\frac{ p \, (1 - m \, p) }{ r - 1 } \, \Big( (1 - 2 \, m \, p) + m \, p \, z^{1-r} - (1 - m \, p) \, z^{r-1} \Big)
\\
& =
\frac{ p \, (1 - m \, p) }{ r - 1 } \, \Big( (1 - z^{r-1}) + m \, p \, z^{1-r} \, (z^{2(r-1)} - 2 \, z^{r-1} + 1) \Big)
\\
& =
\frac{ p \, (1 - m \, p) }{ r - 1 } \, \Big( (1 - z^{r-1}) + m \, p \, z^{1-r} \, (z^{r-1} + 1)^{2} \Big)
\\
& =
\frac{ p \, (1 - m \, p) }{ r - 1 } \, (1 - z^{r-1}) \, \Big( 1 + m \, p \, z^{1-r} \, (1 - z^{r-1}) \Big)
\\
& =
p \, (1 - m \, p) \, \Big( 1 + m \, p \, (z^{1-r} - 1) \Big) \bigg( - z^{r-1} \, \frac{ z^{1-r} - 1 }{ 1 - r } \bigg)
\\
& \overset{\eqref{def:qlog}}{=}
- p \, (1 - m \, p) \, z^{r-1} \, \Big( 1 + m \, p \, (z^{1-r} - 1) \Big) \, (\ln_{r} z)
\\
& \overset{\eqref{def:z2}}{=}
- p \, (1 - m \, p) \, z^{r-1} \, \Big( 1 + m \, \Big( \frac{ 1 }{ m + z } \Big) \, (z^{1-r} - 1) \Big) \, (\ln_{r} z)
\\
& =
- \frac{ p \, (1 - m \, p) }{ m + z } \, z^{r-1} \, \Big( (m+z) + m \, (z^{1-r} - 1) \Big) \, (\ln_{r} z)
\\
& =
- \frac{ p \, (1 - m \, p) }{ m + z } \, z^{r-1} \, \big( z + m \, z^{1-r} \big) \, (\ln_{r} z)
\\
& =
- \frac{ p \, (1 - m \, p) }{ m + z } \, \big( m + z^{r} \big) \, (\ln_{r} z)
\\
& \overset{\eqref{def:z2}}{=}
- \frac{ p^{2} \, (1 - m \, p) }{ m \, p + (1 - m \, p) } \, \big( m + z^{r} \big) \, (\ln_{r} z)
\\
& =
- p^{2} \, (1 - m \, p) \, \big( m + z^{r} \big) \, (\ln_{r} z) .
\end{align}
\end{itemize}
Since $p \in (1/(m+1), 1/m)$ for $t \in \mathcal{J}_{m}( r ) \setminus \{ m^{\theta( r )}, (m+1)^{\theta( r )} \}$ (cf. \eqref{eq:p_t_Jm}), it suffices to consider the range of variable $z$ of \eqref{def:z2} on $z \in (0, 1)$.
A further calculation derives
\begin{align}
&
\sgn \bigg( \frac{ \partial^{2} \| \bvec{v}_{n}( N_{r}^{-1}( \bvec{v}_{n} : t ) ) \|_{s} }{ \partial t^{2} } \bigg)
\notag \\
& \qquad \overset{\eqref{eq:diff2_2norm_w}}{=}
- \underbrace{ \sgn\Big( p^{2} \, (1 - m \, p) \Big) }_{ = 1 } \, \sgn\Big( g(m+1, z; r, s) \Big) \, \sgn\bigg( \frac{ \partial^{2} \| \bvec{w}( p ) \|_{r} }{ \partial p^{2} } \bigg) \, \sgn\bigg( \frac{ \partial^{2} \| \bvec{w}( p ) \|_{s} }{ \partial p^{2} } \bigg) \, \sgn\bigg( \bigg( \frac{ \partial \| \bvec{w}( p ) \|_{r} }{ \partial p } \bigg)^{-3} \bigg)
\\
& \qquad =
- \sgn\Big( g(m+1, z; r, s) \Big) \, \sgn\bigg( \frac{ \partial^{2} \| \bvec{w}( p ) \|_{r} }{ \partial p^{2} } \bigg) \, \sgn\bigg( \frac{ \partial^{2} \| \bvec{w}( p ) \|_{s} }{ \partial p^{2} } \bigg) \, \sgn\bigg( \frac{ \partial \| \bvec{w}( p ) \|_{r} }{ \partial p } \bigg)
\\
& \qquad \overset{\eqref{eq:sgn_norm_w_diff2_r_inf}}{=}
- \sgn\Big( g(m+1, z; r, s) \Big) \, \sgn\bigg( \frac{ \partial^{2} \| \bvec{w}( p ) \|_{s} }{ \partial p^{2} } \bigg)
\\
& \qquad \overset{\eqref{eq:sgn_norm_w_diff2}}{=}
\begin{cases}
\sgn \Big( g(m+1, z; r, s) \Big)
& \mathrm{if} \ s < 1 ,
\\
- \sgn \Big( g(m+1, z; r, s) \Big)
& \mathrm{if} \ s > 1
\end{cases}
\label{eq:sgn_diff2_norm_w_r_s}
\end{align}
for $m \in \mathbb{N}$, distinct $r, s \in (0, 1) \cup (1, \infty)$, and $t \in \mathcal{J}_{m}( r ) \setminus \{ m^{\theta( r )}, (m+1)^{\theta( r )} \}$.
That is, the convexity/concavity of $t \mapsto \| \bvec{w}( N_{r}^{-1}( \bvec{w} : t ) ) \|_{s}$ with respect to $t \in \mathcal{J}_{m}( r )$ depend on the sign of $g(m+1, z; r, s)$.
Combining \eqref{eq:sgn_g_1} of \lemref{lem:sgn_g} and \eqref{eq:sgn_diff2_norm_w_r_s}, we have
\begin{align}
\sgn \bigg( \frac{ \partial^{2} \| \bvec{v}_{n}( N_{r}^{-1}( \bvec{v}_{n} : t ) ) \|_{s} }{ \partial t^{2} } \bigg)
& =
\begin{cases}
-1
& \mathrm{if} \ r < 1 < s \ \mathrm{or} \ 1 < r < s \ \mathrm{or} \ s < r < 1 \ \mathrm{or} \ s < 1 < r ,
\\
1
& \mathrm{if} \ r < s < 1 \ \mathrm{or} \ 1 < s < r
\end{cases}
\\
& =
\begin{cases}
-1
& \mathrm{if} \ \gamma( r, s ) < 1 ,
\\
1
& \mathrm{if} \ \gamma( r, s ) > 1
\end{cases}
\end{align}
for every $m \in \mathbb{N}$, distinct $r, s \in (0, 1) \cup (1, \infty)$, and $t \in \mathcal{J}_{m}( r ) \setminus \{ m^{\theta( r )}, (m+1)^{\theta( r )} \}$, where $\gamma( r, s )$ is defined in \eqref{def:gamma}.
This completes the proof of \lemref{lem:convex_w}.
\end{IEEEproof}

\section{Sharp Bounds on Unconditional R\'{e}nyi Entropy}
\label{sect:unconditional}

In this section, we introduce sharp bounds on the R\'{e}nyi entropy $H_{\beta}( X )$ with a fixed another one $H_{\alpha}( X )$, studied in \cite{itw2016_reject}.
We first show extremality of the distribution $\bvec{v}_{n}( \cdot )$ defined in \eqref{def:v} in terms of the relation between $\ell_{r}$-norm and $\ell_{s}$-norm in the following theorem.

\begin{theorem}[{\cite[Lemma~2]{itw2016_reject}}]
\label{th:v}
Let $P$ be a discrete probability distribution with finite support, and let $n = |\! \supp(P)|$.
For any $r, s \in (0, 1) \cup (1, \infty]$, it holds that
\begin{align}
\| \bvec{v}_{n}( p ) \|_{s}
& \le
\| P \|_{s}
\qquad \mathrm{if} \ \gamma( r, s ) \ge 1 ,
\label{eq:bound_v_1} \\
\| \bvec{v}_{n}( p ) \|_{s}
& \ge
\| P \|_{s}
\qquad \mathrm{if} \ \gamma( r, s ) \le 1
\label{eq:bound_v_2}
\end{align}
with $p = N_{r}^{-1}( \bvec{v}_{n} : \| P \|_{r} )$, where $N_{r}^{-1}( \bvec{v}_{n} : \cdot )$ and $\gamma( r, s )$ are defined in \eqref{def:inv_Nv} and \eqref{def:gamma}, respectively.
\end{theorem}

\begin{IEEEproof}[Proof of \thref{th:v}]
In the original version \cite[Lemma~2]{itw2016_reject}, we wrote this proposition as follows:
Let $P$ be a discrete probability distributions with finite support, i.e., $|\!\supp( P )| = n$ for some $n \in \mathbb{N}_{\ge 2}$.
For any $r \in (0, 1) \cup (1, \infty]$, there exists $p \in [1/n, 1]$ such that
\begin{align}
\| \bvec{v}_{n}( p ) \|_{r}
& =
\| P \|_{r} ,
\label{eq:prev_v_1} \\
\| \bvec{v}_{n}( p ) \|_{s}
& \le
\| P \|_{s}
\quad \mathrm{for} \ \mathrm{all} \ s \in (\min\{ 1, r \}, \max\{ 1, r \}) ,
\label{eq:prev_v_2} \\
\| \bvec{v}_{n}( p ) \|_{s}
& \ge
\| P \|_{s}
\quad \mathrm{for} \ \mathrm{all} \ s \in (0, \min\{ 1, r \}) \cup (\max\{ 1, r \}, \infty] .
\label{eq:prev_v_3} 
\end{align}

It is obvious that the value $p$ satisfying \eqref{eq:prev_v_1} is determined as $p = N_{r}^{-1}( \bvec{v}_{n} : \| P \|_{r} )$.
It follows from the definition \eqref{def:gamma} of $\gamma( r, s )$ that
\begin{align}
s \in (\min\{ 1, r \}, \max\{ 1, r \})
& \iff
\gamma( r, s ) > 1 ,
\\
s \in (0, \min\{ 1, r \}) \cup (\max\{ 1, r \}, \infty]
& \iff
\gamma( r, s ) < 1 .
\end{align}
Moreover, the case $\gamma( r, s ) = 1$ implies $r = s$, i.e., it is a trivial case.
Therefore, the statements of \eqref{eq:prev_v_1}--\eqref{eq:prev_v_3} shown in \cite[Lemma~2]{itw2016_reject} can be rewritten as \thref{th:v}.
\end{IEEEproof}

We second show extremality of the distribution $\bvec{w}( \cdot )$ defined in \eqref{def:w} in terms of the relation between $\ell_{r}$-norm and $\ell_{s}$-norm in the following theorem.

\begin{theorem}[{\cite[Lemma~3]{itw2016_reject}}]
\label{th:w}
Let $P$ be a discrete probability distribution with possibly countably infinite support.
For any $r, s \in (0, 1) \cup (1, \infty]$, it holds that
\begin{align}
\| \bvec{w}( p ) \|_{s}
& \ge
\| P \|_{s}
\qquad \mathrm{if} \ \gamma( r, s ) \ge 1 ,
\\
\| \bvec{w}( p ) \|_{s}
& \le
\| P \|_{s}
\qquad \mathrm{if} \ \gamma( r, s ) \le 1
\end{align}
with $p = N_{r}^{-1}( \bvec{w} : \| P \|_{r} )$, where $N_{r}^{-1}( \bvec{w} : \cdot )$ and $\gamma( r, s )$ are defined in \eqref{def:inv_Nw} and \eqref{def:gamma}, respectively.
\end{theorem}

\begin{IEEEproof}[Proof of \thref{th:w}]
In the proof of \cite[Lemma~2]{itw2016_reject}, we considered only for finite-dimensional probability vectors as follows:
Let $\bvec{p} = (p_{1}, p_{2}, \dots, p_{n})$ be an $n$-dimensional probability vector satisfying
\begin{align}
p_{i}
& \ge
0
\qquad \mathrm{for} \ i = 1, 2, \dots, n,
\\
\sum_{i = 1}^{n} p_{i}
& =
1 .
\end{align}
Since the equiprobable distribution is a trivial case, suppose that $\bvec{p} = (1/n, 1/n, \dots, 1/n)$ is omitted.
Let $k \in \{ 2, 3, \dots, n-1 \}$ and $l \in \{ k+1, k+2, \dots, n \}$ be positive integers chosen so that
\begin{align}
p_{[1]} = \dots = p_{[k-1]} \ge p_{[k]} \ge p_{[k+1]} \ge \dots \ge p_{[l-1]} \ge p_{[l]} > p_{[l+1]} = \dots = p_{[n]} = 0
\quad (p_{[k-1]} > p_{[k+1]}) ,
\label{eq:equal_1_to_k-1}
\end{align}
where
\begin{align}
p_{[1]} \ge p_{[2]} \ge \dots \ge p_{[n]}
\end{align}
denotes the components of $\bvec{p}$ in decreasing order%
\footnote{We used this notation by following the book of Marshall and Olkin \cite{marshall}.}.
Then, total derivatives of the probability vector $\bvec{p}$ was considered in the following assumptions:
\begin{align}
\| \bvec{p} \|_{r}
& =
A
&& \mathrm{for \ some \ constant} \ A \in \mathcal{I}_{n}( r ) ,
\\
\frac{ \mathrm{d} p_{[i]} }{ \mathrm{d} p_{[k]} }
& =
\frac{ \mathrm{d} p_{[1]} }{ \mathrm{d} p_{[k]} }
&& \mathrm{for} \ i \in \{ 2, 3, \dots, k-1 \} ,
\label{eq:hypo1_w} \\
\frac{ \mathrm{d} p_{[j]} }{ \mathrm{d} p_{[k]} }
& =
1
&& \mathrm{for} \ j \in \{ k+1, k+2, \dots, l-1 \} ,
\label{eq:hypo2_w} \\
\frac{ \mathrm{d} p_{[m]} }{ \mathrm{d} p_{[k]} }
& =
0
&& \mathrm{for} \ m \in \{ l+1, l+2, \dots, n \} ,
\label{eq:hypo3_w}
\end{align}
and the parameter $r \in (0, 1) \cup (1, \infty)$ is fixed.
Due to the hypothesis of \eqref{eq:equal_1_to_k-1}, the support size of $\bvec{p}$ is $l \in \mathbb{N}$;
thus, \cite[Lemma~2]{itw2016_reject} only proved for probability distributions with finite support.

Fortunately, considering infinite-dimensional probability vector $\bvec{p} = (p_{1}, p_{2}, \dots)$ and extendind the hypothesis of \eqref{eq:hypo3_w} for every $m \in \{ l+1, l+2, \dots \}$, we can remove the hypothesis of the finite support.
That is, the analyses of the proof of \cite[Lemma~2]{itw2016_reject} can naturally generalized to probability distributions with possibly countably infinite support.

Moreover, in the proof of \cite[Lemma~2]{itw2016_reject}, we examined $\ell_{\infty}$-norm by majorization theory \cite{marshall}.
This analysis can also be extended from finite- to infinite-dimensional probability vectors, as with the proof of \cite[Theorem~10]{verdu} studied by Ho and Verd\'{u}.
\end{IEEEproof}

We now consider the function
\begin{align}
f_{\alpha}( t )
\coloneqq
\frac{ \ln t }{ \theta( \alpha ) } 
\end{align}
for each $\alpha \in (0, 1) \cup (1, \infty]$ and $t > 0$, where $\theta( \cdot )$ is defined in \eqref{def:theta}.
Since
\begin{itemize}
\item
it follows from \eqref{def:renyi} that $H_{\alpha}( P ) = f_{\alpha}( \| P \|_{\alpha} )$ for every $\alpha \in (0, 1) \cup (1, \infty]$,
\item
if $\alpha \in (0, 1)$, then $t \mapsto f_{\alpha}( t )$ is strictly increasing for $t > 0$,
\item
if $\alpha \in (1, \infty]$, then $t \mapsto f_{\alpha}( t )$ is strictly decreasing for $t > 0$,
\end{itemize}
Theorems~\ref{th:v} and~\ref{th:w} can be rewritten from sharp bounds on the $\ell_{s}$-norm $\| P \|_{s}$ to sharp bounds on the R\'{e}nyi entropy $H_{\beta}( P )$ as shown in the following two theorems:

\begin{theorem}[{\cite[Theorem~2]{itw2016_reject}}]
\label{th:renyi_v}
Let $P$ be a discrete probability distribution with finite support, i.e., $|\!\supp(P)| = n$ for some $n \in \mathbb{N}$.
Then, it holds that
\begin{align}
H_{\beta}( P )
& \ge
H_{\beta}( \bvec{v}_{n}( p ) )
\qquad \mathrm{for} \ 0 < \alpha \le \beta \le \infty ,
\label{ineq:renyi_v_1} \\
H_{\beta}( P )
& \le
H_{\beta}( \bvec{v}_{n}( p ) )
\qquad \mathrm{for} \ 0 < \beta \le \alpha \le \infty ,
\label{ineq:renyi_v_2}
\end{align}
with $p = H_{\alpha}^{-1}( \bvec{v}_{n} : H_{\alpha}( P ) )$, where $H_{\alpha}^{-1}( \bvec{v}_{n} : \cdot )$ is defined in \eqref{def:inv_Hv}.
\end{theorem}

\begin{theorem}[{\cite[Theorem~2]{itw2016_reject}}]
\label{th:renyi_w}
Let $P$ be a discrete probability distribution with possibly infinite support.
For any $\alpha \in (0, \infty]$, it holds that
\begin{align}
H_{\beta}( P )
& \le
H_{\beta}( \bvec{w}( p ) )
\qquad \mathrm{for} \ 0 < \alpha \le \beta \le \infty ,
\\
H_{\beta}( P )
& \ge
H_{\beta}( \bvec{w}( p ) )
\qquad \mathrm{for} \ 0 < \beta \le \alpha \le \infty
\end{align}
with $p = H_{\alpha}^{-1}( \bvec{w} : H_{\alpha}( P ) )$, where $H_{\alpha}^{-1}( \bvec{w} : \cdot )$ is defined in \eqref{def:inv_Hw}.
\end{theorem}

Note that \thref{th:renyi_v} has a constraint of finite supports, but \thref{th:renyi_w} enables us to consider countably infinite supports.
If $\alpha = \infty$, then \thref{th:renyi_v} is equivalent to the result by Ben-Bassat and Raviv \cite[Theorem~6]{ben-bassat};
and if $\alpha = \infty$, then \thref{th:renyi_w} is a stronger result than \cite[Theorems~4 and~5]{ben-bassat}.
In addition, Theorems~\ref{th:renyi_v} and~\ref{th:renyi_w} yield same joint ranges of pairs $(H_{\alpha}(P), H_{\beta}(P))$ considered in \cite{harremoes}.
In \cite[Theorem~2]{itw2016_reject}, Theorems~\ref{th:renyi_v} and~\ref{th:renyi_w} are organized in one theorem.
However, in this study, Theorems~\ref{th:w} and~\ref{th:renyi_w} are extended from probability distributions with finite support to countably infinite support.
Due to such extension, Theorems~\ref{th:renyi_v} and~\ref{th:renyi_w} are divided, and \thref{th:renyi_w} is generalized to possibly countably infinite support.
Since Theorems~\ref{th:renyi_v} and~\ref{th:renyi_w} are due to Theorems~\ref{th:v} and~\ref{th:w}, and the strict monotonicity of the logarithm functions, as with Theorems~\ref{th:renyi_v} and~\ref{th:renyi_w}, we can establish sharp bounds on other definitions of entropy \cite{behara, boekee, daroczy, havrda, tsallis2}, which are strictly monotonic for the $\ell_{r}$-norm of a probability distribution (cf. \cite[Table~I]{part1}).

In the next section, using the sharp bounds introduced in this section, we further consider to extend them to sharp bounds on the conditional R\'{e}nyi entropy $H_{\alpha}(X \mid Y)$.

\section{Sharp Bounds on Arimoto's Conditional R\'{e}nyi Entropy}
\label{sect:cond}

\subsection{Bounds Established from Distribution $\bvec{v}_{n}( \cdot )$}
\label{sect:v}

In this subsection, by using the extremality of the distribution $\bvec{v}_{n}( \cdot )$ discussed in \sectref{sect:unconditional}, we derive sharp bounds on $H_{\beta}(X \mid Y)$ with two fixed $H_{\alpha}(X \mid Y)$ and $|\!\supp( P_{X} )|$ in some situations.
We first give the sharp bounds, whose mean interplay between $H_{\alpha}(X \mid Y)$ and $H_{\infty}(X \mid Y)$ in the following theorem.

\begin{theorem}
\label{th:A_renyi_alpha_inf}
Let $X$ be an RV in which $|\!\supp(P_{X})| = n \in \mathbb{N}$, and let $Y$ be an arbitrary RV.
For any $\alpha \in (0, \infty)$, it holds that
\begin{align}
H_{\alpha}(X \mid Y)
& \le
H_{\alpha}( \bvec{v}_{n}( p_{1} ) ) ,
\label{ineq:alpha_inf} \\
H_{\infty}(X \mid Y)
& \ge
H_{\infty}( \bvec{v}_{n}( p_{2} ) )
\label{ineq:inf_alpha}
\end{align}
with $p_{1} = H_{\infty}^{-1}( \bvec{v}_{n} : H_{\infty}(X \mid Y) )$ and $p_{2} = H_{\alpha}^{-1}( \bvec{v}_{n} : H_{\alpha}(X \mid Y) )$, respectively, where $H_{\alpha}^{-1}( \bvec{v}_{n} : \cdot )$ is defined in \eqref{def:inv_Hv}.
\end{theorem}

\begin{IEEEproof}[Proof of \thref{th:A_renyi_alpha_inf}]
Let $X$ be an RV in which $|\!\supp(P_{X})| = n$ for some%
\footnote{If $|\!\supp( P_{X} )| = 1$, it is clear that $H_{\alpha}(X \mid Y) = 0$ for every $\alpha \in [0, \infty]$. That is, we omit such trivial cases in our analyses.}
$n \in \mathbb{N}_{\ge 2}$, and let $Y$ be an arbitrary RV.
If $\alpha = 1$, then \eqref{ineq:alpha_inf} is equivalent to Fano's inequality.
In fact, Inequality~\eqref{ineq:alpha_inf} is equivalent to the right-hand inequalities of \cite[Eq.~(15)]{kovalevsky} and \cite[Eq.~(5)]{tebbe}.
On the other hand, Inequality~\eqref{ineq:inf_alpha} with $\alpha = 1$ can be verified as follows:
\begin{align}
H_{\infty}(X \mid Y)
& \overset{\eqref{eq:cond_infty}}{=}
- \ln N_{\infty}(X \mid Y)
\\
& \overset{\eqref{def:expect_norm}}{=}
- \ln \mathbb{E}[ \| P_{X|Y}(\cdot \mid Y) \|_{\infty} ]
\\
& \overset{\eqref{def:min_entropy}}{=}
- \ln \mathbb{E}[ \exp( - H_{\infty}( P_{X|Y}(\cdot \mid Y) ) ) ]
\\
& \overset{\text{(a)}}{\ge}
- \ln \mathbb{E}[ \exp( - H_{\infty}( \bvec{v}_{n}( H^{-1}( \bvec{v}_{n} : H( P_{X|Y}(\cdot \mid Y) ) ) ) ) ) ]
\\
& \overset{\eqref{def:min_entropy}}{=}
- \ln \mathbb{E}[ \| \bvec{v}_{n}( H^{-1}( \bvec{v}_{n} : H( P_{X|Y}(\cdot \mid Y) ) ) ) \|_{\infty} ]
\\
& \overset{\text{(b)}}{=}
- \ln \mathbb{E}[ H^{-1}( \bvec{v}_{n} : H( P_{X|Y}(\cdot \mid Y) ) ) ]
\\
& \overset{\text{(c)}}{\ge}
- \ln H^{-1}( \bvec{v}_{n} : \mathbb{E}[ H( P_{X|Y}(\cdot \mid Y) ) ] )
\\
& \overset{\eqref{def:cond_shannon}}{=}
- \ln H^{-1}( \bvec{v}_{n} : H(X \mid Y) )
\\
& \overset{\text{(b)}}{=}
- \ln \| \bvec{v}_{n}( H^{-1}( \bvec{v}_{n} : H(X \mid Y) ) ) \|_{\infty}
\\
& \overset{\eqref{def:min_entropy}}{=}
H_{\infty}( \bvec{v}_{n}( H^{-1}( \bvec{v}_{n} : H(X \mid Y) ) )
\\
& =
H_{\infty}( \bvec{v}_{n}( p ) )
\end{align}
with $p = H^{-1}( \bvec{v}_{n} : H(X \mid Y) )$, where (a) follows from \eqref{ineq:renyi_v_1} of \thref{th:renyi_v} with $\alpha = 1$ and $\beta = \infty$, Equalities (b) follow from the fact that $\| \bvec{v}_{n}( p ) \|_{\infty} = p$ for $p \in [1/n, 1]$, and (c) follows from Jensen's inequality and the fact that
$\mu \mapsto H^{-1}( \bvec{v}_{n}: \mu )$ is strictly concave in $\mu \in [0, \ln n]$.
Note that the concavity of $\mu \mapsto H^{-1}( \bvec{v}_{n}: \mu )$ can be verified by the following two facts:
\begin{itemize}
\item
the function%
\footnote{The function $h_{2} : t \mapsto - t \ln t - (1-t) \ln (1-t)$ denotes the binary entropy function.}
$p \mapsto H( \bvec{v}_{n}( p ) ) = h_{2}( p ) + (1-p) \ln (n-1)$ is strictly decreasing for $p \in [1/n, 1]$,
\item
the function $p \mapsto H( \bvec{v}_{n}( p ) ) = h_{2}( p ) + (1-p) \ln (n-1)$ is strictly concave in $p \in [1/n, 1]$.
\end{itemize}
Therefore, both bounds of \thref{th:A_renyi_alpha_inf} hold for $\alpha = 1$.

We next consider to prove \eqref{ineq:alpha_inf} of \thref{th:A_renyi_alpha_inf} for $\alpha \in (0, 1) \cup (1, \infty)$.
Let $s \in (0, 1) \cup (1, \infty)$ be a fixed number.
Note that
\begin{align}
s \in (0, 1)
& \iff
\gamma( \infty, s ) = - \infty < 1 ,
\\
s \in (1, \infty)
& \iff
\gamma( \infty, s ) = \infty > 1 ,
\end{align}
where $\gamma(\cdot, \cdot)$ is defined in
\eqref{def:gamma}.
If $\gamma(\infty, s) < 1$, we have
\begin{align}
N_{s}(X \mid Y)
& \overset{\eqref{def:expect_norm}}{=}
\mathbb{E}\big[ \| P_{X|Y}(\cdot \mid Y) \|_{s} \big]
\\
& \overset{\text{(a)}}{\le}
\mathbb{E}\big[ \big\| \bvec{v}_{n}\big( N_{\infty}^{-1}(\bvec{v}_{n} : \| P_{X|Y}(\cdot \mid Y) \|_{\infty} ) \big) \big\|_{s} \big]
\\
& \overset{\text{(b)}}{\le}
\big\| \bvec{v}_{n}\big( N_{\infty}^{-1}\big(\bvec{v}_{n} : \mathbb{E}[ \| P_{X|Y}(\cdot \mid Y) \|_{\infty} ] \big) \big) \big\|_{s}
\\
& \overset{\eqref{def:expect_norm}}{=}
\big\| \bvec{v}_{n}\big( N_{\infty}^{-1}\big(\bvec{v}_{n} : N_{\infty}(X \mid Y) \big) \big) \big\|_{s}
\\
& \overset{\text{(c)}}{=}
\big\| \bvec{v}_{n}\big( N_{\infty}(X \mid Y) \big) \big\|_{s}
\\
& =
\| \bvec{v}_{n}( p ) \|_{s}
\label{eq:A_renyi_inf_proof1}
\end{align}
with $p = N_{\infty}(X \mid Y)$, where Inequality (a) follows from \eqref{eq:bound_v_2} of \thref{th:v}, Inequality (b) follows from Jensen's inequality and fact that $t \mapsto \| \bvec{v}_{n}( N_{\infty}^{-1}( \bvec{v}_{n} : t ) ) \|_{s}$ is strictly concave in $t \in [1/n, 1]$ (cf. \lemref{lem:convex_v}), and Equality (c) follows from the fact that $N_{\infty}^{-1}( \bvec{v}_{n} : t ) = t$ for $t \in [1/n, 1]$.
Analogously, if $\gamma(\infty, s) > 1$, then we also get
\begin{align}
N_{s}(X \mid Y)
\ge
\| \bvec{v}_{n}( p ) \|_{s}
\label{eq:A_renyi_inf_proof2}
\end{align}
with $p = N_{\infty}(X \mid Y)$.
We now define
\begin{align}
f_{\alpha}( t )
\coloneqq
\frac{ \alpha }{ 1 - \alpha } \ln t
\end{align}
for $\alpha \in (0, 1) \cup (1, \infty)$ and $t > 0$.
Since
\begin{itemize}
\item
it holds that $H_{\alpha}(X \mid Y) = f_{\alpha}( N_{\alpha}(X \mid Y) )$ for every $\alpha \in (0, 1)$,
\item
if $\alpha \in (0, 1)$, then $t \mapsto f_{\alpha}( t )$ is a strictly increasing function of $t > 0$,
\item
if $\alpha \in (1, \infty)$, then $t \mapsto f_{\alpha}( t )$ is a strictly decreasing function of $t > 0$,
\end{itemize}
it follows from \eqref{eq:A_renyi_inf_proof1} and \eqref{eq:A_renyi_inf_proof2} that
\begin{align}
H_{\alpha}(X \mid Y)
& \le
H_{\alpha}( \bvec{v}_{n}( p ) )
\label{eq:A_renyi_inf_proof3}
\end{align}
with $p = N_{\infty}(X \mid Y)$ for every $\alpha \in (0, 1) \cup (1, \infty)$.
In addition, since
\begin{align}
N_{\infty}(X \mid Y)
& =
\exp\Big[ - \Big( - \ln N_{\infty}(X \mid Y) \Big) \Big]
\\
& \overset{\eqref{eq:cond_infty}}{=}
\exp\Big[ - H_{\infty}(X \mid Y) \Big]
\\
& \overset{\eqref{eq:inv_Hv_infty}}{=}
H_{\infty}^{-1}( \bvec{v}_{n} : H_{\infty}(X \mid Y) ) ,
\end{align}
we get from \eqref{eq:A_renyi_inf_proof3} that
\begin{align}
H_{\alpha}(X \mid Y)
\le
H_{\alpha}( \bvec{v}_{n}( p ) )
\end{align}
with $p = H_{\infty}^{-1}( \bvec{v}_{n} : H_{\infty}(X \mid Y) )$ for every $\alpha \in (0, 1) \cup (1, \infty)$ rather than $p = N_{\infty}(X \mid Y)$, which is \eqref{ineq:alpha_inf} of \thref{th:A_renyi_alpha_inf}.

We further consider to prove \eqref{ineq:inf_alpha} of \thref{th:A_renyi_alpha_inf} for $\alpha \in (0, 1) \cup (1, \infty)$.
Note that
\begin{align}
r \in (0, 1) \cup (1, \infty)
\iff
\gamma( r, \infty ) = 0 < 1 .
\end{align}
Thus, we have
\begin{align}
N_{\infty}(X \mid Y)
& \overset{\eqref{def:expect_norm}}{=}
\mathbb{E}\big[ \| P_{X|Y}(\cdot \mid Y) \|_{\infty} \big]
\\
& \overset{\text{(a)}}{\le}
\mathbb{E}\big[ \big\| \bvec{v}_{n}\big( N_{r}^{-1}(\bvec{v}_{n} : \| P_{X|Y}(\cdot \mid Y) \|_{r} ) \big) \big\|_{\infty} \big]
\\
& \overset{\text{(b)}}{\le}
\big\| \bvec{v}_{n}\big( N_{r}^{-1}\big(\bvec{v}_{n} : \mathbb{E}[ \| P_{X|Y}(\cdot \mid Y) \|_{r} ] \big) \big) \big\|_{\infty}
\\
& \overset{\eqref{def:expect_norm}}{=}
\big\| \bvec{v}_{n}\big( N_{r}^{-1}\big(\bvec{v}_{n} : N_{r}(X \mid Y) \big) \big) \big\|_{\infty}
\\
& =
\| \bvec{v}_{n}( p ) \|_{\infty}
\label{eq:A_renyi_inf_proof4}
\end{align}
with $p = N_{r}^{-1}(\bvec{v}_{n} : N_{r}(X \mid Y) )$, where (a) follows from \eqref{eq:bound_v_2} of \thref{th:v}, and (b) follows from Jensen's inequality and fact that $t \mapsto \| \bvec{v}_{n}( N_{r}^{-1}( \bvec{v}_{n} : t ) ) \|_{\infty}$ is strictly concave in $t \in \mathcal{I}_{n}( r )$ (cf. \lemref{lem:convex_v}).
Thus, it holds that
\begin{align}
H_{\infty}(X \mid Y)
& \overset{\eqref{eq:cond_infty}}{=}
- \ln N_{\infty}(X \mid Y)
\\
& \overset{\eqref{eq:A_renyi_inf_proof4}}{\ge}
- \ln \| \bvec{v}_{n}( p ) \|_{\infty}
\\
& \overset{\eqref{def:min_entropy}}{=}
H_{\infty}( \bvec{v}_{n}( p ) )
\label{eq:A_renyi_inf_proof5}
\end{align}
with $p = N_{r}^{-1}(\bvec{v}_{n} : N_{r}(X \mid Y) )$.
Now, it follows from \eqref{def:inv_Nv} and \eqref{def:inv_Hv} that
\begin{align}
H_{\alpha}^{-1}( \bvec{v}_{n} : H_{\alpha}(X \mid Y) )
& \overset{\eqref{def:A_renyi}}{=}
H_{\alpha}^{-1} \bigg( \bvec{v}_{n} : \frac{ \alpha }{ 1 - \alpha } \ln N_{\alpha}(X \mid Y) \bigg)
\\
& \overset{\text{(a)}}{=}
N_{\alpha}^{-1}\Bigg( \bvec{v}_{n} : \exp \bigg( \frac{ 1 - \alpha }{ \alpha } \frac{ \alpha }{ 1 - \alpha} \ln N_{\alpha}(X \mid Y) \bigg) \Bigg)
\\
& =
N_{\alpha}^{-1}( \bvec{v}_{n} : N_{\alpha}(X \mid Y) ) ,
\label{eq:A_renyi_2_proof4}
\end{align}
where (a) follows from the fact that
\begin{align}
\mu
=
H_{\alpha}( \bvec{v}_{n}( p ) )
=
\frac{ \alpha }{ 1 - \alpha } \ln \| \bvec{v}_{n}( p ) \|_{\alpha}
\iff
p
=
H_{\alpha}^{-1}( \bvec{v}_{n} : \mu )
=
N_{\alpha}^{-1}\bigg( \bvec{v}_{n} : \exp\Big( \frac{ 1 - \alpha }{ \alpha } \, \mu \Big) \bigg) .
\end{align}
Thus, Inequality \eqref{eq:A_renyi_inf_proof5} can be restated as
\begin{align}
H_{\infty}(X \mid Y)
\ge
H_{\infty}( \bvec{v}_{n}( p ) )
\end{align}
with $p = H_{\alpha}^{-1}( \bvec{v}_{n} : H_{\alpha}(X \mid Y))$ rather than $p = N_{r}^{-1}(\bvec{v}_{n} : N_{r}(X \mid Y) )$.
This completes the proof of \thref{th:A_renyi_alpha_inf}.
\end{IEEEproof}

Since the minimum average probability of error $P_{\mathrm{e}}(X \mid Y)$ satisfies 
\begin{align}
P_{\mathrm{e}}(X \mid Y)
& \overset{\eqref{def:Pe}}{=}
\min_{f} \Pr( X \neq f(Y) )
\\
& =
1 - \max_{f} \Pr( X = f(Y) )
\\
& =
1 - \mathbb{E} \Big[ \max_{x \in \supp(P_{X|Y}(\cdot \mid Y))} P_{X|Y}(x \mid Y) \Big]
\\
& \overset{\eqref{def:expect_norm}}{=}
1 - N_{\infty}(X \mid Y)
\\
& =
1 - H_{\infty}^{-1}( \bvec{v}_{n} : H_{\infty}(X \mid Y) ) ,
\end{align}
Ineq.~\eqref{ineq:alpha_inf} of \thref{th:A_renyi_alpha_inf} can be seen as a generalization of Fano's inequality from $H(X \mid Y)$ to $H_{\alpha}(X \mid Y)$ (see also \cite{sason}).
Moreover, \thref{th:A_renyi_alpha_inf} is tighter than a generalized Fano's inequality \cite[Theorem~7]{iwamoto}, whose bounds another definition of conditional R\'{e}nyi entropy proposed by Hayashi \cite{hayashi}.
We defer to discuss this comparison until \sectref{sect:fano}.

On the other hand, the following theorem shows sharp bounds on $H_{\beta}(X \mid Y)$ with a fixed $H_{\alpha}(X \mid Y)$ when $X$ is a Bernoulli RV.

\begin{theorem}
\label{th:A_renyi_2}
Let $X$ be a Bernoulli RV, i.e., $|\!\supp( P_{X} )| \le 2$, and let $Y$ be an arbitrary RV.
Then, it holds that
\begin{align}
H_{\beta}(X \mid Y)
& \ge
H_{\beta}( \bvec{v}_{n}( p ) )
\qquad \mathrm{for} \ 1/2 \le \alpha \le \beta \le \infty ,
\label{ineq:A_renyi_2_1} \\
H_{\beta}(X \mid Y)
& \le
H_{\beta}( \bvec{v}_{n}( p ) )
\qquad \mathrm{for} \ 1/2 \le \beta \le \alpha \le \infty ,
\label{ineq:A_renyi_2_2}
\end{align}
with $n = 2$ and $p = H_{\alpha}^{-1}( \bvec{v}_{n} : H_{\alpha}(X \mid Y) )$, where $H_{\alpha}^{-1}( \bvec{v}_{n} : \cdot )$ is defined in \eqref{def:inv_Hv}.
In particular, if $\alpha = 1$, then \eqref{ineq:A_renyi_2_2} also holds for every $0 < \beta < 1/2$.
\end{theorem}

\begin{IEEEproof}[Proof of \thref{th:A_renyi_2}]
We prove this theorem in a similar manner to the proof of \thref{th:A_renyi_alpha_inf}.
If either $\alpha = \infty$ or $\beta = \infty$, then \thref{th:A_renyi_2} comes from \thref{th:A_renyi_alpha_inf}.
If $\alpha = \beta$, then \thref{th:A_renyi_2} is trivial%
\footnote{If $\alpha = \beta$, then both inequalities of \thref{th:A_renyi_2} hold with equality, because $H_{\alpha}( \bvec{v}_{n}( H_{\alpha}^{-1}( \bvec{v}_{n} : H_{\alpha}(X \mid Y) ) ) ) = H_{\alpha}(X \mid Y)$.}.
Hence, we consider otherwise.
Let $X$ be a Bernoulli RV in which $|\!\supp( P_{X} )| = n = 2$, let $Y$ be an arbitrary RV, and let $r, s \in [1/2, 1) \cup (1, \infty)$ be distinct numbers.
If $\gamma(r, s) < 1$, then we have
\begin{align}
N_{s}(X \mid Y)
& \overset{\eqref{def:expect_norm}}{=}
\mathbb{E}\big[ \| P_{X|Y}(\cdot \mid Y) \|_{s} \big]
\\
& \overset{\text{(a)}}{\le}
\mathbb{E}\big[ \big\| \bvec{v}_{2}\big( N_{r}^{-1}(\bvec{v}_{2} : \| P_{X|Y}(\cdot \mid Y) \|_{r} ) \big)  \big\|_{s} \big]
\\
& \overset{\text{(b)}}{\le}
\big\| \bvec{v}_{2}\big( N_{r}^{-1}\big(\bvec{v}_{2} : \mathbb{E}[ \| P_{X|Y}(\cdot \mid Y) \|_{r} ] \big) \big)  \big\|_{s}
\\
& \overset{\eqref{def:expect_norm}}{=}
\big\| \bvec{v}_{2}\big( N_{r}^{-1}\big(\bvec{v}_{2} : N_{r}(X \mid Y) \big) \big)  \big\|_{s}
\\
& =
\| \bvec{v}_{2}( p ) \|_{s}
\label{eq:A_renyi_2_proof1}
\end{align}
with $p = N_{r}^{-1}(\bvec{v}_{2} : N_{r}(X \mid Y))$, where $\gamma( r, s )$ is defined in \eqref{def:gamma}, Inequality (a) follows from \eqref{eq:bound_v_2} of \thref{th:v}, and Inequality (b) follows from Jensen's inequality and fact that $t \mapsto \| \bvec{v}_{2}( N_{r}^{-1}( \bvec{v}_{2} : t ) ) \|_{s}$ is strictly concave in $t \in \mathcal{I}_{2}( r )$ (cf. \lemref{lem:convex_v}).
Analogously, if $\gamma(r, s) > 1$, then we also get
\begin{align}
N_{s}(X \mid Y)
\ge
\| \bvec{v}_{2}( p ) \|_{s}
\label{eq:A_renyi_2_proof2}
\end{align}
with $p = N_{r}^{-1}(\bvec{v}_{2} : N_{r}(X \mid Y))$.
We now define
\begin{align}
f_{\alpha}( t )
\coloneqq
\frac{ \alpha }{ 1 - \alpha } \ln t
\end{align}
for $\alpha \in (0, 1) \cup (1, \infty)$ and $t > 0$.
Since
\begin{itemize}
\item
it holds that $H_{\beta}(X \mid Y) = f_{\beta}( N_{\beta}(X \mid Y) )$ for every $\beta \in (0, 1) \cup (1, \infty)$,
\item
if $\beta \in (0, 1)$, then $t \mapsto f_{\beta}( t )$ is a strictly increasing function of $t > 0$,
\item
if $\beta \in (1, \infty)$, then $t \mapsto f_{\beta}( t )$ is a strictly decreasing function of $t > 0$,
\end{itemize}
it follows from \eqref{eq:A_renyi_2_proof1} and \eqref{eq:A_renyi_2_proof2} that
\begin{align}
H_{\beta}(X \mid Y)
& \ge
H_{\beta}( \bvec{v}_{2}( p ) )
&& \mathrm{for} \ 1/2 \le \alpha < \beta < \infty ,
\label{eq:A_renyi_2_proof3} \\
H_{\beta}(X \mid Y)
& \le
H_{\beta}( \bvec{v}_{2}( p ) )
&& \mathrm{for} \ 1/2 \le \beta < \alpha < \infty
\label{eq:A_renyi_2_proof5}
\end{align}
with $p = N_{\alpha}^{-1}( \bvec{v}_{2} : N_{\alpha}(X \mid Y) )$ for every distinct $\alpha, \beta \in [1/2, 1) \cup (1, \infty)$, where (a) follows from \eqref{eq:A_renyi_2_proof1} and \eqref{eq:A_renyi_2_proof2}.
Combining \eqref{eq:A_renyi_2_proof4}, \eqref{eq:A_renyi_2_proof3}, and \eqref{eq:A_renyi_2_proof5}, we have \thref{th:A_renyi_2} for distinct $\alpha, \beta \in [1/2, 1) \cup (1, \infty)$.

Finally, if $\alpha = 1$, then \thref{th:A_renyi_2} can be proved by employing the concavity of \lemref{lem:Hconvex_v} and the extremality of \thref{th:renyi_v}.
In fact, it holds that for any $\beta \in (0, 1)$,
\begin{align}
H_{\beta}(X \mid Y)
& =
\frac{ \beta }{ 1 - \beta } \ln \mathbb{E}[ \| P_{X|Y}(\cdot \mid Y) \|_{\beta} ]
\\
& \overset{\eqref{def:renyi}}{=}
\frac{ \beta }{ 1 - \beta } \ln \mathbb{E}\bigg[ \exp\bigg( \frac{ 1 - \beta }{ \beta } \, H_{\beta}( P_{X|Y}(\cdot \mid Y) ) \bigg) \bigg]
\\
& \overset{\text{(a)}}{\le}
\frac{ \beta }{ 1 - \beta } \ln \mathbb{E}\bigg[ \exp\bigg( \frac{ 1 - \beta }{ \beta } \, H_{\beta}( \bvec{v}_{2}( H^{-1}( \bvec{v}_{2} : H( P_{X|Y}(\cdot \mid Y) ) ) ) ) \bigg) \bigg]
\\
& \overset{\eqref{def:renyi}}{=}
\frac{ \beta }{ 1 - \beta } \ln \mathbb{E}\Big[ \| \bvec{v}_{2}( H^{-1}( \bvec{v}_{2} : H( P_{X|Y}(\cdot \mid Y) ) ) ) \|_{\beta} \Big]
\\
& \overset{\text{(b)}}{\le}
\frac{ \beta }{ 1 - \beta } \ln \| \bvec{v}_{2}( H^{-1}( \bvec{v}_{2} : \mathbb{E}[ H( P_{X|Y}(\cdot \mid Y) ) ] ) ) \|_{\beta}
\\
& \overset{\eqref{def:cond_shannon}}{=}
\frac{ \beta }{ 1 - \beta } \ln \| \bvec{v}_{2}( H^{-1}( \bvec{v}_{2} : H(X \mid Y) ) ) \|_{\beta}
\\
& \overset{\eqref{def:A_renyi}}{=}
H_{\beta}( \bvec{v}_{2}( H^{-1}( \bvec{v}_{2} : H(X \mid Y) ) ) )
\\
& =
H_{\beta}( \bvec{v}_{2}( p ) )
\end{align}
with $p = H^{-1}(\bvec{v}_{2} : H(X \mid Y) )$, where (a) follows from \eqref{ineq:renyi_v_2} of \thref{th:renyi_v} with $\alpha = 1$, and (b) follows from Jensen's inequality and the fact that $\mu \mapsto \| \bvec{v}_{2}( H^{-1}( \bvec{v}_{2} : \mu ) ) \|_{\beta}$ is strictly concave in $\mu \in [0, \ln 2]$ (cf. \lemref{lem:Hconvex_v}).
Analogously, we also obtain
\begin{align}
H_{\beta}(X \mid Y)
\ge
H_{\beta}( \bvec{v}_{2}( p ) )
\end{align}
with $p = H^{-1}(\bvec{v}_{2} : H(X \mid Y) )$ for every $\beta \in (1, \infty)$.
This completes the proof of \thref{th:A_renyi_2}.
\end{IEEEproof}

In Theorems~\ref{th:A_renyi_alpha_inf} and~\ref{th:A_renyi_2}, we establish bounds on the conditional R\'{e}nyi entropy $H_{\beta}(X \mid Y)$ by another R\'{e}nyi entropy $H_{\beta}(\bvec{v}_{n}( p ))$ of an explicit distribution $\bvec{v}_{n}( \cdot )$.
Namely, these bounds are sharp, i.e., there is no tighter bound than them in these situations.
Theorems~\ref{th:A_renyi_alpha_inf} and~\ref{th:A_renyi_2} are proved by using the convexity/concavity of Lemmas~\ref{lem:Hconvex_v} and~\ref{lem:convex_v}.
However, if $n \in \mathbb{N}_{\ge 3}$ and $r, s \in [1/2, 1) \cup (1, \infty)$, then the convexity/concavity of \lemref{lem:convex_v} is not unique on $\mathcal{I}_{n}( r )$.
Due to this reason, we cannot use same techniques as the proofs of Theorems~\ref{th:A_renyi_alpha_inf} and~\ref{th:A_renyi_2} in the cases of $n \in \mathbb{N}_{\ge 3}$ and $r, s \in [1/2, 1) \cup (1, \infty)$.
In fact, we later show in \thref{th:ST} that $H_{\beta}(X \mid Y)$ cannot be always bounded by $H_{\beta}(\bvec{v}_{n}( p ))$ with a fixed $H_{\beta}(X \mid Y)$ in such situations.
In \cite[Theorem~4 and Corollary~1]{part2}, we established sharp bounds on $H(X \mid Y)$ with a fixed $H_{\alpha}(X \mid Y)$ in which $\supp( P_{X} )$ is finite, by defining a pair of RVs $(X^{\prime\prime}, Y^{\prime\prime})$ \cite[Definition~2]{part2} whose achieves their bounds.
In this study, we also define a specific pair of RVs $(S, T)$ later in \defref{def:ST}, and establish sharp bounds on $H_{\beta}(X \mid Y)$ by using $(S, T)$ in \thref{th:ST}.
To accomplish this goal, we now give the following lemma.

\begin{lemma}
\label{lem:slope_v}
For any fixed $n \in \mathbb{N}_{\ge 3}$ and distinct $r, s \in (0, 1) \cup (1, \infty)$, the equation
\begin{align}
\frac{ \| \bvec{v}_{n}( N_{r}^{-1}( \bvec{v}_{n} : t )) \|_{s} - n^{\theta( s )} }{ t - n^{\theta( r )} }
=
\frac{ \partial \| \bvec{v}_{n}( N_{r}^{-1}( \bvec{v}_{n} : t ) ) \|_{s} }{ \partial t }
\label{eq:slope_v}
\end{align}
has a unique root $t = t^{\ast}(n; r, s) \in \mathcal{I}_{n}( r ) \setminus \{ 1, n^{\theta( r )} \}$, where $\mathcal{I}_{n}( r )$ is defined in \eqref{def:I}.
\end{lemma}

\begin{IEEEproof}[Proof of \lemref{lem:slope_v}]
Suppose that $r < s$ and $\gamma( r, s ) < 1$.
Define
\begin{align}
\chi(n, t, u; r, s)
\coloneqq
\frac{ \| \bvec{v}_{n}( N_{r}^{-1}( \bvec{v}_{n} : u )) \|_{s} - n^{\theta( s )} }{ u - n^{\theta( r )} }
-
\frac{ \partial \| \bvec{v}_{n}( N_{r}^{-1}( \bvec{v}_{n} : t ) ) \|_{s} }{ \partial t } .
\label{def:chi}
\end{align}
We prove this lemma by showing the existence of the value $t^{\ast}(n; r, s) \in \mathcal{I}_{n}^{(1)}( r, s )$ satisfying $\chi(n, t, t^{\ast}(n; r, s); r, s) = 0$, and proving its uniqueness.
Letting $p = N_{r}^{-1}( \bvec{v}_{n} : t )$, the chain rule of derivatives shows
\begin{align}
\frac{ \partial \| \bvec{v}_{n}( N_{r}^{-1}( \bvec{v}_{n} : t ) ) \|_{s} }{ \partial t }
& =
\frac{ \partial N_{r}^{-1}( \bvec{v}_{n} : t ) }{ \partial t } \, \frac{ \partial \| \bvec{v}_{n}( p ) \|_{s} }{ \partial p }
\\
& \overset{\eqref{eq:inverse_norm_v_diff1}}{=}
\bigg( \frac{ \partial \| \bvec{v}_{n}( p ) \|_{r} }{ \partial p } \bigg)^{-1} \, \frac{ \partial \| \bvec{v}_{n}( p ) \|_{s} }{ \partial p }
\\
& \overset{\eqref{diff1:norm_v}}{=}
\Big( p^{r} + (n-1)^{1-r} \, (1-p)^{r} \Big)^{1-(1/r)} \Big( p^{r-1} - (n-1)^{1-r} \, (1-p)^{r-1} \Big)^{-1}
\notag \\
& \qquad \qquad \times
\Big( p^{s} + (n-1)^{1-s} \, (1-p)^{s} \Big)^{(1/s)-1} \Big( p^{s-1} - (n-1)^{1-s} \, (1-p)^{s-1} \Big)
\\
& =
\bigg( \frac{ p^{s-1} - (n-1)^{1-s} \, (1-p)^{s-1} }{ p^{r-1} - (n-1)^{1-r} \, (1-p)^{r-1} } \bigg) \, \Bigg( \frac{ \big( p^{s} + (n-1)^{1-s} \, (1-p)^{s} \big)^{(1/s)-1} }{ \big( p^{r} + (n-1)^{1-r} \, (1-p)^{r} \big)^{(1/r)-1} } \Bigg)
\\
& =
p^{s-r} \, \bigg( \frac{ 1 - (n-1)^{1-s} \, (p/(1-p))^{1-s} }{ 1 - (n-1)^{1-r} \, (p/(1-p))^{1-r} } \bigg) \, \Bigg( \frac{ \big( p^{s} + (n-1)^{1-s} \, (1-p)^{s} \big)^{(1/s)-1} }{ \big( p^{r} + (n-1)^{1-r} \, (1-p)^{r} \big)^{(1/r)-1} } \Bigg)
\\
& \overset{\eqref{def:z}}{=}
p^{s-r} \, \bigg( \frac{ 1 - z^{1-s} }{ 1 - z^{1-r} } \bigg) \, \Bigg( \frac{ \big( p^{s} + (n-1)^{1-s} \, (1-p)^{s} \big)^{(1/s)-1} }{ \big( p^{r} + (n-1)^{1-r} \, (1-p)^{r} \big)^{(1/r)-1} } \Bigg)
\\
& =
p^{s-r} \, \bigg( \frac{ 1 - s }{ 1 - r } \bigg) \, \bigg( \frac{ z^{1-s} - 1 }{ 1 - s } \bigg) \, \bigg( \frac{ z^{1-r} - 1 }{ 1 - r } \bigg)^{-1} \, \Bigg( \frac{ \big( p^{s} + (n-1)^{1-s} \, (1-p)^{s} \big)^{(1/s)-1} }{ \big( p^{r} + (n-1)^{1-r} \, (1-p)^{r} \big)^{(1/r)-1} } \Bigg)
\\
& \overset{\eqref{def:qlog}}{=}
p^{s-r} \, \bigg( \frac{ 1 - s }{ 1 - r } \bigg) \, \bigg( \frac{ \ln_{s} z }{ \ln_{r} z } \bigg) \, \Bigg( \frac{ \big( p^{s} + (n-1)^{1-s} \, (1-p)^{s} \big)^{(1/s)-1} }{ \big( p^{r} + (n-1)^{1-r} \, (1-p)^{r} \big)^{(1/r)-1} } \Bigg)
\\
& \overset{\eqref{def:gamma}}{=}
p^{s-r} \, \gamma(r, s)^{-1} \, \bigg( \frac{ \ln_{s} z }{ \ln_{r} z } \bigg) \, \Bigg( \frac{ \big( p^{s} + (n-1)^{1-s} \, (1-p)^{s} \big)^{(1/s)-1} }{ \big( p^{r} + (n-1)^{1-r} \, (1-p)^{r} \big)^{(1/r)-1} } \Bigg) .
\label{eq:norm_v_diff1_inv_r_s}
\end{align}
It follows from \eqref{def:inv_Nv} and \eqref{def:z} that
\begin{itemize}
\item
if $r \in (0, 1)$, then $p \to 1^{-}$ as $t \to 1^{+}$,
\item
if $r \in (1, \infty)$, then $p \to 1^{-}$ as $t \to 1^{-}$,
\item
it holds that $z \to \infty$ as $p \to 1^{-}$;
\end{itemize}
and thus, if $r \in (0, 1)$, then we obtain
\begin{align}
\lim_{t \to 1^{+}} \frac{ \partial \| \bvec{v}_{n}( N_{r}^{-1}( \bvec{v}_{n} : t ) ) \|_{s} }{ \partial t }
& \overset{\eqref{eq:norm_v_diff1_inv_r_s}}{=}
\lim_{p \to 1^{-}} \Bigg[ p^{s-r} \, \gamma(r, s)^{-1} \, \bigg( \frac{ \ln_{s} z }{ \ln_{r} z } \bigg) \, \Bigg( \frac{ \big( p^{s} + (n-1)^{1-s} \, (1-p)^{s} \big)^{(1/s)-1} }{ \big( p^{r} + (n-1)^{1-r} \, (1-p)^{r} \big)^{(1/r)-1} } \Bigg) \Bigg]
\\
& =
\gamma(r, s)^{-1} \, \lim_{z \to \infty} \bigg( \frac{ \ln_{s} z }{ \ln_{r} z } \bigg)
\\
& \overset{\text{(a)}}{=}
0
\label{limit:t_plus}
\end{align}
for every $n \in \mathbb{N}_{\ge 2}$ and $s \in (0, \infty)$ in which $r < s$, where (a) follows from the limiting value
\begin{align}
\lim_{z \to \infty} \bigg( \frac{ \ln_{s} z }{ \ln_{r} z } \bigg)
& =
\begin{cases}
0
& \mathrm{if} \ r < 1 ,
\\
\gamma( r, s )
& \mathrm{if} \ r > 1
\end{cases}
\end{align}
for every $0 < r < s < \infty$.
Analogously, we also get that if $r \in (1, \infty)$, then
\begin{align}
\lim_{t \to 1^{-}} \frac{ \partial \| \bvec{v}_{n}( N_{r}^{-1}( \bvec{v}_{n} : t ) ) \|_{s} }{ \partial t }
& =
1
\label{limit:t_minus}
\end{align}
for every $n \in \mathbb{N}_{\ge 2}$ and $s \in (0, \infty)$ in which $r < s$.
Therefore, we have the following:
\begin{itemize}
\item
if $r \in (0, 1)$, then
\begin{align}
\lim_{t \to 1^{+}} \chi(n, t, t; r, s)
& \overset{\eqref{def:chi}}{=}
\lim_{t \to 1^{+}} \bigg[ \frac{ \| \bvec{v}_{n}( N_{r}^{-1}( \bvec{v}_{n} : t )) \|_{s} - n^{\theta( s )} }{ t - n^{\theta( r )} } - \frac{ \partial \| \bvec{v}_{n}( N_{r}^{-1}( \bvec{v}_{n} : t ) ) \|_{s} }{ \partial t } \bigg]
\overset{\eqref{limit:t_plus}}{=}
\frac{ 1 - n^{\theta( s )} }{ 1 - n^{\theta( r )} } ,
\label{limig:t_plus_2}
\end{align}
\item
if $r \in (1, \infty)$, then
\begin{align}
\lim_{t \to 1^{-}} \chi(n, t, t; r, s)
& \overset{\eqref{def:chi}}{=}
\lim_{t \to 1^{-}} \bigg[ \frac{ \| \bvec{v}_{n}( N_{r}^{-1}( \bvec{v}_{n} : t )) \|_{s} - n^{\theta( s )} }{ t - n^{\theta( r )} } - \frac{ \partial \| \bvec{v}_{n}( N_{r}^{-1}( \bvec{v}_{n} : t ) ) \|_{s} }{ \partial t } \bigg]
\overset{\eqref{limit:t_minus}}{=}
\frac{ n^{\theta( r )} - n^{\theta( s )} }{ 1 - n^{\theta( r )} }
\label{limig:t_minus_2}
\end{align}
\end{itemize}
for every $n \in \mathbb{N}_{\ge 2}$ and $s \in (0, \infty)$ in which $r < s$, where note that $\| \bvec{v}_{n}( N_{r}^{-1}( \bvec{v}_{n} : 1 ) ) \|_{s} = 1$ because $\| \bvec{v}_{n}( 1 ) \|_{r} = 1$.
Since
\begin{align}
\sgn\Big( 1 - n^{\theta( r )} \Big)
=
\begin{cases}
-1
& \mathrm{if} \ r < 1 ,
\\
0
& \mathrm{if} \ r = 1 ,
\\
1
& \mathrm{if} \ r > 1
\end{cases}
\end{align}
for every $n \in \mathbb{N}_{\ge 2}$ and $r \in (0, \infty]$, it follows from \eqref{limig:t_plus_2} that
\begin{align}
\sgn\Big( \lim_{t \to 1^{+}} \chi(n, t, t; r, s) \Big)
=
\begin{cases}
-1
& \mathrm{if} \ s > 1 ,
\\
0
& \mathrm{if} \ s = 1 ,
\\
1
& \mathrm{if} \ s < 1
\end{cases}
\label{limig:t_plus_3}
\end{align}
for every $n \in \mathbb{N}_{\ge 2}$, $r \in (0, 1)$, and $s \in (r, \infty)$.
Similarly, it also follows from \eqref{limig:t_minus_2} that
\begin{align}
\sgn\Big( \lim_{t \to 1^{-}} \chi(n, t, t; r, s) \Big)
=
1
\label{limig:t_minus_3}
\end{align}
for every $n \in \mathbb{N}_{\ge 2}$ and $1 < r < s < \infty$.

We now verify the sign of the derivative \eqref{eq:norm_v_diff1_inv_r_s} as follows:
\begin{align}
\sgn\bigg( \frac{ \partial \| \bvec{v}_{n}( N_{r}^{-1}( \bvec{v}_{n} : t ) ) \|_{s} }{ \partial t } \bigg)
& \overset{\eqref{eq:norm_v_diff1_inv_r_s}}{=}
\underbrace{ \sgn\Big( p^{s-r} \Big) }_{ = 1 } \, \sgn\Big( \gamma(r, s)^{-1} \Big) \, \underbrace{ \sgn\bigg( \frac{ \ln_{s} z }{ \ln_{r} z } \bigg) }_{ = 1 } \, \underbrace{ \sgn\Bigg( \frac{ \big( p^{s} + (n-1)^{1-s} \, (1-p)^{s} \big)^{(1/s)-1} }{ \big( p^{r} + (n-1)^{1-r} \, (1-p)^{r} \big)^{(1/r)-1} } \Bigg) }_{ = 1 }
\\
& \overset{\eqref{def:gamma}}{=}
\sgn\bigg( \frac{ 1 - s }{ 1 - r } \bigg)
\\
& =
\begin{cases}
-1
& \mathrm{if} \ r < 1 < s ,
\\
0
& \mathrm{if} \ s = 1 ,
\\
1
& \mathrm{if} \ r < s < 1 \ \mathrm{or} \ 1 < r < s
\end{cases}
\end{align}
for every $n \in \mathbb{N}_{\ge 2}$, $0 < r < s < \infty$, and $t \in \mathcal{I}_{n}( r )$, which implies that for each $n \in \mathbb{N}_{\ge 2}$, the following holds:
\begin{itemize}
\item
if $r < 1 < s$, then $t \mapsto \| \bvec{v}_{n}( N_{r}^{-1}( \bvec{v}_{n} : t ) ) \|_{s}$ is strictly decreasing for $t \in \mathcal{I}_{n}( r )$,
\item
if either $r < s <  1$ or $1 < r < s$, then $t \mapsto \| \bvec{v}_{n}( N_{r}^{-1}( \bvec{v}_{n} : t ) ) \|_{s}$ is strictly increasing for $t \in \mathcal{I}_{n}( r )$.
\end{itemize}
Recall from \lemref{lem:convex_v} that if $\gamma( r, s ) < 1$, then $t \mapsto \| \bvec{v}_{n}( N_{r}^{-1}( \bvec{v}_{n} : t )) \|_{s}$ is strictly convex in $t \in \mathcal{I}_{n}^{(2)}( r, s )$ for every fixed $1/2 \le r < s < \infty$, where note that $\mathcal{I}_{n}^{(2)}( r, s ) \subset \mathcal{I}_{n}( r )$ (cf. \eqref{def:I} and \eqref{def:In_2}).
Since
\begin{itemize}
\item
the first term of the right-hand side of \eqref{def:chi} is the slope of the secant line from the point $(n^{\theta( r )}, n^{\theta( s )})$ to the point $(u, \| \bvec{v}_{n}(N_{r}^{-1}( \bvec{v}_{n} : u )) \|_{s})$,
\item
the second term of the right-hand side of \eqref{def:chi} is the slope of the tangent line of the curve $t \mapsto (t, \| \bvec{v}_{n}( N_{r}^{-1}( \bvec{v}_{n} : t ) ) \|_{s})$,
\end{itemize}
it follows from the monotonicity and convexity of $t \mapsto \| \bvec{v}_{n}( N_{r}^{-1}( \bvec{v}_{n} : t )) \|_{s}$ with respect to $t \in \mathcal{I}_{n}^{(2)}( r, s )$ that
\begin{align}
\sgn\Big( \chi(n, t, \tau(n; r, s); r, s) \Big)
=
\begin{cases}
-1
& \mathrm{if} \ r < s < 1 \ \mathrm{or} \ 1 < r < s ,
\\
1
& \mathrm{if} \ r < 1 < s
\end{cases}
\label{eq:sgn_chi_tau}
\end{align}
for every $n \in \mathbb{N}_{\ge 2}$, $1/2 \le r < s < \infty$ in which $\gamma( r, s ) < 1$, and $t \in \mathcal{I}_{n}^{(2)}( r, s )$, where $\tau(n; r, s)$ is the inflection point of $t \mapsto \| \bvec{v}_{n}( N_{r}^{-1}( \bvec{v}_{n} : t )) \|_{s}$ derived in \lemref{lem:convex_v}.
Combining \eqref{limig:t_plus_3}, \eqref{limig:t_minus_3}, and \eqref{eq:sgn_chi_tau}, and applying the intermediate value theorem for the function $u \mapsto \chi(n, t, u; r, s)$, it holds that for any $n \in \mathbb{N}_{\ge 2}$ and any $1/2 \le r < s < \infty$ in which $\gamma( r, s ) < 1$, there exists $t^{\ast}(n; r, s) \in \mathcal{I}_{n}^{(1)}( r, s )$ such that
\begin{align}
\chi(n, t, t^{\ast}(n; r, s); r, s)
=
0 ,
\end{align}
where $\mathcal{I}_{n}^{(1)}( r, s )$ is defined in \eqref{def:In_1}.
Finally, the concavity of $t \mapsto \| \bvec{v}_{n}( N_{r}^{-1}( \bvec{v}_{n} : t ) ) \|_{s}$ with respect to $t \in \mathcal{I}_{n}^{(1)}( r, s )$ implies the uniqueness of the value $t^{\ast}(n; r, s) \in \mathcal{I}_{n}^{(1)}( r, s )$.
Therefore, the assertion of \lemref{lem:slope_v} holds for $1/2 \le r < s < \infty$ in which $\gamma( r, s ) < 1$.
Furthermore, the assertion of \lemref{lem:slope_v} for other situations can also be proved in a similar way to the above discussion.
This completes the proof of \lemref{lem:slope_v}.
\end{IEEEproof}

From the definition \eqref{def:inv_Nv} of $N_{r}^{-1}( \bvec{v}_{n} : \cdot )$, we see that
\begin{align}
\| \bvec{v}_{n}( p ) \|_{r} = t
\iff
N_{r}^{-1}( \bvec{v}_{n} : t ) = p ;
\label{relation:t_p}
\end{align}
thus, it follows from \eqref{eq:inverse_norm_v_diff1} and the chain of derivatives that \eqref{eq:slope_v} of \lemref{lem:slope_v} can be rewritten by
\begin{align}
\frac{ \| \bvec{v}_{n}( p ) \|_{s} - n^{\theta( s )} }{ \| \bvec{v}_{n}( p ) \|_{r} - n^{\theta( r )} }
=
\frac{ \partial \| \bvec{v}_{n}( p ) \|_{s} }{ \partial p } \bigg( \frac{ \partial \| \bvec{v}_{n}( p ) \|_{r} }{ \partial p } \bigg)^{-1}
\label{eq:slope_v_alt}
\end{align}
with the change of variables \eqref{relation:t_p}, where note that the first-order derivatives appeared in \eqref{eq:slope_v_alt} is already derived in \eqref{diff1:norm_v}.
\lemref{lem:slope_v} also ensures that for every $n \in \mathbb{N}_{\ge 3}$ and distinct $r, s \in [1/2, 1) \cup (1, \infty)$, Eq.~\eqref{eq:slope_v_alt} has a unique root with respect to $p = p^{\ast}(n; r, s) \in (1/n, 1)$ under the relation \eqref{relation:t_p}.
Therefore, solving the root $p^{\ast}(n; r, s)$ of \eqref{eq:slope_v_alt}, we can obtain the root of \eqref{eq:slope_v} as $t^{\ast}(n; r, s) = \| \bvec{v}_{n}( p^{\ast}(n; r, s) ) \|_{r}$.
In fact, the root $p^{\ast}(n; r, s)$ of \eqref{eq:slope_v_alt} can be solved via numerical calculations.
However, in general, the root $p^{\ast}(n; r, s)$ is also hard-to-express in closed-forms, as with \eqref{def:inv_Hv} and \eqref{def:inv_Hw}.
Fortunately, if either $r = 1/2$ or $s = 1/2$, then the root $p^{\ast}(n; r, s)$ of \eqref{eq:slope_v_alt} can be written in a simple closed-form, as shown in the following.

\begin{fact}
\label{fact:slope_v_half}
For any $n \in \mathbb{N}_{\ge 3}$ and $t \in (1/2, 1) \cup (1, \infty)$,
\begin{align}
p^{\ast}(n; 1/2, t)
=
p^{\ast}(n; t, 1/2)
& =
\frac{ 1 }{ 1 + ( n - 1 )^{(t-2)/t} } .
\label{eq:slope_v_half}
\end{align}
\end{fact}

\factref{fact:slope_v_half} can be verified by directly substituting \eqref{eq:slope_v_half} into \eqref{eq:slope_v_alt}, as with the proof of \cite[Fact~2]{part2}.
In fact, \factref{fact:slope_v_half} yields the same value $p^{\ast}$ to%
\footnote{Note that the definition of $\bvec{v}_{n}( \cdot )$ used in \cite{part2} is slightly different to \eqref{def:v}; however, these are essentially same.}
\cite[Fact~2]{part2} as $t \to 1$.
Note that $p^{\ast}(n; r, s) = p^{\ast}(n; s, r)$ holds; but $t^{\ast}(n; r, s) = t^{\ast}(n; s, r)$ does not hold in general.
Employing the roots $t^{\ast}(n; r, s)$ and $p^{\ast}(n; r, s)$ of \eqref{eq:slope_v} and \eqref{eq:slope_v_alt}, respectively, we now define the pair of RVs $(S, T)$ as follows:
For $n \in \mathbb{N}_{\ge 3}$ and distinct $r, s \in [1/2, 1) \cup (1, \infty)$, let the real intervals $\mathcal{I}_{n}^{(\mathrm{a})}( r, s )$ and $\mathcal{I}_{n}^{(\mathrm{b})}( r, s )$ be defined by
\begin{align}
\mathcal{I}_{n}^{(\mathrm{a})}( r, s )
& \coloneqq
\begin{cases}
\big( t^{\ast}(n; r, s), n^{\theta( r )} \big]
& \mathrm{if} \ r < 1 ,
\\
\big[ n^{\theta( r )}, t^{\ast}(n; r, s) \big)
& \mathrm{if} \ r > 1 ,
\end{cases}
\label{def:In_3} \\
\mathcal{I}_{n}^{(\mathrm{b})}( r, s )
& \coloneqq
\begin{cases}
\big[ 1, t^{\ast}(n; r, s) \big]
& \mathrm{if} \ r < 1 ,
\\
\big[ t^{\ast}(n; r, s), 1 \big]
& \mathrm{if} \ r > 1 ,
\end{cases}
\label{def:In_4}
\end{align}
respectively, where $\theta( r )$ is defined in \eqref{def:theta}.
Note that $\{ \mathcal{I}_{n}^{(\mathrm{a})}( r, s ), \mathcal{I}_{n}^{(\mathrm{b})}( r, s ) \}$ forms a partition of the interval $\mathcal{I}_{n}( r )$ defined in \eqref{def:I}.
If $r$ and $s$ are clear from the context, for simplicity, we write \eqref{def:In_3} and \eqref{def:In_4} by $\mathcal{I}_{n}^{(\mathrm{a})}$ and $\mathcal{I}_{n}^{(\mathrm{b})}$, respectively.
Using them, we give the definition of the pair of RVs $(S, T)$ as follows.

\begin{definition}
\label{def:ST}
For given distinct $r, s \in [1/2, 1) \cup (1, \infty)$ and pair of RVs $(X, Y) \sim P_{X|Y} P_{Y}$ in which $|\!\supp( P_{X} )| = n \in \mathbb{N}_{\ge 3}$, the pair of RVs $(S, T) \sim P_{S|T} P_{T}$ is defined as follows:
The RV $S$ takes values from $\supp( P_{X} )$; and the RV $T$ takes values from $\{ 0, 1 \}$, i.e., the latter is a Bernoulli RV.
Let $\delta$ be chosen so that
\begin{align}
\delta
& =
\frac{ N_{r}(X \mid Y) - n^{\theta( r )} }{ t^{\ast}(n; r, s) - n^{\theta( r )} } ,
\label{def:delta}
\end{align}
where $\theta( r )$ is defined in \eqref{def:theta}.
Then, the marginal distribution $P_{T}$ is given by
\begin{align}
( P_{T}( 0 ), P_{T}( 1 ) )
& =
\begin{cases}
(1 - \delta, \delta)
& \mathrm{if} \ N_{r}(X \mid Y) \in \mathcal{I}_{n}^{(\mathrm{a})}( r, s ) ,
\\
(0, 1)
& \mathrm{if} \ N_{r}(X \mid Y) \in \mathcal{I}_{n}^{(\mathrm{b})}( r, s ) ,
\end{cases}
\label{def:marginal_T}
\end{align}
and the conditional distribution $P_{S|T}$ is given by
\begin{align}
P_{S|T}(\cdot \mid t)
=
\begin{cases}
\bvec{v}_{n}( 1/n ) = (1/n, 1/n, \dots, 1/n)
& \mathrm{if} \ t = 0 ,
\\
\bvec{v}_{n}( p_{(\mathrm{a})} )
& \mathrm{if} \ t = 1 \ \mathrm{and} \ N_{r}(X \mid Y) \in \mathcal{I}_{n}^{(\mathrm{a})}( r, s ) ,
\\
\bvec{v}_{n}( p_{(\mathrm{b})} )
& \mathrm{if} \ t = 1 \ \mathrm{and} \ N_{r}(X \mid Y) \in \mathcal{I}_{n}^{(\mathrm{b})}( r, s )
\end{cases}
\label{def:cond_ST}
\end{align}
with $p_{(\mathrm{a})} = p^{\ast}(n; r, s)$ and $p_{(\mathrm{b})} = N_{r}^{-1}( \bvec{v}_{n} : N_{r}(X \mid Y) )$, where $\mathcal{I}_{n}^{(\mathrm{a})}( r, s )$ and $\mathcal{I}_{n}^{(\mathrm{b})}( r, s )$ are defined in \eqref{def:In_3} and \eqref{def:In_4}, respectively.
If we want to specify the parameters $(r, s)$ for $(S, T)$, we write $(S_{(r, s)}, T_{(r, s)})$.
\end{definition}

After some algebra, for given pair of RVs $(X, Y)$ in which $|\!\supp( P_{X} )| = n \in \mathbb{N}_{\ge 3}$ and distinct $r, s \in [1/2, 1) \cup (1, \infty)$, the expectation of $\ell_{s}$-norm of $S$ given $T$ can be calculated by
\begin{align}
N_{s}(S_{(r, s)} \mid T_{(r, s)})
& \overset{\eqref{def:expect_norm}}{=}
\mathbb{E}\big[ \big\| P_{S|T}(\cdot \mid T) \big\|_{s} \big]
\\
& =
P_{T}( 0 ) \, \big\| P_{S|T}(\cdot \mid 0) \big\|_{s} + P_{T}( 1 ) \, \big\| P_{S|T}(\cdot \mid 1) \big\|_{s}
\\
& \overset{\eqref{def:marginal_T}}{=}
\begin{cases}
(1-\delta) \, \big\| P_{S|T}(\cdot \mid 0) \big\|_{s} + \delta \, \big\| P_{S|T}(\cdot \mid 1) \big\|_{s}
& \mathrm{if} \ N_{r}(X \mid Y) \in \mathcal{I}_{n}^{(\mathrm{a})}( r, s ) ,
\\
\big\| P_{S|T}(\cdot \mid 1) \big\|_{s}
& \mathrm{if} \ N_{r}(X \mid Y) \in \mathcal{I}_{n}^{(\mathrm{b})}( r, s )
\end{cases}
\\
& \overset{\eqref{def:cond_ST}}{=}
\begin{cases}
(1-\delta) \, \big\| \bvec{v}_{n}(1/n) \|_{s} + \delta \, \big\| \bvec{v}_{n}\big( p^{\ast}(n; r, s) \big) \big\|_{s}
& \mathrm{if} \ N_{r}(X \mid Y) \in \mathcal{I}_{n}^{(\mathrm{a})}( r, s ) ,
\\
\big\| \bvec{v}_{n}\big( N_{r}^{-1}( \bvec{v}_{n} : N_{r}(X \mid Y) ) \big) \big\|_{s}
& \mathrm{if} \ N_{r}(X \mid Y) \in \mathcal{I}_{n}^{(\mathrm{b})}( r, s )
\end{cases}
\\
& =
\begin{cases}
(1-\delta) \, n^{\theta( s )} + \delta \, \big\| \bvec{v}_{n}\big( p^{\ast}(n; r, s) \big) \big\|_{s}
& \mathrm{if} \ N_{r}(X \mid Y) \in \mathcal{I}_{n}^{(\mathrm{a})}( r, s ) ,
\\
\big\| \bvec{v}_{n}\big( N_{r}^{-1}( \bvec{v}_{n} : N_{r}(X \mid Y) ) \big) \big\|_{s}
& \mathrm{if} \ N_{r}(X \mid Y) \in \mathcal{I}_{n}^{(\mathrm{b})}( r, s ) ,
\end{cases}
\\
& =
\begin{cases}
(1-\delta) \, n^{\theta( s )} + \delta \, \| \bvec{v}_{n}( p_{(\mathrm{a})} ) \|_{s}
& \mathrm{if} \ N_{r}(X \mid Y) \in \mathcal{I}_{n}^{(\mathrm{a})}( r, s ) ,
\\
\| \bvec{v}_{n}( p_{(\mathrm{b})} ) \|_{s}
& \mathrm{if} \ N_{r}(X \mid Y) \in \mathcal{I}_{n}^{(\mathrm{b})}( r, s )
\end{cases}
\label{eq:norm_ST}
\end{align}
with
\begin{align}
p_{(\mathrm{a})}
& =
p^{\ast}(n; r, s) ,
\\
p_{(\mathrm{b})}
& =
N_{r}^{-1}( \bvec{v}_{n} : N_{r}(X \mid Y) ) ,
\end{align}
where $\delta$ is given by \eqref{def:delta}, and $p^{\ast}(n; r, s)$ is the root of \eqref{eq:slope_v_alt}.
Letting $(\alpha, \beta) = (r, s)$, for any distinct $\alpha, \beta \in [1/2, 1) \cup (1, \infty)$, the conditional R\'{e}nyi entropy of $S$ given $T$ can be calculated by
\begin{align}
H_{\beta}(S_{(\alpha, \beta)} \mid T_{(\alpha, \beta)})
& \overset{\eqref{def:A_renyi}}{=}
\frac{ \beta }{ 1 - \beta } \ln N_{\beta}(S_{(\alpha, \beta)} \mid T_{(\alpha, \beta)})
\\
& \overset{\eqref{eq:norm_ST}}{=}
\begin{dcases}
\frac{ \beta }{ 1 - \beta } \ln \! \Big[ (1-\delta) \, n^{\theta( \beta )} + \delta \, \| \bvec{v}_{n}( p_{(\mathrm{a})} ) \|_{\beta} \Big]
& \mathrm{if} \ N_{\alpha}(X \mid Y) \in \mathcal{I}_{n}^{(\mathrm{a})}( \alpha, \beta ) ,
\\
\frac{ \beta }{ 1 - \beta } \ln \Big[ \| \bvec{v}_{n}( p_{(\mathrm{b})} ) \|_{\beta} \Big]
& \mathrm{if} \ N_{\alpha}(X \mid Y) \in \mathcal{I}_{n}^{(\mathrm{b})}( \alpha, \beta )
\end{dcases}
\\
& \overset{\eqref{def:renyi}}{=}
\begin{dcases}
\frac{ \beta }{ 1 - \beta } \ln \! \Big[ (1-\delta) \, n^{\theta( \beta )} + \delta \, \| \bvec{v}_{n}( p_{(\mathrm{a})} ) \|_{\beta} \Big]
& \mathrm{if} \ N_{\alpha}(X \mid Y) \in \mathcal{I}_{n}^{(\mathrm{a})}( \alpha, \beta ) ,
\\
H_{\beta}( \bvec{v}_{n}( p_{(\mathrm{b})} ) )
& \mathrm{if} \ N_{\alpha}(X \mid Y) \in \mathcal{I}_{n}^{(\mathrm{b})}( \alpha, \beta )
\end{dcases}
\\
& \overset{\eqref{def:A_renyi}}{=}
\begin{dcases}
\frac{ \beta }{ 1 - \beta } \ln \! \Big[ (1-\delta) \, n^{\theta( \beta )} + \delta \, \| \bvec{v}_{n}( p_{(\mathrm{a})} ) \|_{\beta} \Big]
& \mathrm{if} \ H_{\alpha}(X \mid Y) \in \mathcal{H}_{n}^{(\mathrm{a})}( \alpha, \beta ) ,
\\
H_{\beta}( \bvec{v}_{n}( p_{(\mathrm{b})} ) )
& \mathrm{if} \ H_{\alpha}(X \mid Y) \in \mathcal{H}_{n}^{(\mathrm{b})}( \alpha, \beta )
\end{dcases}
\label{eq:renyi_ST}
\end{align}
with
\begin{align}
p_{(\mathrm{a})}
& =
p^{\ast}(n; \alpha, \beta) ,
\label{eq:p_a_renyi} \\
p_{(\mathrm{b})}
& =
H_{\alpha}^{-1}( \bvec{v}_{n} : H_{\alpha}(X \mid Y) ) ,
\label{eq:p_b_renyi} 
\end{align}
where $\mathcal{H}_{n}^{(\mathrm{a})}( \alpha, \beta )$ and $\mathcal{H}_{n}^{(\mathrm{b})}( \alpha, \beta )$ are two real intervals defined by
\begin{align}
\mathcal{H}_{n}^{(\mathrm{a})}( \alpha, \beta )
& \coloneqq
\big( H_{\alpha}( \bvec{v}_{n}( p_{(\mathrm{a})} ) ), \ln n \big] ,
\label{def:interval_Ha} \\
\mathcal{H}_{n}^{(\mathrm{b})}( \alpha, \beta )
& \coloneqq
\big[ 0, H_{\alpha}( \bvec{v}_{n}( p_{(\mathrm{a})} ) ) \big] ,
\label{def:interval_Hb}
\end{align}
respectively, and $\delta$ is given by \eqref{def:delta} with $(r, s) = (\alpha, \beta)$.
Note that $\{ \mathcal{H}_{n}^{(\mathrm{a})}( \alpha, \beta ), \mathcal{H}_{n}^{(\mathrm{b})}( \alpha, \beta ) \}$ forms a partition of the interval $[0, \ln n]$.
Namely, the quantity $H_{\beta}(S_{(\alpha, \beta)} \mid T_{(\alpha, \beta)})$ is determined by the following three arguments: (i) the number $|\!\supp( P_{X} )| \ge 3$, (ii) the value $H_{\alpha}(X \mid Y)$, and (iii) distinct $\alpha, \beta \in [1/2, 1) \cup (1, \infty)$. 
In fact, for any distinct $\alpha, \beta \in [1/2, 1) \cup (1, \infty)$, we can verify the following:
\begin{itemize}
\item
if $H_{\alpha}(X \mid Y) \in \mathcal{H}_{n}^{(\mathrm{a})}( \alpha, \beta )$, then
\begin{align}
H_{\alpha}(S_{(\alpha, \beta)} \mid T_{(\alpha, \beta)})
& \overset{\eqref{eq:renyi_ST}}{=}
\frac{ \alpha }{ 1 - \alpha } \ln \! \Big[ (1-\delta) \, n^{\theta( \alpha )} + \delta \, \| \bvec{v}_{n}( p_{(\mathrm{a})} ) \|_{\alpha} \Big]
\\
& \overset{\eqref{eq:p_a_renyi}}{=}
\frac{ \alpha }{ 1 - \alpha } \ln \! \Big[ (1-\delta) \, n^{\theta( \alpha )} + \delta \, t^{\ast}(n; \alpha, \beta ) \Big]
\\
& \overset{\eqref{def:delta}}{=}
\frac{ \alpha }{ 1 - \alpha } \ln \! \bigg[ \bigg( \frac{ t^{\ast}(n; \alpha, \beta) - N_{\alpha}(X \mid Y) }{ t^{\ast}(n; \alpha, \beta) - n^{\theta( \alpha )} } \bigg) \, n^{\theta( \alpha )} + \bigg( \frac{ N_{\alpha}(X \mid Y) - n^{\theta( \alpha )} }{ t^{\ast}(n; \alpha, \beta) - n^{\theta( \alpha )} } \bigg) \, t^{\ast}(n; \alpha, \beta ) \bigg]
\\
& =
\frac{ \alpha }{ 1 - \alpha } \ln N_{\alpha}(X \mid Y)
\\
& \overset{\eqref{def:A_renyi}}{=}
H_{\alpha}(X \mid Y) ,
\label{eq:ST_alpha_1}
\end{align}
\item
if $H_{\alpha}(X \mid Y) \in \mathcal{H}_{n}^{(\mathrm{b})}( \alpha, \beta )$, then
\begin{align}
H_{\alpha}(S_{(\alpha, \beta)} \mid T_{(\alpha, \beta)})
& \overset{\eqref{eq:renyi_ST}}{=}
H_{\alpha}( \bvec{v}_{n}( p_{(\mathrm{b})} ) )
\\
& \overset{\eqref{eq:p_b_renyi}}{=}
H_{\alpha}(X \mid Y) .
\label{eq:ST_alpha_2}
\end{align}
\end{itemize}
Hence, the following theorem gives bounds on $H_{\beta}(X \mid Y)$ with a fixed $H_{\alpha}(X \mid Y)$, and the bounds are sharp because these are written by a specific pair of RVs $(S, T)$.

\begin{theorem}
\label{th:ST}
Let $X$ be an RV in which $3 \le |\! \supp( P_{X} )| < \infty$, and let $Y$ be an arbitrary RV.
For any distinct $\alpha, \beta \in [1/2, 1) \cup (1, \infty)$, it holds that
\begin{align}
H_{\beta}(X \mid Y)
\ge
H_{\beta}(S_{(\alpha, \beta)} \mid T_{(\alpha, \beta)})
\qquad
\mathrm{if} \ \alpha < \beta ,
\\
H_{\beta}(X \mid Y)
\le
H_{\beta}(S_{(\alpha, \beta)} \mid T_{(\alpha, \beta)})
\qquad
\mathrm{if} \ \beta < \alpha ,
\end{align}
where the pair of RVs $(S, T)$ is defined in \defref{def:ST}.
\end{theorem}

\begin{IEEEproof}[Proof of \thref{th:ST}]
Let $(X, Y)$ be a pair of RVs in which $|\!\supp( P_{X} )| = n$ for some $n \in \mathbb{N}_{\ge 3}$, and let $r, s \in [1/2, 1) \cup (1, \infty)$ be distinct fixed numbers.
Define
\begin{align}
f_{\mathrm{ST}}(n, t; r, s)
\coloneqq
\begin{cases}
(1-\delta^{\prime}) \, n^{\theta( s )} + \delta^{\prime} \, \big\| \bvec{v}_{n}\big( N_{r}^{-1}( \bvec{v}_{n} : t^{\ast}(n; r, s) ) \big) \big\|_{s}
& \mathrm{if} \ t \in \mathcal{I}_{n}^{(\mathrm{a})}( r, s ) ,
\\
\big\| \bvec{v}_{n}\big( N_{r}^{-1}( \bvec{v}_{n} : t ) \big) \big\|_{s}
& \mathrm{if} \ t \in \mathcal{I}_{n}^{(\mathrm{b})}( r, s ) ,
\end{cases}
\label{def:fST}
\end{align}
where $\mathcal{I}_{n}^{(\mathrm{a})}( r, s )$ and $\mathcal{I}_{n}^{(\mathrm{b})}( r, s )$ are defined in \eqref{def:In_3} and \eqref{def:In_4}, respectively, the value $\delta^{\prime} \in [0, 1)$ is chosen so that
\begin{align}
\delta^{\prime}
=
\frac{ t - n^{\theta( r )} }{ t^{\ast}(n; r, s) - n^{\theta( r )} } ,
\end{align}
and $t^{\ast}(n; r, s)$ is the root of \eqref{eq:slope_v} shown in \lemref{lem:slope_v}.
Note from \eqref{eq:norm_ST} that $f_{\mathrm{ST}}(n, t; r, s)$ is defined to satisfy
\begin{align}
f_{\mathrm{ST}}(n, N_{r}(X \mid Y); r, s)
=
N_{s}(S_{(r, s)} \mid T_{(r, s)}) .
\label{eq:fST_ST}
\end{align}
Then, we can verify the following statements:
\begin{itemize}
\item
the function $t \mapsto f_{\mathrm{ST}}(n, t; r, s)$ is linear in $t \in \mathcal{I}_{n}^{(\mathrm{a})}( r, s )$,
\item
if $\gamma( r, s ) > 1$, then $t \mapsto f_{\mathrm{ST}}(n, t; r, s)$ is strictly convex in $t \in \mathcal{I}_{n}^{(\mathrm{b})}( r, s )$ (cf. \lemref{lem:convex_v}),
\item
if $\gamma( r, s ) < 1$, then $t \mapsto f_{\mathrm{ST}}(n, t; r, s)$ is strictly concave in $t \in \mathcal{I}_{n}^{(\mathrm{b})}( r, s )$ (cf. \lemref{lem:convex_v}),
\end{itemize}
where $\gamma( r, s )$ is defined in \eqref{def:gamma}.
Moreover, since $t^{\ast}(n; r, s)$ used in \defref{def:ST} fulfills \eqref{eq:slope_v} of \lemref{lem:slope_v}, the function $t \mapsto f_{\mathrm{ST}}(n, t; r, s)$ is differentiable $t = t^{\ast}(n; r, s)$.
Therefore, the above convexity/concavity can be modified as follows:
\begin{itemize}
\item
if $\gamma( r, s ) > 1$, then $t \mapsto f_{\mathrm{ST}}(n, t; r, s)$ is convex in $t \in \mathcal{I}_{n}( r )$,
\item
if $\gamma( r, s ) < 1$, then $t \mapsto f_{\mathrm{ST}}(n, t; r, s)$ is concave in $t \in \mathcal{I}_{n}( r )$,
\end{itemize}
where $\mathcal{I}_{n}( r )$ is defined in \eqref{def:I}.

We now consider inequalities between $\| \bvec{v}_{n}( N_{r}^{-1}( \bvec{v}_{n} : t ) ) \|_{s}$ and $f_{\mathrm{ST}}(n, t; r, s)$.
By definition \eqref{def:fST}, it is clear that
\begin{align}
\| \bvec{v}_{n}( N_{r}^{-1}( \bvec{v}_{n} : t ) ) \|_{s}
=
f_{\mathrm{ST}}(n, t; r, s)
\label{eq:fSC_1}
\end{align}
for every $t \in \mathcal{I}_{n}^{(\mathrm{b})}( r, s )$.
On the other hand, the proof of \lemref{lem:slope_v} shows that
\begin{itemize}
\item
if $\gamma( r, s ) > 1$, the curve $t \mapsto (t, \| \bvec{v}_{n}( N_{r}^{-1}( \bvec{v}_{n} : t ) ) \|_{s})$ is bounded from below by the secant line from the point $(n^{\theta( r )}, n^{\theta( s )})$ to the point $(t^{\ast}(n; r, s), \| \bvec{v}_{n}( N_{r}^{-1}( \bvec{v}_{n} : t^{\ast}(n; r, s) ) ) \|_{s})$,
\item
if $\gamma( r, s ) < 1$, the curve $t \mapsto (t, \| \bvec{v}_{n}( N_{r}^{-1}( \bvec{v}_{n} : t ) ) \|_{s})$ is bounded from above by the secant line from the point $(n^{\theta( r )}, n^{\theta( s )})$ to the point $(t^{\ast}(n; r, s), \| \bvec{v}_{n}( N_{r}^{-1}( \bvec{v}_{n} : t^{\ast}(n; r, s) ) ) \|_{s})$.
\end{itemize}
Since the secant line from the point $(n^{\theta( r )}, n^{\theta( s )})$ to the point $(t^{\ast}(n; r, s), \| \bvec{v}_{n}( N_{r}^{-1}( \bvec{v}_{n} : t^{\ast}(n; r, s) ) ) \|_{s})$ can be denoted by $t \mapsto (t, f_{\mathrm{ST}}(n, t; r, s))$ for $t \in \mathcal{I}_{n}^{(\mathrm{a})}( r, s )$, it holds that
\begin{align}
\gamma( r, s ) > 1
\quad & \Longrightarrow \quad
\| \bvec{v}_{n}( N_{r}^{-1}( \bvec{v}_{n} : t ) ) \|_{s}
\ge
f_{\mathrm{ST}}(n, t; r, s) ,
\label{eq:fSC_2} \\
\gamma( r, s ) < 1
\quad & \Longrightarrow \quad
\| \bvec{v}_{n}( N_{r}^{-1}( \bvec{v}_{n} : t ) ) \|_{s}
\le
f_{\mathrm{ST}}(n, t; r, s)
\label{eq:fSC_3}
\end{align}
for every $t \in \mathcal{I}_{n}^{(\mathrm{a})}(r, s)$.
Combining \eqref{eq:fSC_1}, \eqref{eq:fSC_2}, and \eqref{eq:fSC_3}, we get
\begin{align}
\gamma( r, s ) > 1
\quad & \Longrightarrow \quad
\| \bvec{v}_{n}( N_{r}^{-1}( \bvec{v}_{n} : t ) ) \|_{s}
\ge
f_{\mathrm{ST}}(n, t; r, s) ,
\label{eq:fSC_4} \\
\gamma( r, s ) < 1
\quad & \Longrightarrow \quad
\| \bvec{v}_{n}( N_{r}^{-1}( \bvec{v}_{n} : t ) ) \|_{s}
\le
f_{\mathrm{ST}}(n, t; r, s)
\label{eq:fSC_5}
\end{align}
for every $t \in \mathcal{I}_{n}( r )$, because $\mathcal{I}_{n}( r ) = \mathcal{I}_{n}^{(\mathrm{a})}( r, s ) \cup \mathcal{I}_{n}^{(\mathrm{b})}( r, s )$.

According to the above discussion, if $\gamma( r, s ) > 1$, we have
\begin{align}
N_{s}(X \mid Y)
& \overset{\eqref{def:expect_norm}}{=}
\mathbb{E}\big[ \| P_{X|Y}(\cdot \mid Y) \|_{s} \big]
\\
& \overset{\text{(a)}}{\ge}
\mathbb{E}\big[ \big\| \bvec{v}_{n}\big( N_{r}^{-1}(\bvec{v}_{n} : \| P_{X|Y}(\cdot \mid Y) \|_{r} ) \big) \big\|_{s} \big]
\\
& \overset{\eqref{eq:fSC_4}}{\ge}
\mathbb{E}\big[ f_{\mathrm{ST}} \big( n, \| P_{X|Y}(\cdot \mid Y) \|_{r}; r, s \big) \big]
\\
& \overset{\text{(b)}}{\ge}
f_{\mathrm{ST}} \big( n, \mathbb{E}[ \| P_{X|Y}(\cdot \mid Y) \|_{r} ]; r, s \big)
\\
& \overset{\eqref{def:expect_norm}}{=}
f_{\mathrm{ST}} \big( n, N_{r}(X \mid Y); r, s \big)
\\
& \overset{\eqref{eq:fST_ST}}{=}
N_{s}(S_{(r, s)} \mid T_{(r, s)}) ,
\label{eq:ST_proof1}
\end{align}
where (a) follows from \eqref{eq:bound_v_1} of \thref{th:v}, and (b) follows from the convexity of $t \mapsto f_{\mathrm{ST}}( n, t; r, s )$ for $t \in \mathcal{I}_{n}( r )$.
Similarly, if $\gamma( r, s ) < 1$, we also have
\begin{align}
N_{s}(X \mid Y)
\le
N_{s}(S_{(r, s)} \mid T_{(r, s)}) .
\label{eq:ST_proof2}
\end{align}
Finally, we define
\begin{align}
f_{\alpha}( t )
\coloneqq
\frac{ \alpha }{ 1 - \alpha } \ln t
\end{align}
for $\alpha \in (0, 1) \cup (1, \infty)$ and $t > 0$.
Since
\begin{itemize}
\item
it holds that $H_{\beta}(X \mid Y) = f_{\beta}( N_{\beta}(X \mid Y) )$ for every $\beta \in (0, 1) \cup (1, \infty)$,
\item
if $\beta \in (0, 1)$, then $t \mapsto f_{\beta}( t )$ is a strictly increasing function of $t > 0$,
\item
if $\beta \in (1, \infty)$, then $t \mapsto f_{\beta}( t )$ is a strictly decreasing function of $t > 0$,
\end{itemize}
it follows from \eqref{eq:ST_proof1} and \eqref{eq:ST_proof2} that
\begin{align}
H_{\beta}(X \mid Y)
& \ge
H_{\beta}(S_{(\alpha, \beta)} \mid T_{(\alpha, \beta)})
&& \mathrm{if} \ \alpha < \beta ,
\\
H_{\beta}(X \mid Y)
& \le
H_{\beta}(S_{(\alpha, \beta)} \mid T_{(\alpha, \beta)})
&& \mathrm{if} \ \beta < \alpha
\end{align}
for every $\alpha, \beta \in [1/2, 1) \cup (1, \infty)$.
This completes the proof of \thref{th:ST}.
\end{IEEEproof}

Note that if either $\alpha = 1$ or $\beta = 1$, then sharp bounds in a similar situation to \thref{th:ST} were already derived in \cite[Theorem~2 and Corollary~1]{part2}.

In this subsection, we derive sharp bounds on $H_{\beta}(X \mid Y)$ with a fixed $H_{\alpha}(X \mid Y)$, $\alpha \neq \beta$, by employing the extremality of the distribution $\bvec{v}_{n}( \cdot )$ shown in \sectref{sect:unconditional}.
In the next subsection, we further derive sharp bounds on $H_{\beta}(X \mid Y)$ with a fixed $H_{\alpha}(X \mid Y)$, $\alpha \neq \beta$, by employing the extremality of another distribution $\bvec{w}( \cdot )$ shown in \sectref{sect:unconditional}.

\subsection{Bounds Established from Distribution $\bvec{w}( \cdot )$}
\label{sect:w}

In this subsection, by using extremality of the distribution $\bvec{w}( \cdot )$ introduced in \sectref{sect:unconditional}, we derive sharp bounds on $H_{\beta}(X \mid Y)$ with a fixed $H_{\alpha}(X \mid Y)$.
Unlike the bounds established in \sectref{sect:v}, as with \thref{th:renyi_w}, sharp bounds established in this subsection can be considered for every distributions with possibly countably infinite support.
In a similar way to consider a specific pair of RVs $(S, T)$ of \defref{def:ST}, we now define another specific pair of RVs $(U, V)$ in \defref{def:UV}, whose achieves the bounds of \thref{th:UV} as shown later.

\begin{definition}
\label{def:UV}
For given $r \in (0, 1) \cup (1, \infty]$ and pair of RVs $(X, Y) \sim P_{X|Y} P_{Y}$, the pair of RVs $(U, V) \sim P_{U|V} P_{V}$ is defined as follows:
The RV $U$ takes values from $\{ 0, 1, 2, \dots \}$; and the RV $V$ takes values from $\{ 0, 1 \}$, i.e., the latter is a Bernoulli RV.
Let $m \in \mathbb{N}$ and $\lambda \in [0, 1]$ be chosen so that
\begin{align}
m
& =
\Big\lfloor N_{r}(X \mid Y)^{\theta( r )} \Big\rfloor ,
\label{eq:m_UV} \\
\lambda
& =
\frac{ (m+1)^{\theta( r )} - N_{r}(X \mid Y) }{ (m+1)^{\theta( r )} - m^{\theta( r )} } ,
\label{eq:lambda_UV}
\end{align}
respectively, where $\theta( r )$ is defined in \eqref{def:theta}.
Then, the marginal distribution $P_{V}$ is given by
\begin{align}
( P_{V}( 0 ), P_{V}( 1 ) )
& =
(1 - \lambda, \lambda) ,
\label{def:marginal_V}
\end{align}
and the conditional distribution $P_{U|V}$ is given by
\begin{align}
P_{U|V}(\cdot \mid v)
& =
\begin{cases}
\bvec{w} (1/m)
& \mathrm{if} \ v = 0 ,
\\
\bvec{w} ( 1/(m+1) )
& \mathrm{if} \ v = 1 .
\end{cases}
\label{def:cond_UV}
\end{align}
If we want to specify the parameter $r$ for $(U, V)$, we write $(U_{(r)}, V_{(r)})$.
\end{definition}

After some algebra, we see that
\begin{align}
N_{s}(U_{(r)} \mid V_{(r)})
& \overset{\eqref{def:expect_norm}}{=}
\mathbb{E}\big[ \big\| P_{U|V}(\cdot \mid V) \big\|_{s} \big]
\\
& =
P_{V}( 0 ) \, \big\| P_{U|V}(\cdot \mid 0) \big\|_{s} + P_{V}( 1 ) \, \big\| P_{U|V}(\cdot \mid 1) \big\|_{s}
\\
& \overset{\eqref{def:marginal_V}}{=}
\lambda \, \big\| P_{U|V}(\cdot \mid 0) \big\|_{s} + (1-\lambda) \, \big\| P_{U|V}(\cdot \mid 1) \big\|_{s}
\\
& \overset{\eqref{def:cond_UV}}{=}
\lambda \, \big\| \bvec{w}(1/m) \big\|_{s} + (1-\lambda) \, \big\| \bvec{w}(1/(m+1)) \big\|_{s}
\\
& \overset{\eqref{eq:w_1/m}}{=}
\lambda \, m^{\theta( s )} + (1-\lambda) \, (m+1)^{\theta( s )}
\label{eq:expect_norm_UV}
\end{align}
for every pair of RVs $(X, Y)$, $r \in (0, 1) \cup (1, \infty]$, and $s \in (0, \infty]$, where $m \in \mathbb{N}$ and $\lambda \in [0, 1]$ are given by \eqref{eq:m_UV} and \eqref{eq:lambda_UV}, respectively.
Similarly, the conditional R\'{e}nyi entropy of $U$ given $V$ can be calculated as
\begin{align}
H_{\beta}(U_{(\alpha)} \mid V_{(\alpha)})
& \overset{\eqref{def:A_renyi}}{=}
\frac{ 1 }{ \theta( \beta ) } \ln N_{\beta}(U_{(\alpha)} \mid V_{(\alpha)})
\\
& \overset{\eqref{eq:expect_norm_UV}}{=}
\frac{ 1 }{ \theta( \beta ) } \ln \! \Big[ \lambda \, m^{\theta( \beta )} + (1-\lambda) \, (m+1)^{\theta( \beta )} \Big] ,
\label{eq:renyi_UV}
\end{align}
for every pair of RVs $(X, Y)$, $\alpha \in (0, 1) \cup (1, \infty]$, and $\beta \in (0, 1) \cup (1, \infty]$, where $m \in \mathbb{N}$ and $\lambda \in [0, 1]$ are given by \eqref{eq:m_UV} and \eqref{eq:lambda_UV}, respectively, with $r = \alpha$.
Analogously, it follows that
\begin{align}
H(U_{(\alpha)} \mid V_{(\alpha)})
& =
\lambda \ln m + (1 - \lambda) \ln (m + 1)
\label{eq:renyi_UV_1}
\end{align}
for every $\alpha \in (0, 1) \cup (1, \infty]$.
Thus, the quantity $H_{\beta}(U_{(\alpha)} \mid V_{(\alpha)})$ is determined by the following two arguments: (i) the value $H_{\alpha}(X \mid Y)$, and (ii) two orders $\alpha, \beta$.
In fact, as with \eqref{eq:ST_alpha_1} and \eqref{eq:ST_alpha_2}, it also holds that
\begin{align}
N_{r}(U_{(r)} \mid V_{(r)})
& \overset{\eqref{eq:expect_norm_UV}}{=}
\lambda \, m^{\theta( r )} + (1-\lambda) \, (m+1)^{\theta( r )}
\\
& \overset{\eqref{eq:lambda_UV}}{=}
\bigg( \frac{ (m+1)^{\theta( r )} - N_{r}(X \mid Y) }{ (m+1)^{\theta( r )} - m^{\theta( r )} } \bigg) \, m^{\theta( r )} + \bigg( \frac{ N_{r}(X \mid Y) - m^{\theta( r )} }{ (m+1)^{\theta( r )} - m^{\theta( r )} } \bigg) \, (m+1)^{\theta( r )}
\\
& =
\bigg( \frac{ (m+1)^{\theta( r )} - m^{\theta( r )} }{ (m+1)^{\theta( r )} - m^{\theta( r )} } \bigg) \, N_{r}(X \mid Y)
\\
& =
N_{r}(X \mid Y) ,
\label{eq:UV_same_r} \\
H_{\alpha}(U_{(\alpha)} \mid V_{(\alpha)})
& \overset{\eqref{eq:UV_same_r}}{=}
H_{\alpha}(X \mid Y) .
\end{align}
Fortunately, unlike $H_{\beta}(S_{(\alpha, \beta)} \mid T_{(\alpha, \beta)})$, the quantity $H_{\beta}(U_{(\alpha)} \mid V_{(\alpha)})$ can be expressed in closed-forms for every $\alpha \in (0, 1) \cup (1, \infty]$ and $\beta \in (0, \infty]$.
Employing the pair of RVs $(U, V)$, the sharp bounds on $H_{\beta}(X \mid Y)$ with a fixed $H_{\alpha}(X \mid Y)$ can be established for $\alpha \neq \beta$, as shown in the following theorem.

\begin{theorem}
\label{th:UV}
Let $X$ be a discrete RV in which $\supp( P_{X} )$ is possibly countably infinite, and let $Y$ be an arbitrary RV.
For any $\alpha \in (0, 1) \cup (1, \infty]$ and any $\beta \in (0, \infty]$, it holds that
\begin{align}
H_{\beta}(X \mid Y)
& \le
H_{\beta}(U_{(\alpha)} \mid V_{(\alpha)})
\qquad \mathrm{if} \ \alpha \le \beta ,
\label{ineq:UV_renyi_2} \\
H_{\beta}(X \mid Y)
& \ge
H_{\beta}(U_{(\alpha)} \mid V_{(\alpha)})
\qquad \mathrm{if} \ \beta \le \alpha ,
\label{ineq:UV_renyi_1}
\end{align}
where the pair of RVs $(U, V)$ is defined in \defref{def:UV}.
\end{theorem}

\begin{IEEEproof}[Proof of \thref{th:UV}]
Suppose that $0 < r < s \le \infty$.
For given $m \in \mathbb{N}$, $r \in (0, \infty)$ and $t \in \mathcal{J}_{m}( r )$, let $\lambda \in [0, 1]$ be chosen so that
\begin{align}
t
=
\lambda \, m^{\theta( r )} + (1-\lambda) \, (m+1)^{\theta( r )} .
\label{eq:lambda_J}
\end{align}
It follows from \lemref{lem:convex_w} that if $\gamma( r, s ) < 1$, then
\begin{align}
\| \bvec{w}( N_{r}^{-1}( \bvec{w} : t ) ) \|_{s}
& \overset{\eqref{eq:lambda_J}}{=}
\| \bvec{w}( N_{r}^{-1}( \bvec{w} : \lambda \, m^{\theta( r )} + (1-\lambda) \, (m+1)^{\theta( r )} ) ) \|_{s}
\\
& \overset{\text{(a)}}{\ge}
\lambda \, \| \bvec{w}( N_{r}^{-1}( \bvec{w} : m^{\theta( r )} ) ) \|_{s} + (1-\lambda) \, \| \bvec{w}( N_{r}^{-1}( \bvec{w} : (m+1)^{\theta( r )} ) ) \|_{s}
\\
& \overset{\text{(b)}}{=}
\lambda \, \| \bvec{w}( 1/m ) \|_{s} + (1-\lambda) \, \| \bvec{w}( 1/(m+1) ) \|_{s}
\\
& =
\lambda \, m^{\theta( s )} + (1-\lambda) \, (m+1)^{\theta( s )}
\label{previous:phi} \\
& \overset{\eqref{eq:lambda_J}}{=}
\bigg( \frac{ (m+1)^{\theta( r )} - t }{ (m+1)^{\theta( r )} - m^{\theta( r )} } \bigg) \, m^{\theta( s )} + \bigg( \frac{ t - m^{\theta( r )} }{ (m+1)^{\theta( r )} - m^{\theta( r )} } \bigg) \, (m+1)^{\theta( s )}
\label{prev2:phi} \\
& =
\bigg( \frac{ (m+1)^{\theta( r )} \, m^{\theta( s )} - m^{\theta( r )} \, (m+1)^{\theta( s )} }{ (m+1)^{\theta( r )} - m^{\theta( r )} } \bigg) + t \, \bigg( \frac{ (m+1)^{\theta( s )} - m^{\theta( s )} }{ (m+1)^{\theta( r )} - m^{\theta( r )} } \bigg)
\label{prev3:phi} \\
& \eqqcolon
\phi(m, t; r, s)
\label{def:phi}
\end{align}
for every $m \in \mathbb{N}$, $t \in \mathcal{J}_{m}( r )$, and $0 < r < s \le \infty$, where (a) follows by the concavity of \lemref{lem:convex_w} and the definition of concave functions, and (b) follows from the fact that
\begin{align}
\| \bvec{w}( 1/m ) \|_{r}
=
m^{\theta( r )}
\iff
N_{r}^{-1}( \bvec{w} : m^{\theta( r )} )
=
1/m .
\end{align}
Similarly, if $\gamma(r, s) > 1$, it also follows from \lemref{lem:convex_w} that
\begin{align}
\| \bvec{w}( N_{r}^{-1}( \bvec{w} : t ) ) \|_{s}
\le
\phi(m, t; r, s)
\label{eq:phi_another}
\end{align}
for every $m \in \mathbb{N}$, $t \in \mathcal{J}_{m}( r )$, and $0 < r < s \le \infty$.
Note from \eqref{eq:expect_norm_UV} and \eqref{previous:phi} that the function $\phi(m, t; r, s)$ fulfills
\begin{align}
N_{s}(U_{(r)} \mid V_{(r)})
=
\phi\big( \big\lfloor N_{r}(X \mid Y)^{\theta( r )} \big\rfloor, N_{r}(X \mid Y); r, s \big)
\label{eq:UV_phi}
\end{align}
for given pair of RVs $(X, Y)$ and $r, s \in (0, 1) \cup (1, \infty]$.
We now verify the monotonicity of the derivative
\begin{align}
\frac{ \partial \phi(m, t; r, s) }{ \partial t }
& =
\frac{ (m+1)^{\theta( s )} - m^{\theta( s )} }{ (m+1)^{\theta( r )} - m^{\theta( r )} }
\\
& =
\bigg( \frac{ m^{\theta( s )} }{ m^{\theta( r )} } \bigg) \, \bigg( \frac{ ((m+1)/m)^{\theta( s )} - 1 }{ ((m+1)/m)^{\theta( r )} - 1 } \bigg)
\\
& \overset{\eqref{def:qlog}}{=}
m^{\theta( s ) - \theta( r )} \, \frac{ \theta( r ) }{ \theta( s ) } \bigg( \frac{ \ln_{1-\theta(s)} ((m+1)/m) }{ \ln_{1-\theta(r)} ((m+1)/m) } \bigg)
\label{diff:phi_mtrs}
\end{align}
with respect to $m \in \mathbb{N}$.
Since $1 - \theta( r ) < 1 - \theta( s )$ whenever $r < s$ and
\begin{align}
\frac{ \partial }{ \partial x } \, \ln_{q} x
& =
x^{-q} ,
\end{align}
we get that for each fixed $0 < r < s \le \infty$, the function
\begin{align}
m \mapsto \frac{ \ln_{1-\theta(s)} ((m+1)/m) }{ \ln_{1-\theta(r)} ((m+1)/m) }
\end{align}
is strictly decreasing for $m \in \mathbb{N}$.
Moreover, since $\theta( s ) - \theta( r ) < 0$ whenever $r < s$, it also follows that for each fixed $0 < r < s \le \infty$, the function $m \mapsto m^{\theta( s ) - \theta( r )}$ is strictly decreasing for $m \in \mathbb{N}$.
Therefore, we have
\begin{align}
&
\sgn\bigg( \frac{ \partial \phi(m, t; r, s) }{ \partial t } - \frac{ \partial \phi(m+1, t; r, s) }{ \partial t } \bigg)
\notag \\
& =
\sgn\bigg( \frac{ \theta( r ) }{ \theta( s ) } \bigg) \, \underbrace{ \sgn\Bigg( m^{\theta( s ) - \theta( r )} \, \bigg( \frac{ \ln_{1-\theta(s)} ((m+1)/m) }{ \ln_{1-\theta(r)} ((m+1)/m) } \bigg) - (m+1)^{\theta( s ) - \theta( r )} \, \bigg( \frac{ \ln_{1-\theta(s)} ((m+2)/(m+1)) }{ \ln_{1-\theta(r)} ((m+2)/(m+1)) } \bigg) \Bigg) }_{ = 1 }
\\
& \overset{\eqref{def:theta}}{=}
\sgn\bigg( \frac{ s \, (1 - r) }{ r \, (1 - s) } \bigg)
\\
& \overset{\text{(a)}}{=}
\begin{cases}
-1
& \mathrm{if} \ \gamma(r, s) > 1 ,
\\
1
& \mathrm{if} \ \gamma(r, s) < 1
\end{cases}
\label{eq:sgn_phi_m_m+1}
\end{align}
for every $m \in \mathbb{N}$ and $0 < r < s \le \infty$ with $r, s \neq 1$, where (a) follows from the hypothesis: $r < s$.
This implies the strict monotonicity of the derivative
\begin{align}
m \mapsto \frac{ \partial \phi(m, t; r, s) }{ \partial t }
\end{align}
with respect to $m \in \mathbb{N}$.
In addition, it follows from \eqref{prev2:phi} that
\begin{align}
\phi(m, (m+1)^{\theta( r )}; r, s)
& =
\underbrace{ \bigg( \frac{ (m+1)^{\theta( r )} - (m+1)^{\theta( r )} }{ (m+1)^{\theta( r )} - m^{\theta( r )} } \bigg) }_{ = 0 } \, m^{\theta( s )} + \underbrace{ \bigg( \frac{ (m+1)^{\theta( r )} - m^{\theta( r )} }{ (m+1)^{\theta( r )} - m^{\theta( r )} } \bigg) }_{ = 1 } \, (m+1)^{\theta( s )}
\\
& =
(m+1)^{\theta( s )} ,
\\
\phi(m+1, (m+1)^{\theta( r )}; r, s)
& =
\underbrace{ \bigg( \frac{ (m+2)^{\theta( r )} - (m+1)^{\theta( r )} }{ (m+2)^{\theta( r )} - (m+1)^{\theta( r )} } \bigg) }_{ = 1 } \, (m+1)^{\theta( s )} + \underbrace{ \bigg( \frac{ (m+1)^{\theta( s )} - (m+1)^{\theta( r )} }{ (m+2)^{\theta( r )} - (m+1)^{\theta( r )} } \bigg) }_{ = 0 } \, (m+2)^{\theta( s )}
\\
& =
(m+1)^{\theta( s )} ;
\end{align}
i.e, it holds that
\begin{align}
\phi(m, (m+1)^{\theta( r )}; r, s)
=
\phi(m+1, (m+1)^{\theta( r )}; r, s)
\label{eq:phi_m_m+1}
\end{align}
for every $m \in \mathbb{N}$ and $r, s \in (0, 1) \cup (1, \infty]$.
Since $t \mapsto \phi(m, t; r, s)$ is linear in $t$ (cf. \eqref{prev3:phi}), combining \eqref{eq:sgn_phi_m_m+1} and \eqref{eq:phi_m_m+1}, we have that for any fixed $0 < r < s \le \infty$,
\begin{itemize}
\item
if $\gamma( r, s ) > 1$, then $t \mapsto \min \{ \phi(m, t; r, s) \mid m \in \mathbb{N} \}$ is a piecewise linear function of $t \in \mathcal{J}( r )$, whose its slope never increases as $t$ increases, i.e., it is concave in $t \in \mathcal{J}( r )$,
\item
if $\gamma( r, s ) < 1$, then $t \mapsto \min \{ \phi(m, t; r, s) \mid m \in \mathbb{N} \}$ is a piecewise linear function of $t \in \mathcal{J}( r )$, whose its slope never decreases as $t$ increases, i.e., it is convex in $t \in \mathcal{J}( r )$.
\end{itemize}
Moreover, it also follows from \eqref{eq:sgn_phi_m_m+1} and \eqref{eq:phi_m_m+1} that
\begin{align}
\phi( m^{\prime}, t; r, s )
& =
\min \{ \phi(m, t; r, s) \mid m \in \mathbb{N} \}
\qquad \mathrm{if} \ \gamma( r, s ) > 1 ,
\\
\phi( m^{\prime}, t; r, s )
& =
\max \{ \phi(m, t; r, s) \mid m \in \mathbb{N} \}
\qquad \mathrm{if} \ \gamma( r, s ) < 1
\end{align}
for every $m^{\prime} \in \mathbb{N}$ and $t \in \mathcal{J}_{m^{\prime}}( r )$;
thus, we get from \eqref{eq:lambda_J} that
\begin{align}
\phi( \lfloor t^{\theta( r )} \rfloor, t; r, s )
& =
\min \{ \phi(m, t; r, s) \mid m \in \mathbb{N} \}
\qquad \mathrm{if} \ \gamma( r, s ) > 1 ,
\label{eq:t_theta_1} \\
\phi( \lfloor t^{\theta( r )} \rfloor, t; r, s )
& =
\max \{ \phi(m, t; r, s) \mid m \in \mathbb{N} \}
\qquad \mathrm{if} \ \gamma( r, s ) < 1
\label{eq:t_theta_2} 
\end{align}
for every $t \in \mathcal{J}( r )$.
Combining \eqref{def:phi}, \eqref{eq:phi_another}, \eqref{eq:t_theta_1}, and \eqref{eq:t_theta_1}, we obtain
\begin{align}
\| \bvec{w}( N_{r}^{-1}( \bvec{w} : t ) ) \|_{s}
& \le
\min \{ \phi(m, t; r, s) \mid m \in \mathbb{N} \}
\qquad \mathrm{if} \ \gamma(r, s) > 1 ,
\label{ineq:norm_w_phi_1} \\
\| \bvec{w}( N_{r}^{-1}( \bvec{w} : t ) ) \|_{s}
& \ge
\max \{ \phi(m, t; r, s) \mid m \in \mathbb{N} \}
\qquad \mathrm{if} \ \gamma(r, s) < 1
\label{ineq:norm_w_phi_2}
\end{align}
for every $t \in \mathcal{J}( r )$ and $0 < r < s \le \infty$.
Therefore, it $\gamma( r, s ) < 1$, we obtain
\begin{align}
N_{s}(X \mid Y)
& \overset{\eqref{def:expect_norm}}{=}
\mathbb{E} \big[ \| P_{X|Y}(\cdot \mid Y) \|_{s} \big]
\\
& \overset{\text{(a)}}{\ge}
\mathbb{E} \big[ \| \bvec{w}( N_{r}^{-1}( \bvec{w} : \| P_{X|Y}(\cdot \mid Y) \|_{r} ) ) \|_{s} \big]
\\
& \overset{\eqref{ineq:norm_w_phi_2}}{\ge}
\mathbb{E} \Big[ \max_{m \in \mathbb{N}} \phi(m, \| P_{X|Y}(\cdot \mid Y) \|_{r}; r, s) \Big]
\\
& \ge
\max_{m \in \mathbb{N}} \mathbb{E} \Big[ \phi \big(m, \| P_{X|Y}(\cdot \mid Y) \|_{r} ; r, s \big) \Big]
\\
& \overset{\text{(b)}}{=}
\max_{m \in \mathbb{N}} \phi \big(m, \mathbb{E} \big[ \| P_{X|Y}(\cdot \mid Y) \|_{r} \big] ; r, s \big)
\\
& \overset{\eqref{def:expect_norm}}{=}
\max_{m \in \mathbb{N}} \phi \big(m, N_{r}(X \mid Y) ; r, s \big)
\\
& \overset{\eqref{eq:t_theta_2}}{=}
\phi \big( \big\lfloor N_{r}(X \mid Y)^{\theta( r )} \big\rfloor , N_{r}(X \mid Y) ; r, s \big)
\\
& \overset{\eqref{eq:UV_phi}}{=}
N_{s}(U_{(r)} \mid V_{(r)})
\label{bound:expect_norm_UV_1}
\end{align}
for every pair of RVs $(X, Y)$ and $0 < r < s \le \infty$, where (a) follows by \thref{th:w}, and (b) follows by the linearity of $t \mapsto \phi(m, t; r, s)$ (cf. \eqref{prev3:phi}).
Analogously, it can also be verified that if $\gamma( r, s ) > 1$, then
\begin{align}
N_{s}(X \mid Y)
\le
N_{s}(U_{(r)} \mid V_{(r)})
\label{bound:expect_norm_UV_2}
\end{align}
for every pair of RVs $(X, Y)$ and $0 < r < s \le \infty$.

Finally, we define
\begin{align}
f_{\alpha}( t )
\coloneqq
\lim_{u \to \alpha} \frac{ u }{ 1 - u } \ln t
\end{align}
for $\alpha \in (0, \infty]$ and $t > 0$.
Since
\begin{itemize}
\item
it holds that $H_{\alpha}(X \mid Y) = f_{\alpha}( N_{\alpha}(X \mid Y) )$ for every $0 < \alpha \le \infty$,
\item
if $\gamma( \alpha, \beta ) > 1$, then $t \mapsto f_{\beta}( t )$ is a strictly decreasing function of $t > 0$,
\item
if $\gamma( \alpha, \beta ) < 1$, then $t \mapsto f_{\beta}( t )$ is a strictly increasing function of $t > 0$
\end{itemize}
for every $0 < \alpha < \beta \le \infty$, it follows from \eqref{bound:expect_norm_UV_1} and \eqref{bound:expect_norm_UV_2} that
\begin{align}
H_{\beta}(X \mid Y)
\le
H_{\beta}(U_{(\alpha)} \mid V_{(\alpha)})
\end{align}
for every pair of RVs $(X, Y)$ and $0 < \alpha < \beta \le \infty$.
In a similar way to the above discussions, we can also prove that
\begin{align}
H_{\beta}(X \mid Y)
\ge
H_{\beta}(U_{(\alpha)} \mid V_{(\alpha)})
\end{align}
for every pair of RVs $(X, Y)$ and $0 < \beta < \alpha \le \infty$.
This completes the proof of \thref{th:UV}.
\end{IEEEproof}

Note that if $\alpha = 1$, then sharp bounds in a similar situation to \thref{th:UV} were already derived in \cite[Theorem~2 and Corollary~1]{part2}.
We further mention that \thref{th:UV} has no constraint in the size of support $|\!\supp(P_{X})|$, i.e., the RV $X$ may take values from a countably infinite alphabet.

We now compare \thref{th:UV} to the inequality
\begin{align}
H_{\beta}(X \mid Y)
\le
H_{\alpha}(X \mid Y)
\qquad
\mathrm{for} \ 0 \le \alpha \le \beta \le \infty
\label{ineq:order_cond_renyi}
\end{align}
proved by Fehr and Berens \cite[Proposition~5]{fehr}, which shows that $\alpha \mapsto H_{\alpha}(X \mid Y)$ is decreasing for its order $\alpha \in [0, \infty]$.
It follows from \eqref{eq:UV_same_r} and \eqref{ineq:order_cond_renyi} that
\begin{align}
H_{\beta}(U_{(\alpha)} \mid V_{(\alpha)})
\le
H_{\alpha}(X \mid Y)
\qquad
\mathrm{for} \ 0 < \alpha \le \beta \le \infty ,
\end{align}
which implies that \eqref{ineq:UV_renyi_2} of \thref{th:UV} is tighter than \eqref{ineq:order_cond_renyi}.

\section{Applications}
\label{sect:appl}

In \sectref{sect:cond}, we established sharp bounds on $H_{\beta}(X \mid Y)$ with a fixed $H_{\alpha}(X \mid Y)$ for distinct orders $\alpha \neq \beta$.
In this section, we introduce applications of these sharp bounds to other information measures.
If an information measure is a strictly monotone function of $H_{\alpha}(X \mid Y)$, then our results can be applicable to it.

As an example, we can apply Theorems~\ref{th:A_renyi_alpha_inf} and~\ref{th:UV} to the minimum average probability of error
$P_{\mathrm{e}}(X \mid Y) = \exp( - H_{\infty}(X \mid Y) )$ defined in \eqref{def:Pe};
and then, we can obtain a generalization of Fano's inequality from $H(X \mid Y)$ to $H_{\alpha}(X \mid Y)$, as with \cite{sason}.
We organize this discussion in the next subsection.

\subsection{Generalized Fano's Inequality: Interplay Between Conditional R\'{e}nyi Entropy and Average Probability of Error}
\label{sect:fano}

In this subsection, we examine interplay between the conditional R\'{e}nyi entropy and the probability of error, as a generalization of Fano's inequality.
We first show an unconditional version of it in the following corollary.

\begin{corollary}[{Unconditional version of Fano's inequality for the R\'{e}nyi entropy, see also \cite[Corollary~3 and Theorem~2]{sason}}]
\label{cor:unconditional_fano_renyi}
Let $X$ be a discrete RV taking values from a countable alphabet $\mathcal{X}$.
Then, it holds that
\begin{align}
H_{\alpha}( X )
& \ge
\frac{ 1 }{ 1 - \alpha } \ln \! \Bigg[ \bigg\lfloor \frac{ 1 }{ 1 - \varepsilon } \bigg\rfloor \, (1 - \varepsilon)^{\alpha} + \bigg( 1 - \bigg\lfloor \frac{ 1 }{ 1 - \varepsilon } \bigg\rfloor \, (1 - \varepsilon) \bigg)^{\alpha} \Bigg]
\label{ineq:reverse_fano_renyi}
\end{align}
for every $\alpha \in (0, 1) \cup (1, \infty)$ and $\varepsilon \in [0, P_{\mathrm{e}}(X)]$, where the minimum average probability of error $P_{\mathrm{e}}(X)$ for guessing $X$ is defined by
\begin{align}
P_{\mathrm{e}}(X)
\coloneqq
\min_{\hat{x} \in \mathcal{X}} \Pr(X \neq \hat{x}) .
\end{align}
In addition, if $\supp( P_{X} )$ is finite, i.e., $| \! \supp( P_{X} )| \le n$ for some $n \in \mathbb{N}$, then
\begin{align}
H_{\alpha}( X )
& \le
\begin{dcases}
\frac{ 1  }{ 1 - \alpha } \ln \! \Big[ (1 - \varepsilon)^{\alpha} + (n-1)^{1-\alpha} \, \varepsilon^{\alpha} \Big]
& \mathrm{if} \ \varepsilon \le \frac{ n - 1 }{ n } ,
\\
\ln n
& \mathrm{if} \ \varepsilon > \frac{ n - 1 }{ n }
\end{dcases}
\label{ineq:fano_renyi}
\end{align}
for every $\alpha \in (0, 1) \cup (1, \infty)$ and $\varepsilon \in [P_{\mathrm{e}}( X ), 1]$.
\end{corollary}

\begin{IEEEproof}[Proof of \corref{cor:unconditional_fano_renyi}]
Let $X$ be a discrete RV.
It follows from \thref{th:renyi_w} that
\begin{align}
H_{\alpha}(X)
& \ge
H_{\alpha}( \bvec{w}( H_{\infty}^{-1}( \bvec{w} : H_{\infty}(X) ) ) )
\\
& =
H_{\alpha}( \bvec{w}( \| P_{X} \|_{\infty} ) )
\\
& =
H_{\alpha}( \bvec{w}( 1 - P_{\mathrm{e}}( X ) ) )
\\
& =
\frac{ 1 }{ 1 - \alpha } \ln \! \Bigg[ \Big\lfloor \frac{ 1 }{ 1 - P_{\mathrm{e}}( X ) } \Big\rfloor \, (1 - P_{\mathrm{e}}( X ))^{\alpha} + \bigg( 1 - \Big\lfloor \frac{ 1 }{ 1 - P_{\mathrm{e}}( X ) } \Big\rfloor \, (1 - P_{\mathrm{e}}( X )) \bigg)^{\alpha} \Bigg]
\\
& \overset{\text{(a)}}{\ge}
\frac{ 1 }{ 1 - \alpha } \ln \! \Bigg[ \bigg\lfloor \frac{ 1 }{ 1 - \varepsilon } \bigg\rfloor \, (1 - \varepsilon)^{\alpha} + \bigg( 1 - \bigg\lfloor \frac{ 1 }{ 1 - \varepsilon } \bigg\rfloor \, (1 - \varepsilon) \bigg)^{\alpha} \Bigg]
\end{align}
for every $\alpha \in (0, 1) \cup (1, \infty)$ and $\varepsilon \in [0, P_{\mathrm{e}}(X)]$, where (a) follows from the fact that $p \mapsto H_{\alpha}( \bvec{w}( p ) )$ is strictly decreasing%
\footnote{This monotonicity follows from \lemref{lem:mono} and the monotonicity of $t \mapsto (\alpha/(1-\alpha)) \ln t$.}
for $p \in (0, 1]$.

On the other hand, we suppose that $\supp( P_{X} )$ is finite, i.e., $|\!\supp( P_{X} )| = k \in \mathbb{N}$.
It follows from \thref{th:renyi_v} that
\begin{align}
H_{\alpha}(X)
& \le
H_{\alpha}( \bvec{v}_{k}( H_{\infty}^{-1}( \bvec{v}_{k} : H_{\infty}(X) ) ) )
\\
& =
H_{\alpha}( \bvec{v}_{k}( \| P_{X} \|_{\infty} ) )
\\
& =
H_{\alpha}( \bvec{v}_{k}( 1 - P_{\mathrm{e}}(X) ) )
\\
& =
\frac{ 1 }{ 1 - \alpha } \ln \! \Big[ (1 - P_{\mathrm{e}}(X))^{\alpha} + (k - 1)^{1 - \alpha} \, P_{\mathrm{e}}(X)^{\alpha} \Big]
\\
& \overset{\text{(a)}}{\le}
\frac{ 1 }{ 1 - \alpha } \ln \! \Big[ (1 - P_{\mathrm{e}}(X))^{\alpha} + (n - 1)^{1 - \alpha} \, P_{\mathrm{e}}(X)^{\alpha} \Big]
\label{eq:proof_cor_un_fano_UB_1} \\
& \overset{\text{(b)}}{\le}
\frac{ 1 }{ 1 - \alpha } \ln \! \Big[ (1 - \varepsilon)^{\alpha} + (n - 1)^{1 - \alpha} \, \varepsilon^{\alpha} \Big]
\label{eq:proof_cor_un_fano_UB_2}
\end{align}
for every $\alpha \in (0, 1) \cup (1, \infty)$, $n \ge |\!\supp( P_{X} ) |$, and $\varepsilon \in [P_{\mathrm{e}}(X), (n-1)/n]$, where (a) follows from the fact that the right-hand side of \eqref{eq:proof_cor_un_fano_UB_1} is strictly increasing for $n > 1$, and (b) also follows from the fact that the right-hand side of \eqref{eq:proof_cor_un_fano_UB_2} is strictly increasing for $\varepsilon \in [0, 1]$.
Finally, since $0 \le P_{\mathrm{e}}(X \mid Y) \le (k - 1)/ k \le (n - 1)/n$, Inequality \eqref{eq:proof_cor_un_fano_UB_2} can be rewritten by
\begin{align}
H_{\alpha}(X)
\le
\begin{dcases}
\frac{ 1 }{ 1 - \alpha } \ln \! \Big[ (1 - \varepsilon)^{\alpha} + (n - 1)^{1 - \alpha} \, \varepsilon^{\alpha} \Big]
& \mathrm{if} \ \varepsilon \le \frac{ n - 1 }{ n } ,
\\
\ln n
& \mathrm{if} \ \varepsilon > \frac{ n - 1 }{ n }
\end{dcases}
\end{align}
for every $\alpha \in (0, 1) \cup (1, \infty)$, $n \ge |\!\supp( P_{X} ) |$, and $\varepsilon \in [P_{\mathrm{e}}(X), 1]$.
This completes the proof of \corref{cor:unconditional_fano_renyi}.
\end{IEEEproof}

In the following corollary, we give sharp upper and lower bounds on $H_{\alpha}(X \mid Y)$ with a fixed probability of error, i.e., the following corollary shows generalizations of Fano's inequality and the reverse of Fano's inequality.

\begin{corollary}[{Conditional version of Fano's inequality for the R\'{e}nyi entropy, see also \cite[Theorems~3 and~11]{sason}}]
\label{cor:fano_renyi}
Let $X$ be a discrete RV, and let $Y$ be an arbitrary RV.
Then, it holds that
\begin{align}
H_{\alpha}(X \mid Y)
& \ge
\frac{ \alpha }{ 1 - \alpha } \ln \! \Bigg[ \Big( 1 + \Big\lfloor \frac{ 1 }{ 1 - \varepsilon } \Big\rfloor \Big)^{1/\alpha} \bigg( 1 - ( 1 - \varepsilon ) \, \Big\lfloor \frac{ 1 }{ 1 - \varepsilon } \Big\rfloor \bigg) - \Big\lfloor \frac{ 1 }{ 1 - \varepsilon } \Big\rfloor^{1/\alpha} \bigg( \varepsilon - ( 1 - \varepsilon ) \, \Big\lfloor \frac{ 1 }{ 1 - \varepsilon } \Big\rfloor \bigg) \Bigg]
\label{ineq:reverse_fano_renyi}
\end{align}
for every $\alpha \in (0, 1) \cup (1, \infty)$ and $\varepsilon \in [0, P_{\mathrm{e}}(X \mid Y)]$, where the minimum average probability of error $P_{\mathrm{e}}(X \mid Y)$ is defined in \eqref{def:Pe}.
In addition, if $\supp( P_{X} )$ is finite, i.e., $| \! \supp( P_{X} )| \le n$ for some $n \in \mathbb{N}$, then
\begin{align}
H_{\alpha}(X \mid Y)
& \le
\begin{dcases}
\frac{ 1 }{ 1 - \alpha } \ln \! \Big[ (1 - \varepsilon)^{\alpha} + (n-1)^{1-\alpha} \, \varepsilon^{\alpha} \Big]
& \mathrm{if} \ \varepsilon \le \frac{ n - 1 }{ n } ,
\\
\ln n
& \mathrm{if} \ \varepsilon > \frac{ n - 1 }{ n }
\end{dcases}
\label{ineq:fano_renyi}
\end{align}
for every $\alpha \in (0, 1) \cup (1, \infty)$ and $\varepsilon \in [P_{\mathrm{e}}(X \mid Y), 1]$.
\end{corollary}

\begin{IEEEproof}[Proof of \corref{cor:fano_renyi}]
Let $X$ be a discrete RV, and let $Y$ be an arbitrary RV.
It follows from \thref{th:UV} that
\begin{align}
H_{\alpha}(X \mid Y)
& \ge
H_{\alpha}(U_{(\infty)} \mid V_{(\infty)})
\\
& \overset{\eqref{eq:renyi_UV}}{=}
\frac{ \alpha }{ 1 - \alpha } \ln \Big[ \lambda \, m^{(1/\alpha)-1} + (1 - \lambda) \, (1+m)^{(1/\alpha)-1} \Big]
\\
& \overset{\eqref{eq:lambda_UV}}{=}
\frac{ \alpha }{ 1 - \alpha } \ln \Bigg[ \bigg( \frac{ (1+m)^{-1} - N_{\infty}(X \mid Y) }{ (1+m)^{-1} - m^{-1} } \bigg) \, m^{(1/\alpha)-1} + \bigg( \frac{ N_{\infty}(X \mid Y) - m^{-1} }{ (1+m)^{-1} - m^{-1} } \bigg) \, (1+m)^{(1/\alpha)-1} \Bigg]
\\
& =
\frac{ \alpha }{ 1 - \alpha } \ln \Bigg[ \frac{ m^{(1/\alpha)-1} \, (1+m)^{-1} - m^{-1} \, (1+m)^{(1/\alpha)-1} }{ (1+m)^{-1} - m^{-1} } + N_{\infty}(X \mid Y) \bigg( \frac{ (1+m)^{(1/\alpha)-1} - m^{(1/\alpha)-1} }{ (1+m)^{-1} - m^{-1} } \bigg) \Bigg]
\\
& =
\frac{ \alpha }{ 1 - \alpha } \ln \bigg[ \Big( (1+m)^{1/\alpha} - m^{1/\alpha} \Big) + N_{\infty}(X \mid Y) \Big( m^{1/\alpha} \, (1+m) - m \, (1+m)^{1/\alpha} \Big) \bigg]
\\
& =
\frac{ \alpha }{ 1 - \alpha } \ln \bigg[ (1+m)^{1/\alpha} \Big( 1 - m \, N_{\infty}(X \mid Y) \Big) - m^{1/\alpha} \Big( 1 - (1+m) N_{\infty}(X \mid Y) \Big) \bigg]
\\
& \overset{\eqref{eq:m_UV}}{=}
\frac{ \alpha }{ 1 - \alpha } \ln \Bigg[ \Big( 1 + \Big\lfloor \frac{ 1 }{ N_{\infty}(X \mid Y) } \Big\rfloor \Big)^{1/\alpha} \bigg( 1 - \Big\lfloor \frac{ 1 }{ N_{\infty}(X \mid Y) } \Big\rfloor \, N_{\infty}(X \mid Y) \bigg)
\notag \\
& \qquad \qquad \qquad \qquad \qquad \qquad \qquad
- \Big\lfloor \frac{ 1 }{ N_{\infty}(X \mid Y) } \Big\rfloor^{1/\alpha} \bigg( 1 - \Big( 1 + \Big\lfloor \frac{ 1 }{ N_{\infty}(X \mid Y) } \Big\rfloor \Big) N_{\infty}(X \mid Y) \bigg) \Bigg]
\\
& =
\frac{ \alpha }{ 1 - \alpha } \ln \Bigg[ \Big( 1 + \Big\lfloor \frac{ 1 }{ 1 - P_{\mathrm{e}}(X \mid Y) } \Big\rfloor \Big)^{1/\alpha} \bigg( 1 - \Big\lfloor \frac{ 1 }{ 1 - P_{\mathrm{e}}(X \mid Y) } \Big\rfloor \, \Big( 1 - P_{\mathrm{e}}(X \mid Y) \Big) \bigg)
\notag \\
& \qquad \qquad \qquad \qquad \qquad
- \Big\lfloor \frac{ 1 }{ 1 - P_{\mathrm{e}}(X \mid Y) } \Big\rfloor^{1/\alpha} \bigg( 1 - \Big( 1 + \Big\lfloor \frac{ 1 }{ 1 - P_{\mathrm{e}}(X \mid Y) } \Big\rfloor \Big) \Big( 1 - P_{\mathrm{e}}(X \mid Y) \Big) \bigg) \Bigg]
\\
& \overset{\text{(a)}}{\ge}
\frac{ \alpha }{ 1 - \alpha } \ln \Bigg[ \Big( 1 + \Big\lfloor \frac{ 1 }{ 1 - \varepsilon } \Big\rfloor \Big)^{1/\alpha} \bigg( 1 - ( 1 - \varepsilon ) \, \Big\lfloor \frac{ 1 }{ 1 - \varepsilon } \Big\rfloor \bigg) - \Big\lfloor \frac{ 1 }{ 1 - \varepsilon } \Big\rfloor^{1/\alpha} \bigg( 1 - ( 1 - \varepsilon ) \, \Big( 1 + \Big\lfloor \frac{ 1 }{ 1 - \varepsilon } \Big\rfloor \Big) \bigg) \Bigg]
\label{eq:proof_cor_fano_LB}
\end{align}
for every $\alpha \in (0, 1) \cup (1, \infty)$ and $\varepsilon \in [0, P_{\mathrm{e}}(X \mid Y)]$, where (a) follows from the fact that the right-hand side of \eqref{eq:proof_cor_fano_LB} is strictly increasing for $\varepsilon \in [0, 1)$.
Note that this monotonicity can be verified as with the proof of \lemref{lem:mono}.

On the other hand, we suppose that $\supp( P_{X} )$ is finite, i.e., $|\!\supp( P_{X} )| = k \in \mathbb{N}$.
It follows from \thref{th:A_renyi_alpha_inf} that
\begin{align}
H_{\alpha}(X \mid Y)
& \le
H_{\alpha}( \bvec{v}_{k}( H_{\infty}^{-1}( \bvec{v}_{k} : H_{\infty}(X \mid Y) ) ) )
\\
& =
H_{\alpha}( \bvec{v}_{k}( N_{\infty}(X \mid Y) ) )
\\
& =
H_{\alpha}( \bvec{v}_{k}( 1 - P_{\mathrm{e}}(X \mid Y) ) )
\\
& =
\frac{ 1 }{ 1 - \alpha } \ln \! \Big[ (1 - P_{\mathrm{e}}(X \mid Y))^{\alpha} + (k - 1)^{1 - \alpha} \, P_{\mathrm{e}}(X \mid Y)^{\alpha} \Big]
\\
& \overset{\text{(a)}}{\le}
\frac{ 1 }{ 1 - \alpha } \ln \! \Big[ (1 - P_{\mathrm{e}}(X \mid Y))^{\alpha} + (n - 1)^{1 - \alpha} \, P_{\mathrm{e}}(X \mid Y)^{\alpha} \Big]
\label{eq:proof_cor_fano_UB_1} \\
& \overset{\text{(b)}}{\le}
\frac{ 1 }{ 1 - \alpha } \ln \! \Big[ (1 - \varepsilon)^{\alpha} + (n - 1)^{1 - \alpha} \, \varepsilon^{\alpha} \Big]
\label{eq:proof_cor_fano_UB_2}
\end{align}
for every $\alpha \in (0, 1) \cup (1, \infty)$, $n \ge |\!\supp( P_{X} ) |$, and $\varepsilon \in [P_{\mathrm{e}}(X \mid Y), (n-1)/n]$, where (a) follows from the fact that the right-hand side of \eqref{eq:proof_cor_fano_UB_1} is strictly increasing for $n > 1$, and (b) also follows from the fact that the right-hand side of \eqref{eq:proof_cor_fano_UB_2} is strictly increasing for $\varepsilon \in [0, 1]$.
Finally, since $0 \le P_{\mathrm{e}}(X \mid Y) \le (k - 1)/ k \le (n - 1)/n$, Inequality \eqref{eq:proof_cor_fano_UB_2} can be rewritten by
\begin{align}
H_{\alpha}(X \mid Y)
\le
\begin{dcases}
\frac{ 1 }{ 1 - \alpha } \ln \! \Big[ (1 - \varepsilon)^{\alpha} + (n - 1)^{1 - \alpha} \, \varepsilon^{\alpha} \Big]
& \mathrm{if} \ \varepsilon \le \frac{ n - 1 }{ n } ,
\\
\ln n
& \mathrm{if} \ \varepsilon > \frac{ n - 1 }{ n }
\end{dcases}
\end{align}
for every $\alpha \in (0, 1) \cup (1, \infty)$, $n \ge |\!\supp( P_{X} ) |$, and $\varepsilon \in [P_{\mathrm{e}}(X \mid Y), 1]$.
This completes the proof of \corref{cor:fano_renyi}.
\end{IEEEproof}

\begin{figure}[!t]
\centering
\subfloat[Case of $\alpha = 1/3$.]{
\begin{overpic}[width = 0.45\hsize, clip]{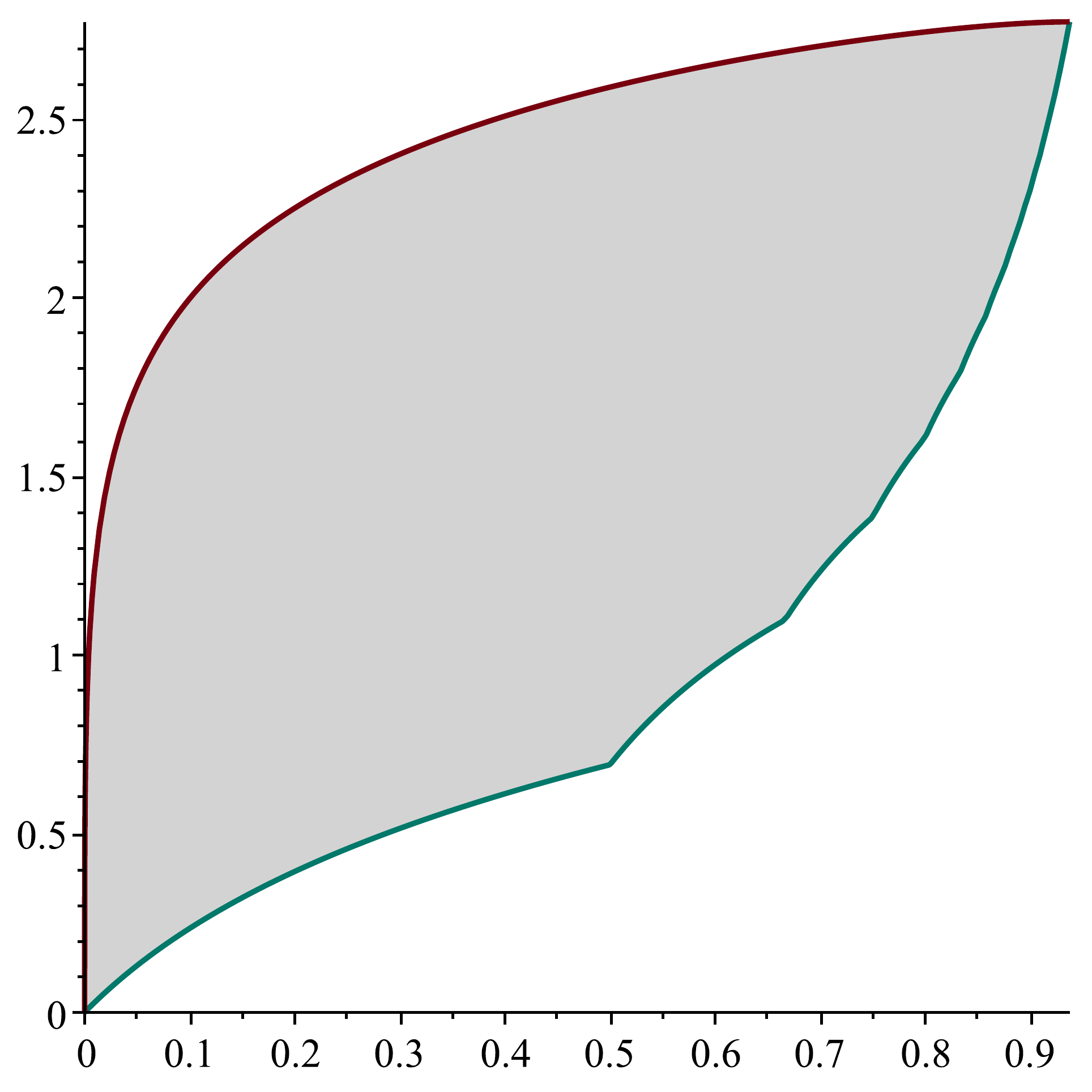}
\put(75, 10){$P_{\mathrm{e}}(X \mid Y)$}
\put(-3, 98){\scriptsize [nats]}
\put(-4, 45){\rotatebox{90}{$H_{1/3}(X \mid Y)$}}
\put(60, 23){\rotatebox{42}{\color{bluegreen} lower bound \eqref{ineq:reverse_fano_renyi}}}
\put(25, 87){\rotatebox{20}{\color{burgundy} upper bound \eqref{ineq:fano_renyi}}}
\put(35, 45){\rotatebox{42}{\Large \textbf{feasible region}}}
\end{overpic}
}\hfill
\subfloat[Case of $\alpha = 3$.]{
\begin{overpic}[width = 0.45\hsize, clip]{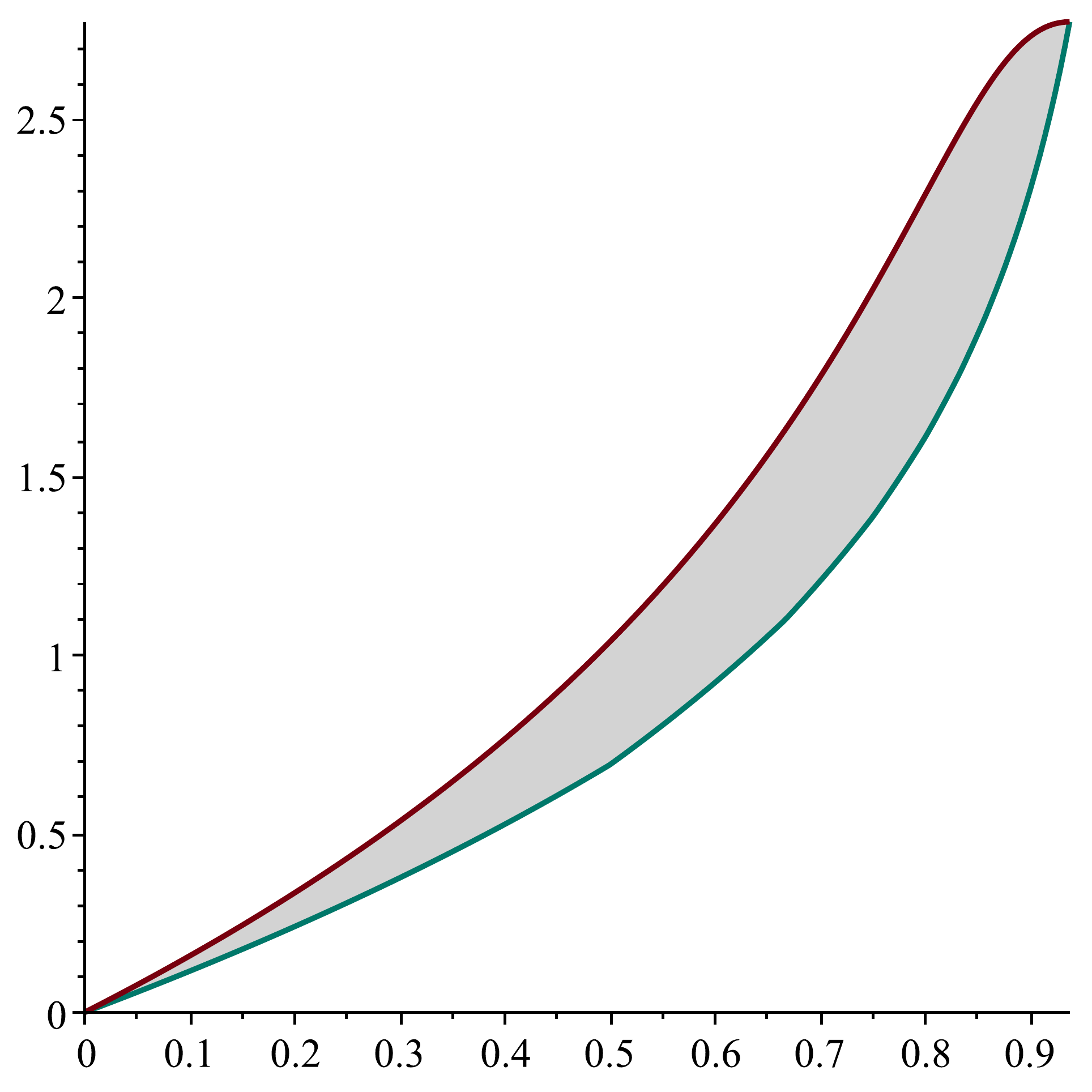}
\put(75, 10){$P_{\mathrm{e}}(X \mid Y)$}
\put(-3, 98){\scriptsize [nats]}
\put(-4, 45){\rotatebox{90}{$H_{3}(X \mid Y)$}}
\put(65, 26){\rotatebox{43}{\color{bluegreen} lower bound \eqref{ineq:reverse_fano_renyi}}}
\put(25, 25){\rotatebox{43}{\color{burgundy} upper bound \eqref{ineq:fano_renyi}}}
\put(65, 43){\rotatebox{55}{\large \textbf{feasible region}}}
\end{overpic}
}
\caption{Plot of the upper and lower bounds on $H_{\alpha}(X \mid Y)$ with a fixed $P_{\mathrm{e}}(X \mid Y)$ in the case of $|\!\supp( P_{X} )| \le n = 16$ (cf. \corref{cor:fano_renyi}).}
\label{fig:H_Pe}
\end{figure}

In \figref{fig:H_Pe}, we illustrate feasible regions of pairs $(P_{\mathrm{e}}(X \mid Y), H_{\alpha}(X \mid Y))$ established by the upper and lower bounds of \corref{cor:fano_renyi}.
The well-known bounds $0 \le H_{\alpha}(X \mid Y) \le \ln | \!\supp( P_{X} ) |$, e.g., \cite[Proposition~3]{fehr}, immediately follow by \corref{cor:fano_renyi}.
In addition, \corref{cor:fano_renyi} also implies that
\begin{align}
&
H_{\alpha}(X \mid Y)
\to
0
& \iff
& \qquad \qquad \qquad
P_{\mathrm{e}}(X \mid Y)
\to
0 ,
\\
&
H_{\alpha}(X \mid Y)
\to
\ln |\! \supp( P_{X} )|
& \iff
& \qquad \qquad \qquad
P_{\mathrm{e}}(X \mid Y)
\to
\frac{ |\! \supp( P_{X} )| - 1 }{ |\! \supp( P_{X} )| }
\end{align}
for every $\alpha \in (0, 1) \cup (1, \infty)$.

We now consider RVs $X$ and $Y$ taking values from same finite alphabet $\mathcal{X}$.
Since $P_{\mathrm{e}}(X \mid Y) \le \Pr(X \neq Y)$, note that \eqref{ineq:fano_renyi} also holds with $\varepsilon = \Pr(X \neq Y)$.
If $\varepsilon = \Pr(X \neq Y) \le 1 - 1/|\mathcal{X}|$, then \eqref{ineq:fano_renyi} approaches to
\begin{align}
H(X \mid Y)
\le
h_{2} \big( \! \Pr(X \neq Y) \big) + \Pr(X \neq Y) \ln \! \big( |\mathcal{X}| - 1 \big)
\end{align}
as $\alpha \to 1$, where $h_{2} : t \mapsto - t \ln t - (1-t) \ln (1-t)$ denotes the binary entropy function.
Thus, Ineq. \eqref{ineq:fano_renyi} is a part of generalized Fano's inequality in terms of Arimoto's conditional R\'{e}nyi entropy.
Unlike \eqref{ineq:fano_renyi}, since $P_{\mathrm{e}}(X \mid Y) \le \Pr(X \neq Y)$, note that \eqref{ineq:reverse_fano_renyi} does not hold with $\varepsilon = \Pr(X \neq Y)$ in general.
In fact, the reverse of Fano's inequality (cf.  \cite[Theorem~1]{feder}, \cite[Eq.~(15)]{kovalevsky}, \cite[Eq.~(6)]{tebbe}) is a sharp lower bound on the conditional Shannon entropy $H(X \mid Y)$ with not fixed $\Pr( X \neq Y )$ but fixed $\varepsilon = P_{\mathrm{e}}(X \mid Y)$ as
\begin{align}
H(X \mid Y)
\ge
\bigg( 1 - (1 - \varepsilon) \Big\lfloor \frac{ 1 }{ 1 - \varepsilon } \Big\rfloor \bigg) \Big( 1 + \Big\lfloor \frac{ 1 }{ 1 - \varepsilon } \Big\rfloor \Big) \ln \! \Big( 1 + \Big\lfloor \frac{ 1 }{ 1 - \varepsilon } \Big\rfloor \Big) - \bigg( \varepsilon - (1 - \varepsilon) \Big\lfloor \frac{ 1 }{ 1 - \varepsilon } \Big\rfloor \bigg) \Big\lfloor \frac{ 1 }{ 1 - \varepsilon } \Big\rfloor \ln \! \Big\lfloor \frac{ 1 }{ 1 - \varepsilon } \Big\rfloor .
\label{ineq:reverse_fano}
\end{align}
Since \eqref{ineq:reverse_fano_renyi} approaches to \eqref{ineq:reverse_fano} with $\varepsilon = P_{\mathrm{e}}(X \mid Y)$ as $\alpha \to 1$, Inequality~\eqref{ineq:reverse_fano_renyi} can be seen as a generalized reverse of Fano's inequality in terms of Arimoto's conditional R\'{e}nyi entropy.
Indeed, it can be verified that the right-hand side of \eqref{ineq:reverse_fano} is same as the right-hand side of \eqref{eq:renyi_UV_1}.

We now compare \corref{cor:fano_renyi} with another generalized Fano's inequality, which is an upper bound on another definition of conditional R\'{e}nyi entropy
\begin{align}
H_{\alpha}^{\mathrm{H}}(X \mid Y)
\coloneqq
\frac{ 1 }{ 1 - \alpha } \ln \mathbb{E} \bigg[ \sum_{x \in \supp( P_{X|Y}(\cdot \mid Y) )} P_{X|Y}(x \mid Y)^{\alpha} \bigg]
\end{align}
proposed by Hayashi \cite{hayashi}.
Iwamoto and Shikata \cite{iwamoto} investigated many information theoretic properties of $H_{\alpha}^{\mathrm{H}}(X \mid Y)$.
Then, they derived a different type of Fano's inequality, as shown in the following theorem.

\begin{theorem}[{\cite[Theorem~7]{iwamoto}}]
\label{th:fano_hayashi_renyi}
Let $X$ and $Y$ be RVs taking values from same finite alphabet $\mathcal{X}$.
Define
\begin{align}
g_{1}(\alpha, \varepsilon, n)
& \coloneqq
\frac{ 1 }{ 1 - \alpha } \ln \Big[ (1 - \varepsilon)^{\alpha} + (n - 1)^{1-\alpha} \, \varepsilon^{\alpha} \Big] ,
\\
g_{2}(\alpha, \varepsilon, n)
& \coloneqq
\frac{ 1 }{ 1 - \alpha } \ln \Big[ (1 - \varepsilon) + \varepsilon^{\alpha-1} \, (1 - (1-\varepsilon)^{2-\alpha}) \, (n - 1)^{1-\alpha} \Big] .
\end{align}
Then, it holds that
\begin{align}
H_{\alpha}^{\mathrm{H}}(X \mid Y)
\le
\max \! \Big\{ g_{1} \big( \alpha, \Pr(X \neq Y), |\mathcal{X}| \big), g_{2} \big(\alpha, \Pr(X \neq Y), |\mathcal{X}| \big) \Big\}
\label{ineq:fano_hayashi_renyi}
\end{align}
for every $\alpha \in (0, 1) \cup (1, \infty)$ whenever%
\footnote{%
Note that $g_{2}(\alpha, 0, n)$ is undefined if $\alpha \in (0, 1)$ and $g_{2}(\alpha, 1, n)$ is also undefined if $\alpha \ge 2$.
In \cite[Theorem~7]{iwamoto}, the limiting value was considered as $\Pr(X \neq Y) \to 0$.%
}
$0 < \Pr(X \neq Y) < 1$.
\end{theorem}

Since
\begin{align}
H_{\alpha}^{\mathrm{H}}(X \mid Y)
\le
H_{\alpha}(X \mid Y)
\end{align}
(cf. \cite[Theorem~1]{iwamoto}), Inequality~\eqref{ineq:fano_renyi} of \corref{cor:fano_renyi} can be relaxed by replacing $H_{\alpha}(X \mid Y)$ by $H_{\alpha}^{\mathrm{H}}(X \mid Y)$.
Moreover, since the right-hand side of \eqref{ineq:fano_renyi} is equal to
$
g_{1}(\alpha, \varepsilon, n)
$
for $0 \le \varepsilon \le (n-1)/n$, Inequality~\eqref{ineq:fano_renyi} of \corref{cor:fano_renyi} can also be relaxed by replacing the right-hand side of \eqref{ineq:fano_renyi} by the right-hand side of \eqref{ineq:fano_hayashi_renyi} for $0 \le \varepsilon \le (n-1)/n$.
Thus, Inequality~\eqref{ineq:fano_renyi} of \corref{cor:fano_renyi} is tighter than \eqref{ineq:fano_hayashi_renyi} of \thref{th:fano_hayashi_renyi} when $0 \le \Pr(X \neq Y) \le (|\mathcal{X}|-1)/|\mathcal{X}|$.

Finally, we give sharp bounds on $P_{\mathrm{e}}(X \mid Y)$ with a fixed $H_{\alpha}(X \mid Y)$ by using the results of \sectref{sect:cond}, as shown in the following corollary.

\begin{corollary}[{see also \cite[Theorems~5 and~12]{sason}}]
\label{cor:bound_Pe}
Let $X$ be a discrete RV, and let $Y$ be an arbitrary RV.
Then, it holds that
\begin{align}
P_{\mathrm{e}}(X \mid Y)
\le
1 - \frac{ \Big(1 + \Big\lfloor \exp \big( H_{\alpha}(X \mid Y) \big) \Big\rfloor \Big)^{1/\alpha} - \Big\lfloor \exp \big( H_{\alpha}(X \mid Y) \big) \Big\rfloor^{1/\alpha} - \exp \Big( \dfrac{ 1 - \alpha }{ \alpha } \, H_{\alpha}(X \mid Y) \Big) }{ \Big\lfloor \exp \big( H_{\alpha}(X \mid Y) \big) \Big\rfloor \, \Big(1 + \Big\lfloor \exp \big( H_{\alpha}(X \mid Y) \big) \Big\rfloor\Big)^{1/\alpha} - \Big\lfloor \exp \big( H_{\alpha}(X \mid Y) \big) \Big\rfloor^{1/\alpha} \, \Big(1 + \Big\lfloor \exp \big( H_{\alpha}(X \mid Y) \big) \Big\rfloor\Big) }
\label{eq:Pe_UB}
\end{align}
for every $\alpha \in (0, 1) \cup (1, \infty)$.
In addition, if $\supp( P_{X} )$ is finite, i.e., $|\!\supp( P_{X} )| = n$ for some $n \in \mathbb{N}$, then
\begin{align}
P_{\mathrm{e}}(X \mid Y)
\ge
1 - H_{\alpha}^{-1}( \bvec{v}_{n} : H_{\alpha}(X \mid Y) )
\label{eq:Pe_LB}
\end{align}
for every $\alpha \in (0, \infty)$, where $H_{\alpha}^{-1}( \bvec{v}_{n} : \cdot )$ is defined in \eqref{def:inv_Hv}.
In particular, if either $\alpha = 1/2$ or $\alpha = 2$, then the following closed-form bounds hold:
\begin{align}
P_{\mathrm{e}}(X \mid Y)
& \ge
1 - \frac{ n \, (n - 1) - (n-2) \, \exp\big( H_{1/2}(X \mid Y) \big) + 2 \, \sqrt{ \exp\big( H_{1/2}(X \mid Y) \big) \, (n-1) \, \big(n - \exp\big( H_{1/2}(X \mid Y) \big) \big) } }{ n^{2} } ,
\label{eq:Pe_LB_order_half} \\
P_{\mathrm{e}}(X \mid Y)
& \ge
1 - \frac{ 1 + \sqrt{ \exp\big( - H_{2}(X \mid Y) \big) \, (n-1) \, \big( n - \exp\big( H_{2}(X \mid Y) \big) \big) } }{ n }
\label{eq:Pe_LB_order_2}
\end{align}
with $n = |\!\supp( P_{X} )|$.
\end{corollary}

\begin{IEEEproof}[Proof of \corref{cor:bound_Pe}]
Let $X$ be a discrete RV, and let $Y$ be an arbitrary RV.
It follows from \thref{th:UV} that
\begin{align}
H_{\infty}(X \mid Y)
& \le
H_{\infty}(U_{(\alpha)} \mid V_{(\alpha)})
\\
& =
- \ln \! \Big[ \lambda \, m^{-1} + (1-\lambda) \, (1+m)^{-1} \Big]
\\
& =
- \ln \! \bigg[ \bigg( \frac{ (1+m)^{(1/\alpha)-1} - N_{\alpha}(X \mid Y) }{ (1+m)^{(1/\alpha)-1} - m^{(1/\alpha)-1} } \bigg) \, m^{-1} + \bigg( \frac{ N_{\alpha}(X \mid Y) - m^{(1/\alpha)-1} }{ (1+m)^{(1/\alpha)-1} - m^{(1/\alpha)-1} } \bigg) \, (1+m)^{-1} \bigg]
\\
& =
- \ln \! \bigg[ \bigg( \frac{ m^{-1} \, (1+m)^{(1/\alpha)-1} - m^{(1/\alpha)-1} \, (1+m)^{-1} }{ (1+m)^{(1/\alpha)-1} - m^{(1/\alpha)-1} } \bigg) + N_{\alpha}(X \mid Y) \, \bigg( \frac{ (1+m)^{-1} - m^{-1} }{ (1+m)^{(1/\alpha)-1} - m^{(1/\alpha)-1} } \bigg) \bigg]
\\
& =
- \ln \! \bigg[ \bigg( \frac{ (1+m)^{1/\alpha} - m^{1/\alpha} }{ m \, (1+m)^{1/\alpha} - m^{1/\alpha} \, (1+m) } \bigg) - N_{\alpha}(X \mid Y) \, \bigg( \frac{ 1 }{ m \, (1+m)^{1/\alpha} - m^{1/\alpha} \, (m+1) } \bigg)\bigg]
\\
& =
\ln \! \bigg[ \frac{ m \, (1+m)^{1/\alpha} - m^{1/\alpha} \, (1+m) }{ (1+m)^{1/\alpha} - m^{1/\alpha} - N_{\alpha}(X \mid Y) } \bigg]
\\
& =
\ln \! \left[ \frac{ \Big\lfloor \exp \big( H_{\alpha}(X \mid Y) \big) \Big\rfloor \, \Big(1 + \Big\lfloor \exp \big( H_{\alpha}(X \mid Y) \big) \Big\rfloor\Big)^{1/\alpha} - \Big\lfloor \exp \big( H_{\alpha}(X \mid Y) \big) \Big\rfloor^{1/\alpha} \, \Big(1 + \Big\lfloor \exp \big( H_{\alpha}(X \mid Y) \big) \Big\rfloor\Big) }{ \Big(1 + \Big\lfloor \exp \big( H_{\alpha}(X \mid Y) \big) \Big\rfloor \Big)^{1/\alpha} - \Big\lfloor \exp \big( H_{\alpha}(X \mid Y) \big) \Big\rfloor^{1/\alpha} - \exp \Big( \dfrac{ 1 - \alpha }{ \alpha } \, H_{\alpha}(X \mid Y) \Big) } \right]
\label{eq:proof_Pe_UB}
\end{align}
for every $\alpha \in (0, 1) \cup (1, \infty)$.
Along with \eqref{eq:proof_Pe_UB}, the equation
\begin{align}
H_{\infty}(X \mid Y)
=
\ln \bigg[ \frac{ 1 }{ 1 - P_{\mathrm{e}}(X \mid Y) } \bigg]
\label{eq:minH_Pe}
\end{align}
yields \eqref{eq:Pe_UB}.

On the other hand, we suppose that $|\!\supp( P_{X} )| = n$ for some $n \in \mathbb{N}$.
It follows from \thref{th:A_renyi_alpha_inf} that
\begin{align}
H_{\infty}(X \mid Y)
& \ge
H_{\infty}( \bvec{v}_{n}( H_{\alpha}^{-1}( \bvec{v}_{n} : H_{\alpha}(X \mid Y) ) ) )
\\
& =
- \ln H_{\alpha}^{-1}( \bvec{v}_{n} : H_{\alpha}(X \mid Y) )
\label{eq:proof_Pe_LB}
\end{align}
for every $\alpha \in (0, \infty)$.
Combining \eqref{eq:minH_Pe} and \eqref{eq:proof_Pe_LB}, we have \eqref{eq:Pe_LB}.
Finally, Inequalities~\eqref{eq:Pe_LB_order_half} and~\eqref{eq:Pe_LB_order_2} can be obtained by substituting \eqref{eq:Pe_LB} into the closed-forms of \factref{ex:renyi}.
This completes the proof of \corref{cor:bound_Pe}.
\end{IEEEproof}

\begin{figure}[!t]
\centering
\subfloat[Case of $\alpha = 1/2$.]{
\begin{overpic}[width = 0.45\hsize, clip]{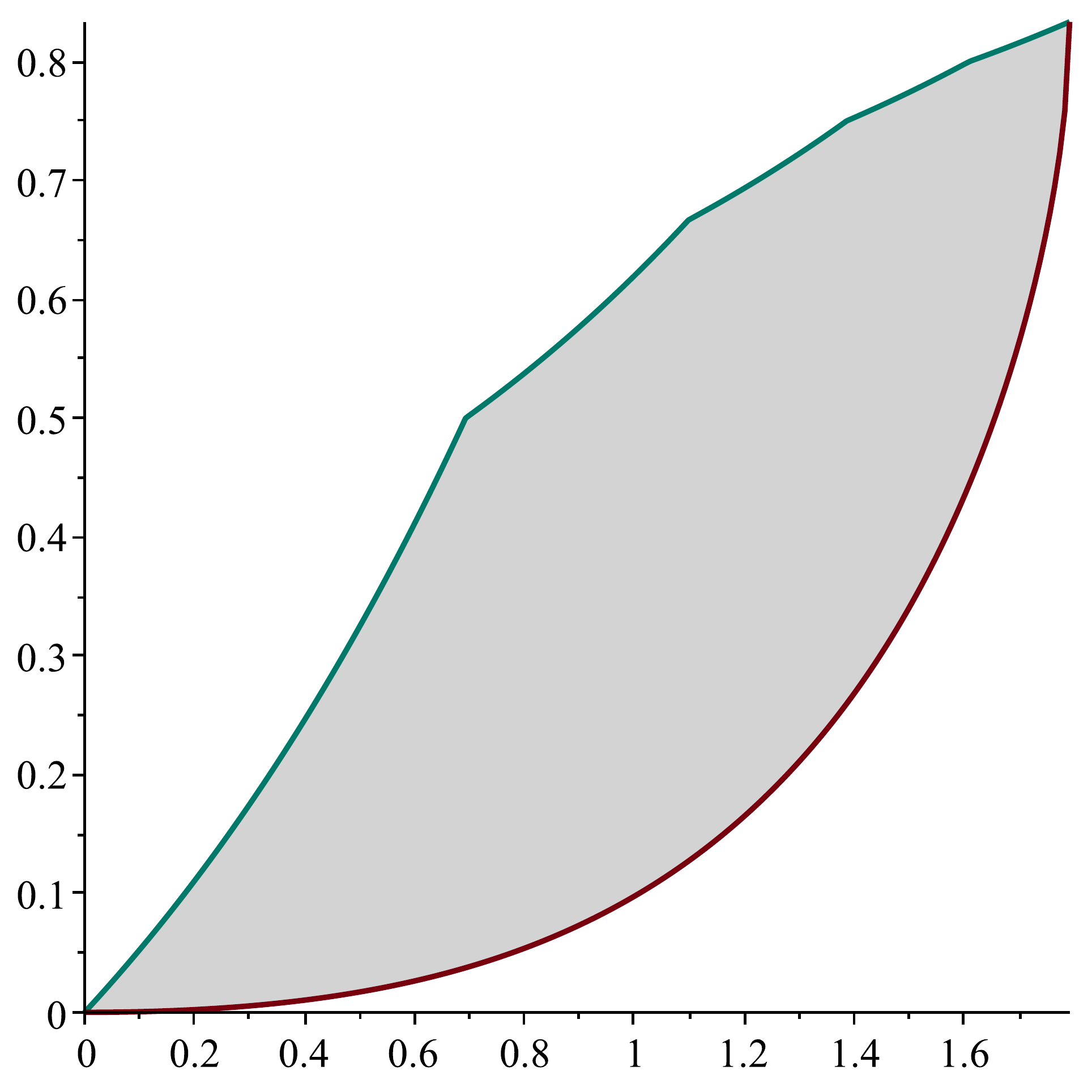}
\put(75, 10){$H_{1/2}(X \mid Y)$}
\put(98, 2){\scriptsize [nats]}
\put(-4, 45){\rotatebox{90}{$P_{\mathrm{e}}(X \mid Y)$}}
\put(35, 64){\rotatebox{40}{\color{bluegreen} upper bound \eqref{eq:Pe_UB}}}
\put(83, 30){\rotatebox{65}{\color{burgundy} lower bound \eqref{eq:Pe_LB_order_half}}}
\put(50, 37){\rotatebox{47}{\Large \textbf{feasible region}}}
\end{overpic}
}\hfill
\subfloat[Case of $\alpha = 2$.]{
\begin{overpic}[width = 0.45\hsize, clip]{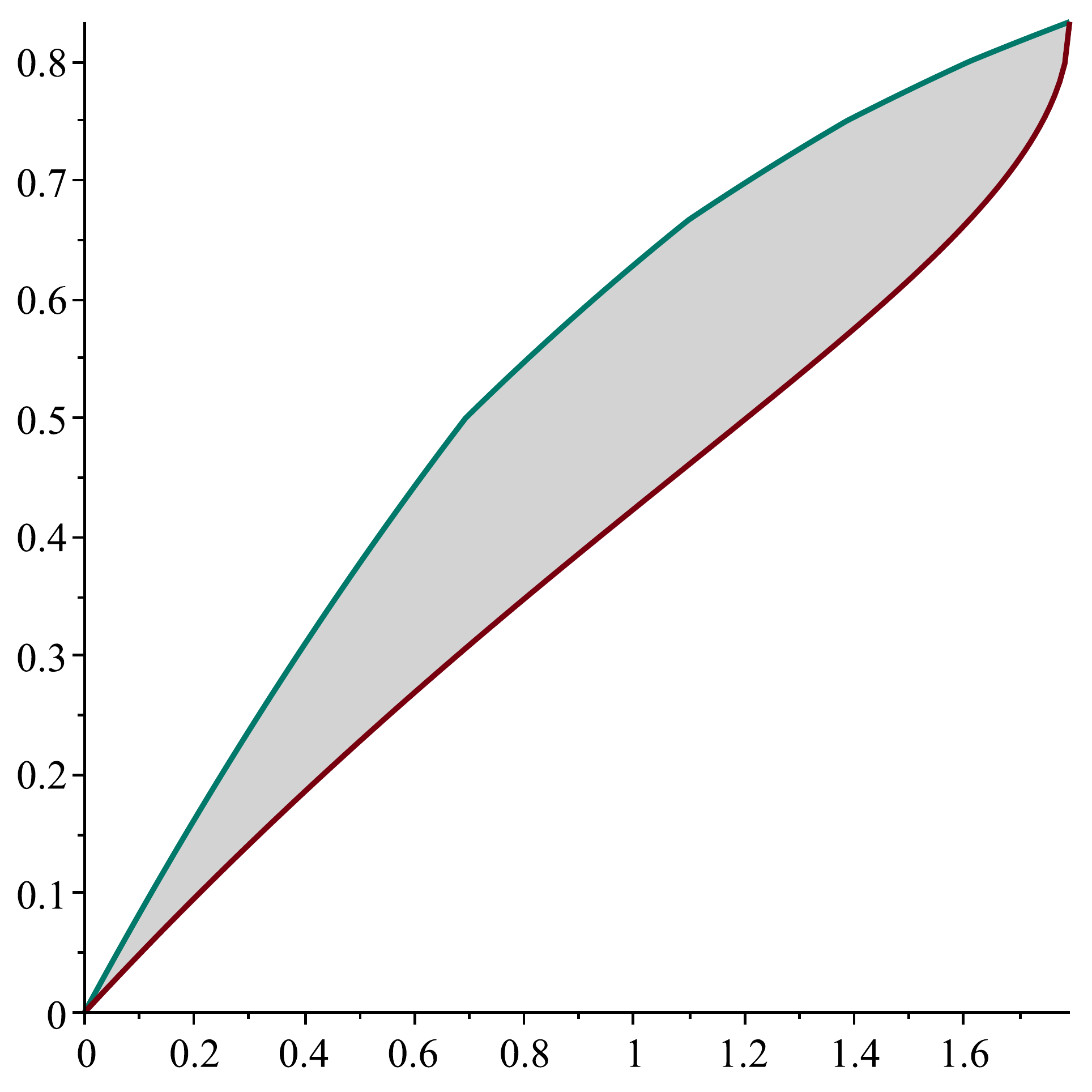}
\put(75, 10){$H_{2}(X \mid Y)$}
\put(98, 2){\scriptsize [nats]}
\put(-4, 45){\rotatebox{90}{$P_{\mathrm{e}}(X \mid Y)$}}
\put(35, 64){\rotatebox{40}{\color{bluegreen} upper bound \eqref{eq:Pe_UB}}}
\put(55, 40){\rotatebox{42}{\color{burgundy} lower bound \eqref{eq:Pe_LB_order_2}}}
\put(50, 55){\rotatebox{41}{\Large \textbf{feasible region}}}
\end{overpic}
}
\caption{Plot of the upper and lower bounds on $P_{\mathrm{e}}(X \mid Y)$ with a fixed $H_{\alpha}(X \mid Y)$ in the case of $|\!\supp( P_{X} )| \le n = 6$ (cf. \corref{cor:bound_Pe}).}
\label{fig:Pe_H}
\end{figure}

In \figref{fig:Pe_H}, we illustrate feasible regions of pairs $(H_{\alpha}(X \mid Y), P_{\mathrm{e}}(X \mid Y))$ established by the upper and lower bounds on \corref{cor:bound_Pe}.
In this subsection, we examined interplay between $H_{\alpha}(X \mid Y)$ and $P_{\mathrm{e}}(X \mid Y)$ as a generalization of Fano's inequality.
In the next subsection, we further consider applications of the results of this study to other information measures.

\subsection{Other Related Information Measures}

We now consider the Bhattacharrya parameter \cite[Definition~17]{mori} of $X$ given $Y$, defined by
\begin{align}
Z(X \mid Y)
\coloneqq
\frac{ 1 }{ |\mathcal{X}|-1 } \sum_{\substack{ x, x^{\prime} \in \mathcal{X} : \\ x \neq x^{\prime}}} \mathbb{E} \bigg[ \sqrt{ P_{X|Y}(x \mid Y) \, P_{X|Y}(x^{\prime} \mid Y) } \bigg] ,
\label{def:bhattacharrya}
\end{align}
where $X$ is an RV taking values from a finite alphabet $\mathcal{X}$, and $Y$ is an arbitrary RV.
This quantity $Z(X \mid Y)$ is useful to analyze rate of polarization for $|\mathcal{X}|$-ary polar codes \cite[Section~VII-B]{mori}, \cite[Section~4.1.2]{sasoglu}.
After some algebra, we have
\begin{align}
H_{1/2}(X \mid Y)
& \overset{\eqref{def:A_renyi}}{=}
\ln N_{1/2}(X \mid Y)
\\
& \overset{\eqref{def:expect_norm}}{=}
\ln \mathbb{E} \Big[ \| P_{X|Y}(\cdot \mid Y) \|_{1/2} \Big]
\\
& \overset{\eqref{def:norm}}{=}
\ln \mathbb{E} \bigg[ \bigg( \sum_{x \in \mathcal{X}} \sqrt{ P_{X|Y}(x \mid Y) } \bigg)^{2} \bigg]
\\
& =
\ln \mathbb{E} \bigg[ \sum_{x, x^{\prime} \in \mathcal{X}} \sqrt{ P_{X|Y}(x \mid Y) \, P_{X|Y}(x^{\prime} \mid Y) } \bigg]
\\
& =
\ln \mathbb{E} \bigg[ \sum_{x \in \mathcal{X}} \bigg( P_{X|Y}(x \mid Y) + \sum_{x^{\prime} \in \mathcal{X} : x^{\prime} \neq x} \sqrt{ P_{X|Y}(x \mid Y) \, P_{X|Y}(x^{\prime} \mid Y) } \bigg) \bigg]
\\
& =
\ln \bigg( 1 + \mathbb{E} \bigg[ \sum_{x, x^{\prime} \in \mathcal{X} : x^{\prime} \neq x} \sqrt{ P_{X|Y}(x \mid Y) \, P_{X|Y}(x^{\prime} \mid Y) } \bigg] \bigg)
\\
& \overset{\eqref{def:bhattacharrya}}{=}
\ln \! \Big( 1 + (|\mathcal{X}| - 1) \, Z(X \mid Y) \Big) ;
\label{eq:Z_H}
\end{align}
therefore, our results can be applicable to $Z(X \mid Y)$.
Fortunately, if $\alpha = 1/2$, i.e., in the case of \eqref{eq:Z_H}, our results can be expressed in closed-forms by Facts~\ref{ex:renyi} and~\ref{fact:slope_v_half}.
In the following corollary, we show sharp upper and lower bounds on $Z(X \mid Y)$ with fixed $P_{\mathrm{e}}(X \mid Y)$ and $|\mathcal{X}|$.

\begin{corollary}[Sharp bounds on Bhattacharrya parameter with a fixed average probability of error]
\label{cor:bhattacharrya_Pe}
Let $X$ be an RV taking values from a finite alphabet $\mathcal{X}$, and let $Y$ be an arbitrary RV.
Then, it holds that
\begin{align}
\frac{ 1 }{ |\mathcal{X}| - 1 } \Bigg( \Big\lfloor \frac{ 1 }{ 1 - \varepsilon_{1} } \Big\rfloor + \Big( 1 + \Big\lfloor \frac{ 1 }{ 1 - \varepsilon_{1} } \Big\rfloor \Big) \bigg( 1 - (1-\varepsilon_{1}) \Big\lfloor \frac{ 1 }{ 1 - \varepsilon_{1} } \Big\rfloor \bigg) - 1 \Bigg)
\le
Z(X \mid Y)
\le
\bigg( \frac{ |\mathcal{X}| - 2 }{ |\mathcal{X}| - 1 } \bigg) \, \varepsilon_{2} + 2 \, \sqrt{ \frac{ \varepsilon_{2} \, (1-\varepsilon_{2}) }{ |\mathcal{X}| - 1 } }
\label{bound:Z_Pe}
\end{align}
for every $0 \le \varepsilon_{1} \le P_{\mathrm{e}}(X \mid Y) \le \varepsilon_{2} \le (|\mathcal{X}| - 1) / |\mathcal{X}|$.
\end{corollary}

\begin{IEEEproof}[Proof of \corref{cor:bhattacharrya_Pe}]
It follows from \eqref{ineq:reverse_fano_renyi} of \corref{cor:fano_renyi} and \eqref{eq:Z_H} that
\begin{align}
Z(X \mid Y)
& \ge
\frac{ 1 }{ |\mathcal{X}| - 1 } \Bigg[ \Big( 1 + \Big\lfloor \frac{ 1 }{ 1 - \varepsilon } \Big\rfloor \Big)^{2} \bigg( 1 - ( 1 - \varepsilon ) \, \Big\lfloor \frac{ 1 }{ 1 - \varepsilon } \Big\rfloor \bigg) - \Big\lfloor \frac{ 1 }{ 1 - \varepsilon } \Big\rfloor^{2} \bigg( 1 - ( 1 - \varepsilon ) \, \Big( 1 + \Big\lfloor \frac{ 1 }{ 1 - \varepsilon } \Big\rfloor \Big) \bigg) - 1 \Bigg]
\\
& =
\frac{ 1 }{ |\mathcal{X}| - 1 } \Bigg[ \Big( 1 + 2 \Big\lfloor \frac{ 1 }{ 1 - \varepsilon } \Big\rfloor \Big) - (1-\varepsilon) \Big\lfloor \frac{ 1 }{ 1 - \varepsilon } \Big\rfloor \Big( 1 + \Big\lfloor \frac{ 1 }{ 1 - \varepsilon } \Big\rfloor \Big) - 1 \Bigg]
\\
& =
\frac{ 1 }{ |\mathcal{X}| - 1 } \Bigg[ \Big\lfloor \frac{ 1 }{ 1 - \varepsilon } \Big\rfloor + \Big( 1 + \Big\lfloor \frac{ 1 }{ 1 - \varepsilon } \Big\rfloor \Big) \bigg( 1 - (1-\varepsilon) \Big\lfloor \frac{ 1 }{ 1 - \varepsilon } \Big\rfloor \bigg) - 1 \Bigg]
\end{align}
for every $\varepsilon \in [0, P_{\mathrm{e}}(X \mid Y)]$.
In addition, it also follows from \eqref{ineq:fano_renyi} of \corref{cor:fano_renyi} and \eqref{eq:Z_H} that
\begin{align}
Z(X \mid Y)
& \le
\frac{ 1 }{ |\mathcal{X}| - 1 } \bigg[ \Big( \sqrt{ 1-\varepsilon } + \sqrt{ (|\mathcal{X}|-1) \, \varepsilon } \Big)^{2} - 1 \bigg]
\\
& =
\frac{ 1 }{ |\mathcal{X}| - 1 } \bigg[ \Big( (1-\varepsilon) + 2 \sqrt{ (|\mathcal{X}|-1) \, \varepsilon \, (1-\varepsilon) } + (|\mathcal{X}|-1) \, \varepsilon \Big) - 1 \bigg]
\\
& =
\frac{ 1 }{ |\mathcal{X}| - 1 } \bigg[ 2 \sqrt{ (|\mathcal{X}|-1) \, \varepsilon \, (1-\varepsilon) } + (|\mathcal{X}|-2) \, \varepsilon \bigg]
\\
& =
\bigg( \frac{ |\mathcal{X}| - 2 }{ |\mathcal{X}| - 1 } \bigg) \, \varepsilon + 2 \, \sqrt{ \frac{ \varepsilon \, (1-\varepsilon) }{ |\mathcal{X}| - 1 } }
\end{align}
for every $\varepsilon \in [P_{\mathrm{e}}(X \mid Y), (|\mathcal{X}|-1)/|\mathcal{X}|]$.
This completes the proof of \corref{cor:bhattacharrya_Pe}.
\end{IEEEproof}

\begin{figure}[!t]
\centering
\begin{overpic}[width = 0.6\hsize, clip]{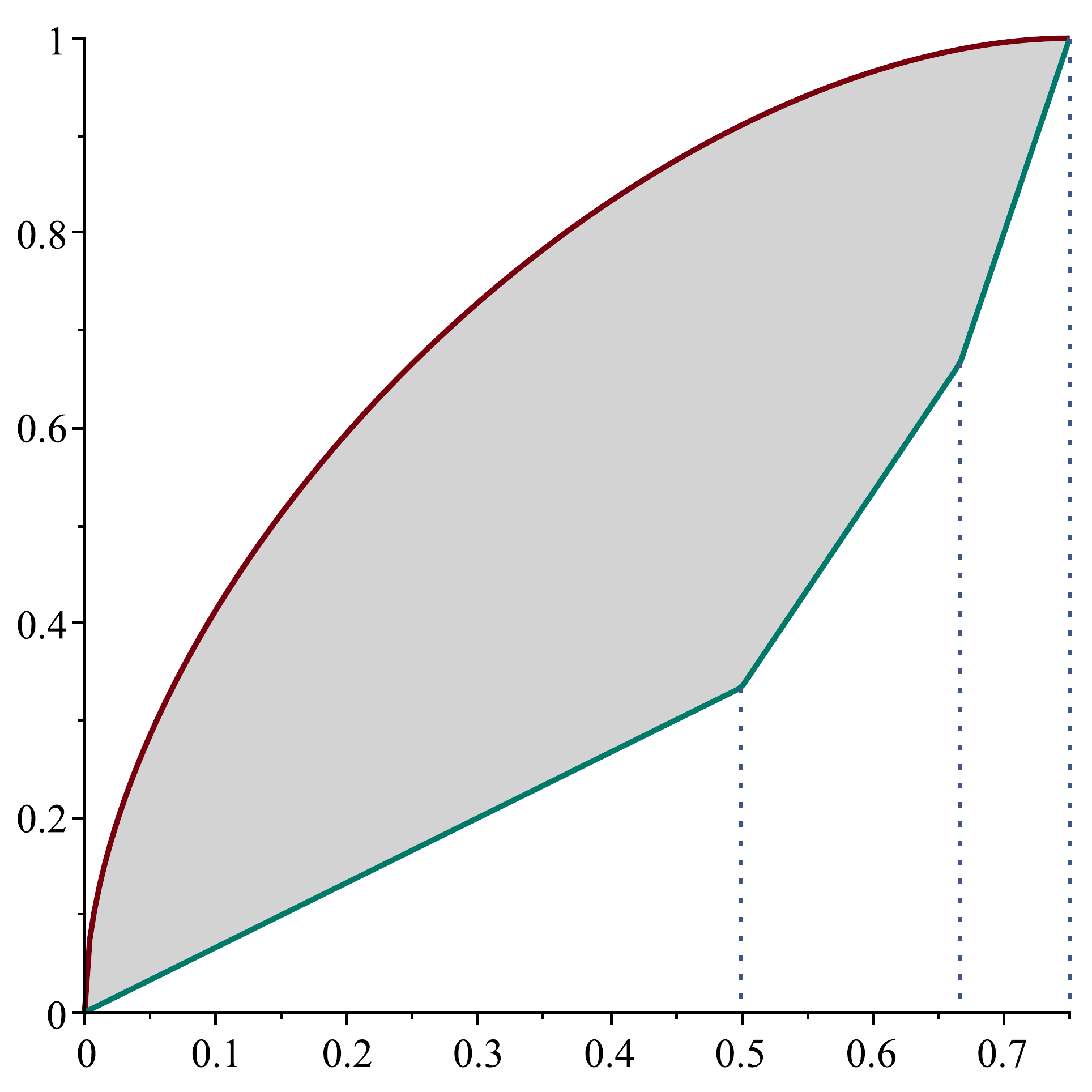}
\put(50, -2){$P_{\mathrm{e}}(X \mid Y)$}
\put(-2, 45){\rotatebox{90}{$Z(X \mid Y)$}}
\put(56.5, 11.5){$1/2$}
\put(61.5, 10.75){\vector(2, -1){6}}
\put(78.5, 12){$2/3$}
\put(83.5, 11.5){\vector(1, -1){4.25}}
\put(104, 12){$3/4$}
\put(103, 12){\vector(-1, -1){4.5}}
\put(30, 68){\rotatebox{44}{\color{burgundy} upper bound of \eqref{bound:Z_Pe}}}
\put(70, 33){\rotatebox{53.5}{\color{bluegreen} lower bound of \eqref{bound:Z_Pe}}}
\put(44, 44){\rotatebox{45}{\Large \textbf{feasible region}}}
\end{overpic}
\caption{Plot of the upper and lower bounds on $Z(X \mid Y)$ with a fixed $P_{\mathrm{e}}(X \mid Y)$ in the case of $|\mathcal{X}| = 4$ (cf. \corref{cor:bhattacharrya_Pe}).}
\label{fig:Z_Pe}
\end{figure}

In \figref{fig:Z_Pe}, we illustrate a feasible region of pairs $(P_{\mathrm{e}}(X \mid Y), Z(X \mid Y))$ established by the upper and lower bounds of \corref{cor:bhattacharrya_Pe}.
In a similar way to the proof of \corref{cor:bhattacharrya_Pe}, we can also derive sharp upper and lower bounds on $P_{\mathrm{e}}(X \mid Y)$ with fixed $Z(X \mid Y)$ and $|\mathcal{X}|$, as shown in the following corollary.

\begin{corollary}[{Sharp bounds on minimum average probability of error with a fixed Bhattacharrya parameter, see also \cite[Lemma~22]{mori}}]
\label{cor:Pe_bhattacharrya}
Let $X$ be an RV taking values from a finite alphabet $\mathcal{X}$, and let $Y$ be an arbitrary RV.
Then, it holds that
\begin{align}
&
\frac{ |\mathcal{X}| - 1 }{ |\mathcal{X}|^{2} } \bigg( 2 + (|\mathcal{X}|-2) Z(X \mid Y) - 2 \sqrt{(1 - Z(X \mid Y)) (1 + (|\mathcal{X}|-1) Z(X \mid Y)) } \bigg)
\notag \\
& \qquad \qquad \qquad \qquad \qquad \le
P_{\mathrm{e}}(X \mid Y)
\le
1 + \frac{ (|\mathcal{X}| - 1) \, Z(X \mid Y) - 2 \Big\lfloor 1 + (|\mathcal{X}| - 1) \, Z(X \mid Y) \Big\rfloor }{ \Big\lfloor 1 + (|\mathcal{X}| - 1) \, Z(X \mid Y) \Big\rfloor \, \Big(1 + \Big\lfloor 1 + (|\mathcal{X}| - 1) \, Z(X \mid Y) \Big\rfloor\Big) } .
\label{bound:Pe_Z}
\end{align}
\end{corollary}

\begin{IEEEproof}[Proof of \corref{cor:Pe_bhattacharrya}]
Let $X$ be an RV taking values from a finite alphabet $\mathcal{X}$, and let $Y$ be an arbitrary RV.
For simplicity, let $Z = Z(X \mid Y)$, let $\varepsilon = P_{\mathrm{e}}(X \mid Y)$, and let $n = |\mathcal{X}|$.
Substituting $\alpha = 1/2$ and $\exp( H_{1/2}(X \mid Y) ) = 1 + (|\mathcal{X}|-1) Z(X \mid Y)$ (see \eqref{eq:Z_H}) into \eqref{eq:Pe_UB}, we have
\begin{align}
\varepsilon
& \le
1 - \frac{ (1 + \lfloor 1 + (n - 1) \, Z \rfloor )^{2} - \lfloor 1 + (n - 1) \, Z \rfloor^{2} - ( 1 + (n - 1) \, Z ) }{ \lfloor 1 + (n - 1) \, Z \rfloor \, (1 + \lfloor 1 + (n - 1) \, Z \rfloor )^{2} - \lfloor 1 + (n - 1) \, Z  \rfloor^{2} \, (1 + \lfloor 1 + (n - 1) \, Z \rfloor ) }
\\
& =
1 + \frac{ (n - 1) \, Z - 2 \lfloor 1 + (n - 1) \, Z \rfloor }{ \lfloor 1 + (n - 1) \, Z \rfloor \, (1 + \lfloor 1 + (n - 1) \, Z \rfloor ) } ,
\end{align}
which is the upper bound of \eqref{bound:Pe_Z}.

On the other hand, consider the right-hand inequality of \eqref{bound:Z_Pe}.
We readily see that
\begin{align}
&&
Z
& \le
\bigg( \frac{ n - 2 }{ n - 1 } \bigg) \, \varepsilon + 2 \, \sqrt{ \frac{ \varepsilon \, (1-\varepsilon) }{ n - 1 } }
\\
& \iff &
Z - \bigg( \frac{ n - 2 }{ n - 1 } \bigg) \, \varepsilon
& \le
2 \, \sqrt{ \frac{ \varepsilon \, (1-\varepsilon) }{ n - 1 } }
\\
& \iff &
Z^{2} - 2 \, Z \, \bigg( \frac{ n - 2 }{ n - 1 } \bigg) \, \varepsilon + \bigg( \frac{ n - 2 }{ n - 1 } \bigg)^{2} \, \varepsilon^{2}
& \le
\frac{ 4 \, \varepsilon \, (1-\varepsilon) }{ n - 1 } 
\\
& \iff &
\Bigg( \bigg( \frac{ n - 2 }{ n - 1 } \bigg)^{2} + \frac{ 4 }{ n - 1 } \Bigg) \, \varepsilon^{2} - \Bigg( 2 \, Z \, \bigg( \frac{ n - 2 }{ n - 1 } \bigg) + \frac{ 4 }{ n - 1 } \Bigg) \, \varepsilon + Z^{2}
& \le
0 .
\end{align}
By the quadratic formula, we have
\begin{align}
\frac{ n - 1 }{ n^{2} } \bigg( 2 + (n-2) Z - 2 \sqrt{(1 - Z) (1 + (n-1) Z) } \bigg)
\le
\varepsilon
\le
\frac{ n - 1 }{ n^{2} } \bigg( 2 + (n-2) Z + 2 \sqrt{(1 - z) (1 + (n-1) Z) } \bigg) ;
\end{align}
and the left-hand inequality is indeed the lower bound of \eqref{bound:Pe_Z}.
This completes the proof of \corref{cor:Pe_bhattacharrya}.
\end{IEEEproof}

\corref{cor:Pe_bhattacharrya} is equivalent to \cite[Lemma~22]{mori}; and thus, this study gives an alternative proof of it.
Note that \corref{cor:Pe_bhattacharrya} also shows same feasible regions as \figref{fig:Z_Pe}.

So far, in this section, we presented applications of the sharp bounds on $H_{\beta}(X \mid Y)$ with a fixed $H_{\alpha}(X \mid Y)$ in the case of either $\alpha = \infty$ or $\beta = \infty$.
However, the results of this study enable us to consider the sharp bounds on $H_{\beta}(X \mid Y)$ with a fixed $H_{\alpha}(X \mid Y)$ in the case of that both $\alpha$ and $\beta$ are finite orders.
As an example, the following corollary shows sharp bounds on $H_{2}(X \mid Y)$ with a fixed $H_{1/2}(X \mid Y)$.

\begin{corollary}
\label{cor:2_half}
Let $X$ be a discrete RV, and let $Y$ be an arbitrary RV.
Then, it holds that
\begin{align}
H_{2}(X \mid Y)
& \le
\ln \! \bigg[ \Big\lfloor \exp \big( H_{1/2}(X \mid Y) \big) \Big\rfloor \, \Big( 1 + \Big\lfloor \exp \big( H_{1/2}(X \mid Y) \big) \Big\rfloor \Big) \bigg]
\notag \\
& \qquad
- 2 \ln \! \Bigg[ \Big( 1 + \Big\lfloor \exp \big( H_{1/2}(X \mid Y) \big) \Big\rfloor \Big)^{3/2} - \Big\lfloor \exp \big( H_{1/2}(X \mid Y) \big) \Big\rfloor^{3/2}
\notag \\
& \qquad \qquad \qquad
+ \exp\big( H_{1/2}(X \mid Y) \big) \bigg( \sqrt{ \Big\lfloor \exp \big( H_{1/2}(X \mid Y) \big) \Big\rfloor } - \sqrt{ 1 + \Big\lfloor \exp \big( H_{1/2}(X \mid Y) \big) \Big\rfloor } \bigg) \Bigg] .
\label{eq:2_half_UB}
\end{align}
In addition, if $|\!\supp( P_{X} )| = n$ for some $n \in \mathbb{N}$, then the following lower bounds hold:
\begin{itemize}
\item
if $0 \le H_{1/2}(X \mid Y) \le 2 \ln (1 + \sqrt{n-1}) - \ln 2$, then
\begin{align}
H_{2}(X \mid Y)
\ge
\ln \Bigg( \frac{ n - 1 }{ n \, H_{1/2}^{-1}( \bvec{v}_{n} : H_{1/2}(X \mid Y) )^{2} - 2 \, H_{1/2}^{-1}( \bvec{v}_{n} : H_{1/2}(X \mid Y) ) + 1 } \Bigg) ,
\label{eq:2_half_LB1}
\end{align}
where $H_{1/2}^{-1}( \bvec{v}_{n} : \cdot )$ is already shown in \factref{ex:renyi}, and
\item
if $2 \ln (1 + \sqrt{n-1}) - \ln 2 < H_{1/2}(X \mid Y) \le \ln n$, then
\begin{align}
H_{2}(X \mid Y)
& \ge
2 \ln \! \Big[ n - 2 \, \sqrt{n-1} \Big] + \ln \! \Big[ n \, (n - 1) \Big]
\notag \\
& \qquad \qquad \qquad
- 2 \ln \! \bigg[ 2 + \exp\big( H_{1/2}(X \mid Y) \big) \Big( 2 \, \sqrt{ n - 1 } - n \Big) + n \, \Big( n - \sqrt{ n - 1 } - 2 \Big) \bigg] .
\label{eq:2_half_LB2}
\end{align}
\end{itemize}
\end{corollary}

\begin{figure}[!t]
\centering
\begin{overpic}[width = 0.6\hsize, clip]{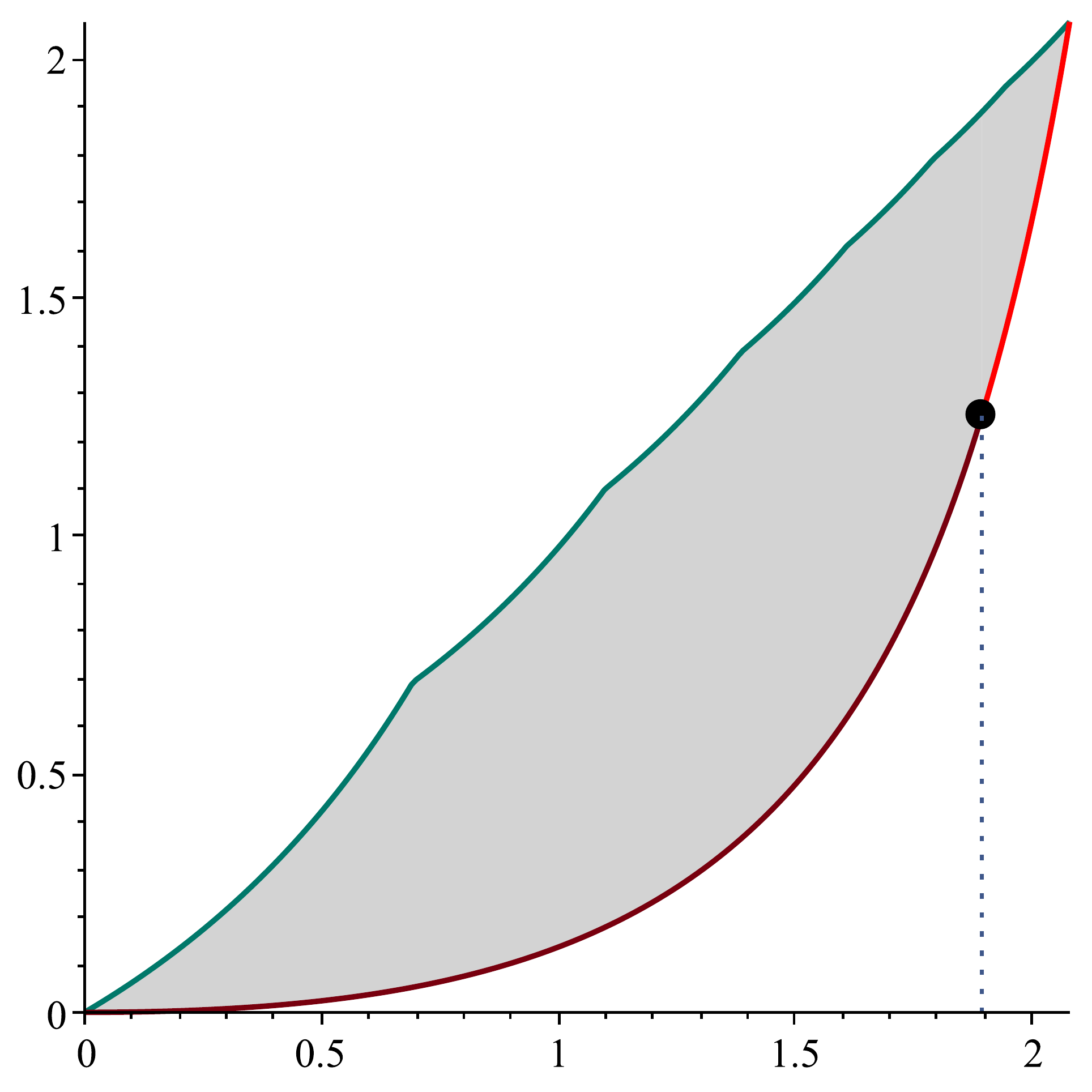}
\put(45, -1){$H_{1/2}(X \mid Y)$}
\put(-2, 43){\rotatebox{90}{$H_{2}(X \mid Y)$}}
\put(100, 2){\scriptsize [nats]}
\put(-3, 98){\scriptsize [nats]}
\put(98, 25){$2 \ln (1 + \sqrt{n-1}) - \ln 2$}
\put(120, 20){$\approx 1.89398$}
\put(100.5, 23){\vector(-2, -3){10}}
\put(40, 48){\rotatebox{46}{\color{bluegreen} upper bound \eqref{eq:2_half_UB}}}
\put(60, 10){\rotatebox{40}{\color{burgundy} lower bound \eqref{eq:2_half_LB1}}}
\put(96, 65){\rotatebox{80}{\color{red} lower bound \eqref{eq:2_half_LB2}}}
\put(52, 30){\rotatebox{45}{\Large \textbf{feasible region}}}
\end{overpic}
\caption{Plot of upper and lower bounds on $H_{2}(X \mid Y)$ with a fixed $H_{1/2}(X \mid Y)$ in the case of $|\! \supp( P_{X} ) | \le n = 8$ (cf. \corref{cor:2_half}).}
\end{figure}

\begin{IEEEproof}[Proof of \corref{cor:2_half}]
Let $X$ be a discrete RV, and let $Y$ be an arbitrary RV.
It follows from \thref{th:UV} that
\begin{align}
H_{2}(X \mid Y)
& \le
H_{2}(U_{(1/2)} \mid V_{(1/2)})
\\
& =
- 2 \ln \! \Big[ \lambda \, m^{-1/2} + (1-\lambda) \, (1+m)^{-1/2} \Big]
\\
& =
- 2 \ln \! \bigg[ \Big( (1+m) - N_{1/2}(X \mid Y) \Big) \, m^{-1/2} + \Big( N_{1/2}(X \mid Y) - m \Big) \, (1+m)^{-1/2} \bigg]
\\
& =
- 2 \ln \! \bigg[ \Big( (1+m) - N_{1/2}(X \mid Y) \Big) \, m^{-1/2} + \Big( N_{1/2}(X \mid Y) - m \Big) \, (1+m)^{-1/2} \bigg]
\\
& =
2 \ln \! \Bigg[ \frac{ \sqrt{ m \, (1+m) } }{ \big( (1+m) - N_{1/2}(X \mid Y) \big) \sqrt{ 1 + m } + \big( N_{1/2}(X \mid Y) - m \big) \sqrt{ m } } \Bigg]
\\
& =
\ln \! \Big[ m \, (1+m) \Big] - 2 \ln \! \bigg[ \Big( (1+m) - N_{1/2}(X \mid Y) \Big) \sqrt{ 1 + m } + \Big( N_{1/2}(X \mid Y) - m \Big) \sqrt{ m } \bigg]
\\
& =
\ln \! \Big[ m \, (1+m) \Big] - 2 \ln \! \bigg[ (1+m)^{3/2} - m^{3/2} + N_{1/2}(X \mid Y) \Big( \sqrt{ m } - \sqrt{ 1 + m } \Big) \bigg] .
\label{eq:2_half_UB_proof}
\end{align}
Substituting \eqref{eq:m_UV} into \eqref{eq:2_half_UB_proof}, we obtain \eqref{eq:2_half_UB}.

On the other hand, we suppose that $\supp( P_{X} )$ is finite.
Let $n = |\!\supp( P_{X} )|$.
By \factref{fact:slope_v_half}, the following identities hold:
\begin{align}
p^{\ast}(n; 1/2, 2)
& =
\frac{ 1 }{ 2 } ,
\\
t^{\ast}(n; 1/2, 2)
& =
\| \bvec{v}_{n}( p^{\ast}(n; 1/2, 2) ) \|_{1/2}
\\
& =
\| \bvec{v}_{n}( 1/2 ) \|_{1/2}
\\
& =
\bigg( \sqrt{ \frac{1}{2} } + \sqrt{ \frac{ n-1 }{ 2 } } \bigg)^{2}
\\
& =
\frac{ 1 }{ 2 } \Big( 1 + \sqrt{ n - 1 } \Big)^{2} ,
\\
\| \bvec{v}_{n}( p^{\ast}(n; 1/2, 2) ) \|_{2}
& =
\| \bvec{v}_{n}( 1/2 ) \|_{2}
\\
& =
\sqrt{ \frac{ 1 }{ 2^{2} } + \frac{ 1 }{ 2^{2} (n-1) } }
\\
& =
\frac{ 1 }{ 2 } \sqrt{ \frac{ n }{ n - 1 } } ,
\\
H_{1/2}( \bvec{v}_{n}( p^{\ast}(n; 1/2, 2) ) )
& =
H_{1/2}( \bvec{v}_{n}( 1/2 ) )
\\
& =
\ln \! \bigg( \frac{ 1 }{ 2 } \Big( 1 + \sqrt{ n - 1 } \Big)^{2} \bigg)
\\
& =
2 \ln \Big( 1 + \sqrt{ n - 1 } \Big) - \ln 2 .
\end{align}
Hence, it follows from \eqref{def:interval_Ha} and \eqref{def:interval_Hb} that
\begin{align}
\mathcal{H}_{n}^{(\mathrm{a})}( 1/2, 2 )
& =
\big( 2 \ln ( 1 + \sqrt{ n - 1 } ) - \ln 2, \ln n \big] ,
\\
\mathcal{H}_{n}^{(\mathrm{b})}( 1/2, 2 )
& =
\big[ 0, 2 \ln ( 1 + \sqrt{ n - 1 } ) - \ln 2 \big] ,
\end{align}
respectively, where note that if $n = 2$, then
\begin{align}
\mathcal{H}_{2}^{(\mathrm{a})}( 1/2, 2 )
& =
\emptyset ,
\\
\mathcal{H}_{2}^{(\mathrm{b})}( 1/2, 2 )
& =
[0, \ln 2] .
\end{align}
If $n \ge 3$ and $H_{1/2}(X \mid Y) \in \mathcal{H}_{n}^{(\mathrm{a})}( 1/2, 2 )$, then it follows from \thref{th:ST} that
\begin{align}
H_{2}(X \mid Y)
& \ge
H_{2}(S_{(1/2, 2)} \mid T_{(1/2, 2)})
\\
& =
- 2 \ln \! \Big[ (1-\delta) \, n^{-1/2} + \delta \, \| \bvec{v}_{n}( p_{(\mathrm{a})} ) \|_{2} \Big]
\\
& =
- 2 \ln \! \bigg[ (1-\delta) \, \sqrt{ \frac{ 1 }{ n } } + \delta \, \frac{ 1 }{ 2 } \sqrt{ \frac{ n }{ n - 1 } } \bigg]
\\
& =
- 2 \ln \! \bigg[ \bigg( \frac{ 2 \, N_{1/2}(X \mid Y) - 2 \, \sqrt{n-1} - n }{ n - 2 \, \sqrt{n-1} } \bigg) \, \sqrt{ \frac{ 1 }{ n } } + \bigg( \frac{ n - N_{1/2}(X \mid Y) }{ n - 2 \, \sqrt{n-1} } \bigg)  \sqrt{ \frac{ n }{ n - 1 } } \bigg]
\\
& =
2 \ln \! \Big[ n - 2 \, \sqrt{n-1} \Big] - 2 \ln \! \bigg[ \frac{ 2 \, N_{1/2}(X \mid Y) - 2 \, \sqrt{n-1} - n }{ \sqrt{ n } } +  \frac{ \big( n - N_{1/2}(X \mid Y) \big) \sqrt{ n } }{ \sqrt{ n - 1 } } \bigg]
\\
& =
2 \ln \! \Big[ n - 2 \, \sqrt{n-1} \Big] + 2 \ln \! \Big[ \sqrt{ n \, (n - 1) } \Big]
\notag \\
& \qquad \qquad \qquad
- 2 \ln \! \bigg[ \Big( 2 \, N_{1/2}(X \mid Y) - 2 \, \sqrt{n-1} - n \Big) \, \sqrt{ n - 1 } + \Big( n - N_{1/2}(X \mid Y) \Big) \, n \bigg]
\\
& =
2 \ln \! \Big[ n - 2 \, \sqrt{n-1} \Big] + \ln \! \Big[ n \, (n - 1) \Big]
\notag \\
& \qquad \qquad \qquad
- 2 \ln \! \bigg[ N_{1/2}(X \mid Y) \Big( 2 \, \sqrt{ n - 1 } - n \Big) - 2 \, (n-1) + n \, \Big( n - \sqrt{ n - 1 } \Big) \bigg]
\\
& =
2 \ln \! \Big[ n - 2 \, \sqrt{n-1} \Big] + \ln \! \Big[ n \, (n - 1) \Big]
\notag \\
& \qquad \qquad \qquad
- 2 \ln \! \bigg[ 2 + \exp\big( H_{1/2}(X \mid Y) \big) \Big( 2 \, \sqrt{ n - 1 } - n \Big) + n \, \Big( n - \sqrt{ n - 1 } - 2 \Big) \bigg] ;
\end{align}
thus, we have \eqref{eq:2_half_LB2}.
Moreover, if $n \ge 3$ and $H_{1/2}(X \mid Y) \in \mathcal{H}_{n}^{(\mathrm{b})}( 1/2, 2 )$, then it also follows from \thref{th:ST} that
\begin{align}
H_{2}(X \mid Y)
& \ge
H_{2}(S_{(1/2, 2)} \mid T_{(1/2, 2)})
\\
& =
H_{2}( \bvec{v}_{n}( H_{1/2}^{-1}( \bvec{v}_{n} : H_{1/2}(X \mid Y) ) ) )
\\
& =
- \ln \Bigg( H_{1/2}^{-1}( \bvec{v}_{n} : H_{1/2}(X \mid Y) )^{2} + \frac{ \big(1 - H_{1/2}^{-1}( \bvec{v}_{n} : H_{1/2}(X \mid Y) ) \big)^{2} }{ n - 1 } \Bigg)
\\
& =
\ln \Bigg( \frac{ n - 1 }{ n \, H_{1/2}^{-1}( \bvec{v}_{n} : H_{1/2}(X \mid Y) )^{2} - 2 \, H_{1/2}^{-1}( \bvec{v}_{n} : H_{1/2}(X \mid Y) ) + 1 } \Bigg) .
\label{eq:2_half_LB1_proof}
\end{align}
Finally, if $n = 2$, then \thref{th:A_renyi_2} also yields \eqref{eq:2_half_LB1_proof}.
Hence, we have \eqref{eq:2_half_LB1}.
This completes the proof of \corref{cor:2_half}.
\end{IEEEproof}

Analogously, we can also establish closed-form sharp bounds on $H_{1/2}(X \mid Y)$ with a fixed $H_{2}(X \mid Y)$ in a similar way to the proof of \corref{cor:2_half}.

Furthermore, since $H_{\alpha}(X \mid Y)$ is closely related to Gallager's reliability function $E_{0}$ \cite{gallager} and $\alpha$-mutual information \cite{sibson, ho} (cf. \cite{arimoto, ita}), we can also establish sharp bounds on them in some situation, as with \cite[Theorem~5]{part2}.

\appendices

\section{Proof of \lemref{lem:sgn_g}}
\label{app:sgn_g}

\begin{IEEEproof}[Proof of \lemref{lem:sgn_g}]
The identity $g(n, z; r, s) = - g(n, z; s, r)$ is trivial from the definition \eqref{def:g}.
Hence, suppose throughout this proof that $0 < r < s < \infty$, and we only consider the function $g(n, z; r, s)$.
We first prove the first assertion of \lemref{lem:sgn_g} for $z \in (0, 1)$.
It is clear that if $z \in (0, 1)$, then $r \mapsto z^{r}$ is strictly decreasing for $r \in \mathbb{R}$.
In addition, for each fixed $z \in (0, 1) \cup (1, \infty)$, the function $r \mapsto \ln_{r} z$ is strictly decreasing for $r \in \mathbb{R}$ (cf. \cite[Lemma~1]{part2}).
Therefore, we obtain
\begin{align}
g(n, z; r, s)
& \overset{\eqref{def:g}}{=}
\big( z^{r} + (n-1) \big) \, \ln_{r} z - \big( z^{s} + (n-1) \big) \, \ln_{s} z
\\
& >
\big( z^{r} + (n-1) \big) \, \ln_{r} z - \big( z^{r} + (n-1) \big) \, \ln_{r} z
\\
& =
0
\end{align}
for every $n \in \mathbb{N}$, $z \in (0, 1)$, and $0 < r < s < \infty$, which is the first assertion of \lemref{lem:sgn_g} for $z \in (0, 1)$.

We next consider the second and third assertions of \lemref{lem:sgn_g} for $z \in (1, \infty)$.
Consider two functions
\begin{align}
f_{1}(z; r, s)
& \coloneqq
z^{r} \ln_{r} z - z^{s} \ln_{s} z ,
\label{def:f1} \\
f_{2}(n, z; r, s)
& \coloneqq
(n-1) \big( \ln_{r} z - \ln_{s} z \big)
\label{def:f2}
\end{align}
satisfying
\begin{align}
g(n, z; r, s)
=
f_{1}(z; r, s) + f_{2}(n, z; r, s) .
\label{eq:g_f1_f2}
\end{align}
Direct calculations show the following derivatives:
\begin{align}
\frac{ \partial f_{1}(z; r, s) }{ \partial z }
& \overset{\eqref{def:f1}}{=}
\frac{ \partial }{ \partial z } \Big( z^{r} \ln_{r} z - z^{s} \ln_{s} z \Big)
\\
& =
\bigg( \frac{ \partial z^{r} }{ \partial z } \bigg) \ln_{r} z + z^{r} \bigg( \frac{ \partial \ln_{r} z }{ \partial z } \bigg) - \bigg( \frac{ \partial z^{s} }{ \partial z } \bigg) \ln_{s} z - z^{s} \bigg( \frac{ \partial \ln_{s} z }{ \partial z } \bigg)
\\
& =
r \, z^{r-1} \, \ln_{r} z + z^{r} \, z^{-r} - s \, z^{s-1} \, \ln_{s} z - z^{s} \, z^{-s}
\\
& =
r \, z^{r-1} \, \ln_{r} z + 1 - s \, z^{s-1} \, \ln_{s} z - 1
\\
& =
r \, z^{r-1} \, \ln_{r} z - s \, z^{s-1} \, \ln_{s} z
\\
& =
r \, z^{r-1} \, \frac{ z^{1-r} - 1 }{ 1 - r } - s \, z^{s-1} \, \frac{ z^{1-s} - 1 }{ 1 - s }
\\
& =
\frac{ r }{ r-1 } \big( z^{r-1} - 1 \big) - \frac{ s }{ s-1 } \big( z^{s-1} - 1 \big)
\label{eq:diff1_f1} \\
\frac{ \partial^{2} f_{1}(z; r, s) }{ \partial z^{2} }
& \overset{\eqref{eq:diff1_f1}}{=}
\frac{ \partial }{ \partial z } \bigg( \frac{ r }{ r-1 } \big( z^{r-1} - 1 \big) - \frac{ s }{ s-1 } \big( z^{s-1} - 1 \big) \bigg)
\\
& =
\frac{ r }{ r-1 } \bigg( \frac{ \partial z^{r-1} }{ \partial z } \bigg) - \frac{ s }{ s-1 } \bigg( \frac{ \partial z^{s-1} }{ \partial z } \bigg)
\\
& =
r \, z^{r-2} - s \, z^{s-2} ,
\label{eq:diff2_f1} \\
\frac{ \partial f_{2}(n, z; r, s) }{ \partial z }
& \overset{\eqref{def:f2}}{=}
\frac{ \partial }{ \partial z } \Big( (n-1) \big( \ln_{r} z - \ln_{s} z \big) \Big)
\\
& =
(n-1) \bigg[ \bigg( \frac{ \partial \ln_{r} z }{ \partial z } \bigg) - \bigg( \frac{ \partial \ln_{s} z }{ \partial z } \bigg) \bigg]
\\
& =
(n-1) \, \big( z^{-r} - z^{-s} \big) ,
\label{eq:diff1_f2} \\
\frac{ \partial^{2} f_{2}(n, z; r, s) }{ \partial z^{2} }
& \overset{\eqref{eq:diff1_f2}}{=}
\frac{ \partial }{ \partial z } \Big( (n-1) \, \big( z^{-r} - z^{-s} \big) \Big)
\\
& =
(n-1) \bigg[ \bigg( \frac{ \partial z^{-r} }{ \partial z } \bigg) - \bigg( \frac{ \partial z^{-s} }{ \partial z } \bigg) \bigg]
\\
& =
(n-1) \Big( s \, z^{-(1+s)} - r \, z^{-(1+r)} \Big) .
\label{eq:diff2_f2}
\end{align}
We readily see that
\begin{align}
g(n, 1; r, s)
& \overset{\eqref{def:g}}{=}
\Big( \big( z^{r} + (n-1) \big) \, \ln_{r} z - \big( z^{s} + (n-1) \big) \, \ln_{s} z \Big) \Big|_{z = 1}
\\
& =
(1 + (n-1)) \, \underbrace{ (\ln_{r} 1) }_{ = 0 } \, - \, (1 + (n-1)) \, \underbrace{ (\ln_{s} 1) }_{ = 0 }
\\
& =
0 ,
\label{eq:g_z1} \\
\lim_{z \to \infty} g(n, z; r, s)
& \overset{\eqref{def:g}}{=}
\lim_{z \to \infty} \Big( \big( z^{r} + (n-1) \big) \ln_{r} z - \big( z^{s} + (n-1) \big) \ln_{s} z \Big)
\\
& =
\lim_{z \to \infty} z \, \bigg( z^{r-1} \, \ln_{r} z - (n-1) \, \bigg( \frac{ \ln_{r} z }{ z } \bigg) - z^{s-1} \, \ln_{s} z + (n-1) \, \bigg( \frac{ \ln_{s} z }{ z } \bigg) \bigg)
\\
& \overset{\text{(a)}}{=}
\lim_{z \to \infty} z \, \Big( z^{r-1} \, \ln_{r} z - z^{s-1} \, \ln_{s} z \Big)
\\
& \overset{\text{(b)}}{=}
\lim_{z \to \infty} z \, \Big( \ln_{s} \Big( \frac{1}{z} \Big) - \ln_{r} \Big( \frac{1}{z} \Big) \Big)
\\
& =
\lim_{u \to 0^{+}} \bigg( \frac{ \ln_{s} u - \ln_{r} u }{ u } \bigg)
\\
& \overset{\text{(c)}}{=}
- \infty ,
\label{eq:lim_g_z_infty}
\\
\frac{ \partial g(n, z; r, s) }{ \partial z } \bigg|_{z = 1}
& \overset{\eqref{eq:g_f1_f2}}{=}
\frac{ \partial f_{1}(z; r, s) }{ \partial z } \bigg|_{z = 1} + \frac{ \partial f_{2}(n, z; r, s) }{ \partial z } \bigg|_{z = 1}
\\
& \overset{\eqref{eq:diff1_f1}}{=}
\bigg( \frac{ r }{ r-1 } \big( z^{r-1} - 1 \big) - \frac{ s }{ s-1 } \big( z^{s-1} - 1 \big) \bigg) \bigg|_{z = 1} + \frac{ \partial f_{2}(n, z; r, s) }{ \partial z } \bigg|_{z = 1}
\\
& \overset{\eqref{eq:diff1_f2}}{=}
\bigg( \frac{ r }{ r-1 } \big( z^{r-1} - 1 \big) - \frac{ s }{ s-1 } \big( z^{s-1} - 1 \big) \bigg) \bigg|_{z = 1} + \Big( (n-1) \, \big( z^{-r} - z^{-s} \big) \Big) \Big|_{z = 1}
\\
& =
\frac{ r }{ r-1 } (1 - 1) - \frac{ s }{ s-1 } (1 - 1) + (n-1) \, (1 - 1)
\\
& =
0 ,
\label{eq:sgn_g_diff1_1} \\
\sgn \bigg( \frac{ \partial^{2} g(n, z; r, s) }{ \partial z^{2} } \bigg|_{z = 1} \bigg)
& \overset{\eqref{eq:g_f1_f2}}{=}
\sgn \bigg( \frac{ \partial^{2} f_{1}(z; r, s) }{ \partial z^{2} } + \frac{ \partial^{2} f_{2}(n, z; r, s) }{ \partial z^{2} } \bigg) \bigg|_{z = 1}
\\
& \overset{\eqref{eq:diff2_f1}}{=}
\sgn \bigg( \Big( r \, z^{r-2} - s \, z^{s-2} \Big) + \frac{ \partial^{2} f_{2}(n, z; r, s) }{ \partial z^{2} } \bigg) \bigg|_{z = 1}
\\
& \overset{\eqref{eq:diff2_f2}}{=}
\sgn \bigg( \Big( r \, z^{r-2} - s \, z^{s-2} \Big) + (n-1) \, \Big( s \, z^{-(1+s)} - r \, z^{-(1+r)} \Big) \bigg) \bigg|_{z = 1}
\\
& =
\sgn \Big( \big( r - s \big) + (n-1) \, (s - r) \Big)
\\
& =
\sgn(n-2) \, \sgn(s-r)
\\
& =
\begin{cases}
0
& \mathrm{if} \ n = 2 ,
\\
1
& \mathrm{if} \ n \ge 3
\end{cases}
\label{eq:sgn_g_diff2_1}
\end{align}
for every $n \in \mathbb{N}_{\ge 2}$ and $0 < r < s < \infty$, where (a) follows from the limiting value
\begin{align}
\lim_{x \to \infty} \bigg( \frac{ \ln_{q} x }{ x } \bigg)
& =
\begin{cases}
0
& \mathrm{if} \ q > 0 ,
\\
1
& \mathrm{if} \ q = 0 ,
\\
\infty
& \mathrm{if} \ q < 0 ,
\end{cases}
\end{align}
(b) follows from the fact that
\begin{align}
\ln_{q} x
& =
- x^{1-q} \, \ln_{q} \Big( \frac{1}{x} \Big) ,
\end{align}
and (c) follows from the limiting value
\begin{align}
\lim_{u \to 0^{+}} \Big( \ln_{s} u - \ln_{r} u \Big)
& =
\begin{dcases}
- \infty
& \mathrm{if} \ s > 1 ,
\\
\frac{ r-s }{ (1-r) \, (1-s) }
& \mathrm{if} \ s < 1
\end{dcases}
\end{align}
for every $0 < r < s < \infty$.

In particular, if $n = 2$, then we get
\begin{align}
\frac{ \partial^{2} g(2, z; r, s) }{ \partial z^{2} }
& \overset{\eqref{eq:g_f1_f2}}{=}
\frac{ \partial^{2} f_{1}(z; r, s) }{ \partial z^{2} } + \frac{ \partial^{2} f_{2}(2, z; r, s) }{ \partial z^{2} }
\\
& \overset{\eqref{eq:diff2_f1}}{=}
\Big( r \, z^{r-2} - s \, z^{s-2} \Big) + \frac{ \partial^{2} f_{2}(2, z; r, s) }{ \partial z^{2} }
\\
& \overset{\eqref{eq:diff2_f2}}{=}
\Big( r \, z^{r-2} - s \, z^{s-2} \Big) + \Big( s \, z^{-(1+s)} - r \, z^{-(1+r)} \Big)
\\
& =
\frac{ r \, (z^{r} - z^{1-r}) - s \, (z^{s} - z^{1-s}) }{ z^{2} }
\\
& \overset{\text{(a)}}{<}
\frac{ (r-s) \, (z^{r} - z^{1-r}) }{ z^{2} }
\\
& \overset{\text{(b)}}{\le}
0 \qquad \qquad \mathrm{if} \ r \ge 1/2
\label{eq:g_diff2_n2}
\end{align}
for every $z \in (1, \infty)$ and $1/2 \le r < s < \infty$, where (a) and (b) follow from the facts that
\begin{itemize}
\item
for each fixed $z \in (1, \infty)$, the function $t \mapsto z^{t} - z^{1-t}$ is strictly increasing for $t \in \mathbb{R}$,
\item
$r - s < 0$ whenever $r < s$,
\item
$(z^{t} - z^{1-t})|_{t = 1/2} = \sqrt{z} - \sqrt{z} = 0$ for every $z \in [0, \infty)$.
\end{itemize}
It follows from \eqref{eq:sgn_g_diff1_1}, \eqref{eq:sgn_g_diff2_1}, and \eqref{eq:g_diff2_n2} that for each fixed $1/2 \le r < s < \infty$, the function $z \mapsto g(2, z; r, s)$ is strictly decreasing for $z \in [1, \infty)$;
and therefore, we observe from \eqref{eq:g_z1} that
\begin{align}
g(2, z; r, s)
<
0
\end{align}
for every $z \in (1, \infty)$ and $1/2 \le r < s < \infty$, which is the second assertion of \lemref{lem:sgn_g}.

We further consider the third assertion of \lemref{lem:sgn_g}, i.e., the case: $n \in \mathbb{N}_{\ge 3}$.
It follows from \eqref{eq:sgn_g_diff1_1} and \eqref{eq:sgn_g_diff2_1} that for any $n \in \mathbb{N}_{\ge 3}$ and $0 < r < s < \infty$, there exists $\eta(n; r, s) \in (1, \infty)$ such that
\begin{align}
\sgn\bigg( \frac{ \partial g(n, z; r, s) }{ \partial z } \bigg)
& =
\begin{cases}
0
& \mathrm{if} \ z = 1 ,
\\
1
& \mathrm{if} \ 1 < z < \eta(n; r, s) ,
\end{cases}
\label{eq:eta1}
\end{align}
which implies that $z \mapsto g(n, z; r, s)$ is strictly increasing for $z \in [1, \eta(z; r, s)]$.
By this strict monotonicity, it follows from \eqref{eq:g_z1} that
\begin{align}
\sgn\Big( g(n, z; r, s) \Big)
& =
\begin{cases}
0
& \mathrm{if} \ z = 1 ,
\\
1
& \mathrm{if} \ 1 < z \le \eta(n; r, s) .
\end{cases}
\label{eq:eta}
\end{align}
for every $n \in \mathbb{N}_{\ge 3}$ and $0 < r < s < \infty$.
From \eqref{eq:lim_g_z_infty} and \eqref{eq:eta}, the intermediate value theorem shows that for any $n \in \mathbb{N}_{\ge 3}$ and $0 < r < s < \infty$, there exists $\zeta(n; r, s) \in (\eta_{1}(n; r, s), \infty)$ such that 
\begin{align}
\sgn\Big( g(n, z; r, s) \Big)
& =
\begin{cases}
0
& \mathrm{if} \ z = 1 \ \mathrm{or} \ z = \zeta(n; r, s) ,
\\
1
& \mathrm{if} \ 1 < z < \zeta(n; r, s) .
\end{cases}
\label{eq:zeta}
\end{align}
It is clear from \eqref{eq:zeta} that
\begin{align}
\frac{ \partial g(n, z; r, s) }{ \partial z } \bigg|_{z = \zeta(n; r, s)}
\le
0
\label{eq:diff1_g_at_zeta}
\end{align}
for every $n \in \mathbb{N}_{\ge 3}$ and $0 < r < s < \infty$.
Since
\begin{align}
g(n, z; r, s)
& \overset{\eqref{eq:g_f1_f2}}{=}
f_{1}(z; r, s) + f_{2}(n, z; r, s)
\\
& \overset{\eqref{def:f2}}{=}
f_{1}(z; r, s) + (n-1) \, \Big( \ln_{r} z - \ln_{s} z \Big)
\\
& \overset{\eqref{def:f2}}{=}
f_{1}(z; r, s) + (n-1) \, f_{2}(2, z; r, s) ,
\label{eq:g_f1_f2_n2}
\end{align}
we get from \eqref{eq:zeta} and \eqref{eq:diff1_g_at_zeta} that
\begin{align}
f_{1} \big( \zeta(n; r, s); r, s \big)
& =
- (n-1) \, f_{2} \big( 2, \zeta(n; r, s); r, s \big) ,
\label{eq:f1f2_zeta} \\
\frac{ \partial f_{1}(z; r, s) }{ \partial z } \bigg|_{z = \zeta(n; r, s)}
& \le
- (n-1) \, \frac{ \partial f_{2}(2, z; r, s) }{ \partial z } \bigg|_{z = \zeta(n; r, s)}
\label{ineq:f1f2_diff1_zeta}
\end{align}
for every $n \in \mathbb{N}_{\ge 3}$ and $0 < r < s < \infty$.
Since \eqref{eq:g_diff2_n2} shows
\begin{align}
\frac{ \partial^{2} f_{1}(z; r, s) }{ \partial z^{2} }
& <
- \frac{ \partial^{2} f_{2}(2, z; r, s) }{ \partial z^{2} }
\end{align}
for every $z \in (1, \infty)$ and $1/2 \le r < s < \infty$, it follows from \eqref{ineq:f1f2_diff1_zeta} that
\begin{align}
\frac{ \partial f_{1}(z; r, s) }{ \partial z }
<
- (n-1) \, \frac{ \partial f_{2}(2, z; r, s) }{ \partial z }
\end{align}
for every $n \in \mathbb{N}_{\ge 3}$, $z > \zeta_{1}(n; r, s)$, and $1/2 \le r < s < \infty$;
thus, we have from \eqref{eq:f1f2_zeta} that
\begin{align}
f_{1} \big( z; r, s \big)
& <
- (n-1) \, f_{2} \big( 2, z; r, s \big)
\label{ineq:f1f2_zeta}
\end{align}
for every $n \in \mathbb{N}_{\ge 3}$, $z > \zeta(n; r, s)$, and $1/2 \le r < s < \infty$. 
Therefore, combining \eqref{eq:zeta}, \eqref{eq:g_f1_f2_n2}, and \eqref{ineq:f1f2_zeta}, we have
\begin{align}
\sgn \Big( g(n, z; r, s) \Big)
=
\begin{cases}
-1
& \mathrm{if} \ \zeta(n; r, s) < z < \infty ,
\\
0
& \mathrm{if} \ z = 1 \ \mathrm{or} \ z = \zeta(n; r, s) ,
\\
1
& \mathrm{if} \ 1 < z < \zeta(n; r, s)
\end{cases}
\end{align}
for every $n \in \mathbb{N}_{\ge 3}$, $z \in [1, \infty)$, and $1/2 \le r < s < \infty$, which is the third assertion of \lemref{lem:sgn_g}.
This completes the proof of \lemref{lem:sgn_g}.
\end{IEEEproof}


\begin{thebibliography}{99}

\bibitem{arikan}
E.~Ar{\i}kan,
``An inequality on guessing and its application to sequential decoding,''
\emph{IEEE\ Trans.\ Inf.\ Theory},
vol.~42, no.~1, pp.~99--105, Jan.~1996.

\bibitem{arimoto}
S.~Arimoto,
``Information measures and capacity of order $\alpha$ for discrete memoryless channels,''
in \emph{Topics in Information Theory, 2nd Colloq. Math. Soc. J. Bolyai},
Keszthely, Hungary, vol.~16, pp.~41--52, 1977.

\bibitem{behara}
M.~Behara and J.~S.~Chawla,
``Generalized $\gamma$-entropy,''
\emph{Entropy and Ergodic Theory: Selecta Statistica Canadiana.}
vol.~2, pp.~15--38, 1974.

\bibitem{ben-bassat}
M.~Ben-Bassat and J.~Raviv,
``R\'{e}nyi entropy and probability of error,''
\emph{IEEE\ Trans.\ Inf.\ Theory},
vol.~24, no.~3, pp.~324--331, May~1978.

\bibitem{boekee}
D.~E.~Boekee and J.~C.~A.~Van~der~Lubbe,
``The $R$-norm information measure,''
\emph{Inf.\ Control},
vol.~45, no.~2, pp.~136--155, May 1980.

\bibitem{bunte}
C.~Bunte and A.~Lapidoth,
``Encoding tasks and R\'{e}nyi entropy,''
\emph{IEEE\ Trans.\ Inf.\ Theory},
vol.~60, no.~9, pp.~5065--5076, Sept.~2014.

\bibitem{campbell}
L.~L.~Campbell,
``A coding theorem and R\'{e}nyi entropy,
\emph{Inf.\ Control},
vol.~8, no.~4, pp.~423--429, Aug.~1965.

\bibitem{cover}
T.~M.~Cover and J.~A.~Thomas,
\emph{Elements of Information Theory.}
2nd ed., New~York: Wiley, 2006.

\bibitem{csiszar}
I.~Csisz\'{a}r,
``Generalized cutoff rates and R\'{e}nyi information measures,''
\emph{IEEE\ Trans.\ Inf.\ Theory},
vol.~41, no.~1, pp.~26--34, Jan.~1995.

\bibitem{daroczy}
Z.~Dar\'{o}czy,
``Generalized information functions,''
\emph{Inf.\ Control},
vol.~16, no.~1, pp.~36--51, Mar. 1970.

\bibitem{fano}
R.~M.~Fano,
``Class notes for Transmission of Information,''
Course~6.574, MIT, Cambridge, MA, 1952.

\bibitem{feder}
M.~Feder and N.~Merhav,
``Relations between entropy and error probability,''
\emph{IEEE\ Trans.\ Inf.\ Theory},
vol.~40, no.~1, pp.~259--266, Jan. 1994.

\bibitem{fehr}
S.~Fehr and S.~Berens,
``On the conditional R\'{e}nyi entropy,''
\emph{IEEE\ Trans.\ Inf.\ Theory},
vol.~60, no.~11, pp.~6801--6810, Nov.~2014.

\bibitem{gallager}
R.~G.~Gallager,
\emph{Information Theory and Reliable Communication.}
New~York: Wiley, 1968.

\bibitem{fabregas}
A.~Guill\'{e}n~i~F\`{a}bregas, I.~Land, and A.~Martinez,
``Extremes of error exponents,''
\emph{IEEE\ Trans.\ Inf.\ Theory},
vol.~59, no.~4, pp.~2201--2207, Apr. 2013.

\bibitem{harremoes}
P.~Harremo\"{e}s,
``Joint range of R\'{e}nyi entropies,''
\emph{Kybernetika},
vol.~45, no.~6, pp.~901--911, 2009.

\bibitem{topsoe}
P.~Harremo\"{e}s and F.~Tops{\o}e,
``Inequalities between entropy and index of coincidence derived from information diagrams,''
\emph{IEEE\ Trans.\ Inf.\ Theory},
vol.~47, no.~7, pp.~2944--2960, Nov. 2001.

\bibitem{havrda}
J.~Havrda and F.~Charv\'{a}t,
``Quantification method of classification processes. Concept of structural $a$-entropy,''
\emph{Kybernetika},
vol.~3, no.~1, pp.~30--35, 1967.

\bibitem{hayashi}
M.~Hayashi,
``Exponential decreasing rate of leaked information in universal random privacy amplification,''
\emph{IEEE\ Trans.\ Inf.\ Theory},
vol.~57, no.~6, pp.~3989--4001, June~2011.

\bibitem{verdu}
S.-W.~Ho and S.~Verd\'{u},
``On the interplay between conditional entropy and error probability,''
\emph{IEEE\ Trans.\ Inf.\ Theory}
vol.~56, no.~12, pp.~5930--5942, Dec. 2010.

\bibitem{ho}
---------,
``Convexity/concavity of R\'{e}nyi entropy and $\alpha$-mutual information,''
\emph{Proc. 2015\ IEEE\ Int.\ Symp.\ Inf.\ Theory} (ISIT'2015),
Hong~Kong, pp.~745--749, June~2015.

\bibitem{iwamoto}
M.~Iwamoto and J.~Shikata,
``Information theoretic security for encryption based on conditional R\'{e}nyi entropies,''
\emph{Proc.\ 9th\ Int.\ Conf.\ Inf.\ Theoretic Security} (ICITS),
LNSC8317, pp.~103--121, Springer-Verlag, Jan.~2014.

\bibitem{kovalevsky}
V.~A.~Kovalevsky,
``The problem of character recognition from the point of view of mathematical statistics,''
\emph{Character\ Readers\ and\ Pattern\ Recognition.}
New York: Spartan, pp.~3--30, 1968. (Russian edition in 1965).

\bibitem{marshall}
A.~W.~Marshall and I.~Olkin,
\emph{Inequalities: Theory of Majorization and Its Applications.}
New York: Academic, 1979.

\bibitem{mori}
R.~Mori and T.~Tanaka,
``Source and channel polarization over finite fields and Reed--Solomon matrices,''
\emph{IEEE\ Trans.\ Inf.\ Theory},
vol.~60, no.~5, pp.~2720--2736, May~2014.

\bibitem{renyi}
A.~R\'{e}nyi,
``On measures of information and entropy,''
\emph{Proc.\ 4th\ Berkeley\ Symp.\ Math.\ Statist.\ Prob.},
Berkeley, Calif., vol.~1, Univ.\ of\ Calif. Press, pp.~547--561, 1961.

\bibitem{isit2015}
Y.~Sakai and K.~Iwata,
``Feasible regions of symmetric capacity and Gallager's $E_{0}$ function for ternary-input discrete memoryless channels,''
\emph{Proc.\ 2015\ IEEE\ Int.\ Symp.\ Inf.\ Theory} (ISIT'2015),
Hong~Kong, pp.~81--85, June 2015.

\bibitem{itw2016_reject}
---------,
``Sharp bounds between two R\'{e}nyi entropies of distinct positive orders,''
May~2016.
[Online]. Available at \url{https://arxiv.org/abs/1605.00019}.

\bibitem{part2}
---------,
``Relations between conditional Shannon entropy and expectation of $\ell_{\alpha}$-norm,''
\emph{Proc. 2016 IEEE\ Int.\ Symp.\ Inf.\ Theory}, (ISIT'2016)
Barcelona, Spain, pp.~1641--1645, July~2016.

\bibitem{part1}
---------,
``Extremal relations between shannon entropy and $\ell_{\alpha}$-norm,''
\emph{Proc.\ 2016\ Int.\ Symp.\ Inf.\ Theory\ Appl.}, (ISITA'2016),
Monterey, CA, USA, pp.~433--437, Jan.~2016.

\bibitem{sasoglu}
E.~\c{S}a\c{s}o\u{g}lu,
``Polarization and polar codes,''
\emph{Found.\ Trends\ Commum.\ Inf.\ Theory},
vol.~8, no.~4, pp.~259--381, Oct.~2012.

\bibitem{coupling}
I.~Sason,
``Entropy bounds for discrete random variables via maximal coupling,''
\emph{IEEE\ Trans.\ Inf.\ Theory},
vol.~59, no.~11, pp.~7118--7131, Nov.~2013.

\bibitem{sason}
I.~Sason and S.~Verd\'{u},
``Arimoto--R\'{e}nyi conditional entropy and Bayesian $M$-ary hypothesis testing,''
submitted to \emph{2017\ IEEE\ Int.\ Symp.\ Inf.\ Theory},
Jan.~2017.
[Online]. Available at \url{https://arxiv.org/abs/1701.01974}.

\bibitem{shannon}
C.~E.~Shannon,
``A mathematical theory of communication,''
\emph{Bell\ Syst.\ Tech.\ J.},
vol.~27, pp.~379--423 and 623--656, July and Oct. 1948.

\bibitem{sibson}
R.~Sibson,
``Information radius,''
\emph{Z.\ Wahrsch.\ Verw.\ Geb.},
vol.~14, no.~2, pp.~149--160, June~1969.

\bibitem{tebbe}
D.~L.~Tebbe and S.~J.~Dwyer~III,
``Uncertainty and probability of error,''
\emph{IEEE\ Trans.\ Inf.\ Theory},
vol.~14, no.~3, pp.~516--518, May 1968.

\bibitem{teixeira}
A.~Teixeira, A.~Matos, and L.~Antunes,
``Conditional R\'{e}nyi entropies,''
\emph{IEEE\ Trans.\ Inf.\ Theory},
vol.~58, no.~7, pp.~4273--4277, July~2012.

\bibitem{tomamichel}
M.~Tomamichel and M.~Hayashi,
``Operational interpretation of R\'{e}nyi conditional mutual information via composite hypothesis testing against Markov distributions,
\emph{Proc. 2016 IEEE\ Int.\ Symp.\ Inf.\ Theory}, (ISIT'2016)
Barcelona, Spain, pp.~585--589, July~2016.

\bibitem{tsallis2}
C.~Tsallis,
``Possible generalization of Boltzmann-Gibbs statistics,''
\emph{J.~Statist.~Phys.},
vol.~52, no.~1--2, pp.~479--487, 1988.

\bibitem{tsallis}
---------,
``What are the numbers that experiments provide?''
\emph{Qu\'{i}mica Nova},
vol.~17, no.~6, pp.~468--471, 1994.

\bibitem{ita}
S.~Verd\'{u},
``$\alpha$-mutual information,''
\emph{Proc.\ 2015\ Inf.\ Theory\ Appl.\ Workshop} (ITA'2015),
San Diego, CA, USA, pp.~1--6, Feb.~2015.

\end{thebibliography}
\end{document}